\documentclass[11pt,letterpaper]{article}

\usepackage[T1]{fontenc}
\usepackage[utf8]{inputenc}
\usepackage{textcomp}
\usepackage{amsmath}
\usepackage{mathptmx}                 %
\usepackage[scaled=0.92]{helvet}

\usepackage[letterpaper,margin=1in,footskip=0.55in]{geometry}
\usepackage{microtype}
\usepackage{graphicx}
\usepackage{xcolor}
\usepackage{booktabs}
\usepackage{array}
\usepackage{tabularx}
\usepackage{longtable}
\usepackage{framed}

\usepackage[super,comma,sort&compress]{natbib}

\usepackage[font=small,labelfont=bf,labelsep=period]{caption}
\usepackage{titlesec}
\titleformat{\section}{\large\bfseries}{\thesection}{0.6em}{}
\titlespacing*{\section}{0pt}{20pt plus 4pt minus 2pt}{8pt plus 2pt}
\titleformat{\subsection}{\normalsize\bfseries\itshape}{\thesubsection}{0.6em}{}
\titlespacing*{\subsection}{0pt}{14pt plus 3pt minus 2pt}{5pt plus 1pt}

\definecolor{linknavy}{RGB}{18,58,122}
\usepackage[colorlinks=true,linkcolor=linknavy,citecolor=linknavy,
            urlcolor=linknavy,breaklinks=true]{hyperref}
\usepackage{xurl}
\hypersetup{
  pdftitle={Reducing belief in conspiracy theories as they unfold using large language models},
  pdfauthor={Thomas H. Costello, Nathaniel Rabb, Michael Nicholas Stagnaro, Gordon Pennycook, David G. Rand},
  pdfsubject={Forthcoming in PNAS},
  pdfkeywords={conspiracy theories, artificial intelligence, belief change}
}

\definecolor{shadecolor}{gray}{0.95}
\newenvironment{promptbox}
  {\begin{snugshade}\footnotesize\ttfamily\frenchspacing\raggedright
   \setlength{\parindent}{0pt}\hbadness=10000}
  {\end{snugshade}}
\newenvironment{infobox}
  {\begin{snugshade}\small\setlength{\parindent}{0pt}\sloppy\hbadness=10000}
  {\end{snugshade}}

\begin{document}

\begin{center}
  {\small\itshape Forthcoming in PNAS\par}
  \vspace{12pt}
  {\LARGE\bfseries\boldmath
   Reducing belief in conspiracy theories as they unfold\\
   using large language models\par}

  \vspace{18pt}

  {\large
   Thomas H. Costello$^{1,\ast}$, Nathaniel Rabb$^{2,\ast}$,
   Michael Nicholas Stagnaro$^{2}$,\\[2pt]
   Gordon Pennycook$^{3}$, \& David G. Rand$^{2,3,4,5}$\par}

  \vspace{6pt}
  {\small $^{\ast}$Contributed equally\par}

  \vspace{10pt}
  {\small\itshape
   $^{1}$Department of Social and Decision Sciences, Carnegie Mellon University, Pittsburgh, PA 15213\\
   $^{2}$Sloan School of Management, Massachusetts Institute of Technology, Cambridge, MA 02142, USA\\
   $^{3}$Department of Psychology, Cornell University, Ithaca, NY 14853, USA\\
   $^{4}$Department of Information Science, Cornell University, Ithaca, NY 14853, USA\\
   $^{5}$Marketing and Management Communications, SC Johnson School of Business, Cornell University,\\
   Ithaca, NY 14853, USA\par}

  \vspace{10pt}
  {\small Corresponding author: David Rand, 208 Gates Hall, 107 Hoy Rd, Ithaca, NY 14850.
   Email: \href{mailto:dgr7@cornell.edu}{dgr7@cornell.edu}\par}
\end{center}

\vspace{6pt}
\noindent\rule{\textwidth}{0.4pt}
\vspace{2pt}

{\small
\noindent\textbf{Author Contributions:} All authors conceived of the study. All authors designed
the experiments. THC, NR, and DGR conducted the analysis. All authors contributed to the writing.

\smallskip
\noindent\textbf{Competing Interest Statement:} The authors have no conflicts of interest to declare.

\smallskip
\noindent\textbf{Data and Code Availability:} Code and data are available at \url{https://osf.io/dmu9k}.

\smallskip
\noindent\textbf{Classification:} Social Sciences; Psychological and Cognitive Sciences

}

\vspace{2pt}
\noindent\rule{\textwidth}{0.4pt}

\section{Abstract}

The emergence of conspiracy theories in the wake of major events is a significant societal
challenge. Here we test whether conversational dialogues with a large language model (LLM) can
reduce belief in immediately unfolding conspiracies. In experiments conducted in the days
following the July 2024 assassination attempt on Donald Trump and the September 2025
assassination of Charlie Kirk, U.S. adults (Experiment~1: $N = 472$; Experiment~2: $N = 1042$)
holding conspiratorial views about the crisis event engaged in a multi-turn conversation with an
LLM prompted to reduce their conspiracy belief. Compared to control participants who either
discussed an irrelevant topic with an LLM or viewed a static fact sheet, participants in the LLM
treatment showed significantly reduced conspiracy beliefs in both experiments. We also found
some suggestive evidence of downstream effects of the LLM treatment, observing reduced belief in
some---but not all---analyses of different conspiratorial beliefs one to two months later.
These results shed
light on the psychology of emerging conspiracies and highlight the potential for scalable,
cognitively-focused interventions to counteract misinformation in the immediate aftermath of
high-profile societal events.

\clearpage

\section{Significance Statement}

The rapid emergence of conspiracy theories following major events poses urgent challenges for
democracy and public trust. Using the 2024 assassination attempts on Donald Trump and the 2025
assassination of Charlie Kirk as case studies, we show that short, targeted conversations with a
large language model (LLM) can meaningfully reduce belief in conspiracies as they emerge, with
some suggestive evidence of carry-over to how people evaluated related subsequent events. These
results demonstrate the potential for AI dialogues to be deployed in real time to counter
emerging conspiracies.

\section{Introduction}

Many of the best-known unsubstantiated conspiracy theories are historical, having circulated for
decades (e.g., JFK assassination, government UFO cover-up) or centuries (e.g., Illuminati, Jewish
control of finance). As a result, large bodies of evidence debunking these theories have
accumulated. Recent work has shown that large language model (LLM) AI systems, which functionally
serve as a means of querying humanity's collective knowledge,\cite{farrell2025} can leverage
evidence from their training data to substantially and durably reduce conspiracy beliefs via
brief human-AI dialogues.\cite{costello2024science} While many previous interventions meant to
reduce belief in conspiracy theories are largely ineffective,\cite{omahony2023} AI models can
quickly and legibly provide compelling rebuttals to any given conspiracy topic, a task which is
extremely challenging and time-consuming for even the most experienced human would-be debunkers.
Experimental variants of this paradigm that alter the human-AI interaction framing or the AI
system prompts have specifically identified the provision of relevant facts and evidence as the
key ingredient for the AI's success.\cite{costello2025facts} Furthermore, even when participants
believe they are talking to a human expert, the debunking evidence and facts presented by the AI
is effective at undermining conspiracy belief.\cite{boissin2025}

This evidence-based approach is well suited to addressing well-known conspiracy theories because
the models' training data contains the results of pre-existing (typically human) efforts to, for
example, surface compelling evidence, write investigatory reports, and identify logical gaps in
conspiratorial reasoning. However, new influential conspiracy theories often originate and spread
rapidly in the days following societal emergencies or surprising (and typically distressing)
events. Because of the emergent nature of such events, little is typically known publicly in the
immediate aftermath. Thus, it would seem that the AI models' fact-based approach to debunking
would be comparatively ineffective in such settings---even if the model is provided with whatever
information is available, it will have little to work with when little is known. On the other
hand, it may be that even in the absence of clear counter-evidence, the AI is able to effectively
reduce conspiracy beliefs, perhaps by leveraging different approaches such as highlighting the
lack of clear evidence in support of the conspiracy or encouraging critical thinking.
Adjudicating between these possibilities is of both theoretical and practical significance,
shedding light on the psychology of conspiracy beliefs and crisis events and assessing whether or
not AI tools may be useful in combatting emerging conspiracies.

Therefore, we investigate this issue by examining the effects of human-AI debunking dialogues on
conspiracy beliefs in the immediate aftermath of two crisis events: the July 2024 near-miss
assassination attempt on President Donald Trump and the September 2025 assassination of Charlie
Kirk, an American political activist.

Virtually nothing was publicly known about the would-be Trump assassin or his motives in the days
following the assassination attempt. He left no manifesto, had a highly isolated social life, and
had a digital footprint that was sparse and contradictory. Yet, within a week, 50\% of a
nationally representative US sample had heard that the event was staged, and 38\% had heard that
it was planned by Democratic operatives,\cite{ognyanova2024} with 11\% and 12\%, respectively,
believing such theories to be true.

Somewhat more information was publicly available about the man who assassinated Kirk and his
motives in the days after the event, including cryptic messages he sent to his partner and
engraved on bullets. Officials also claimed that he had been ``deeply indoctrinated with leftist
ideology'', although direct evidence of this was not presented, leaving substantial ambiguity.
Accordingly, conspiracy theories proliferated widely in the wake of Kirk's
death,\cite{thompson2025nyt} arguing for example that there was a law enforcement
cover-up,\cite{power2025} that Israel and the Mossad were responsible,\cite{splc2025} or that it
was a false-flag operation.\cite{isd2025}

In the immediate aftermath of each event, we conducted an experiment assessing an LLM's ability
to reduce belief in related conspiracies and descriptively characterizing the persuasion
strategies used by the model. In both experiments, US adults first provided both open-ended and
closed-ended measures of conspiratorial thinking about the event. Participants were then randomly
assigned to (1) a Debunking Dialogue condition in which the LLM tried to reduce their
conspiratorial belief, (2) an Informational List condition in which they were shown a static list
of the facts currently available about the event, or (3) an Irrelevant Dialogue control in which
they discussed a topic unrelated to the event with the LLM. All participants then again completed
the open- and closed-ended measures of conspiratorial thinking. We also tested whether these
debunking conversations had durable effects by examining participants' beliefs about other
conspiracies one to two months later in the context of a subsequent crisis event.

\section{Results}

\subsection{Main effects}

To test the effect of the LLM dialogue intervention on conspiratorial thinking about the Trump
and Kirk assassination attempts, we follow Costello et al.\cite{costello2024science} and
conducted OLS regression with robust standard errors predicting the post-treatment rating using
condition while controlling for pre-treatment rating for each of the four measures (note that
this is mathematically identical to predicting pre-to-post change using condition while
controlling for pre-treatment rating); we also report Cohen's $d$ standardizing effects by pooled
pre-treatment standard deviation. The Trump study, conducted 3--11 days after the event, was not
pre-registered; the primary analyses for the Kirk study, conducted 6--8 days after the event, were
pre-registered.

\begin{figure}[tb]
  \centering
  \includegraphics[width=\textwidth]{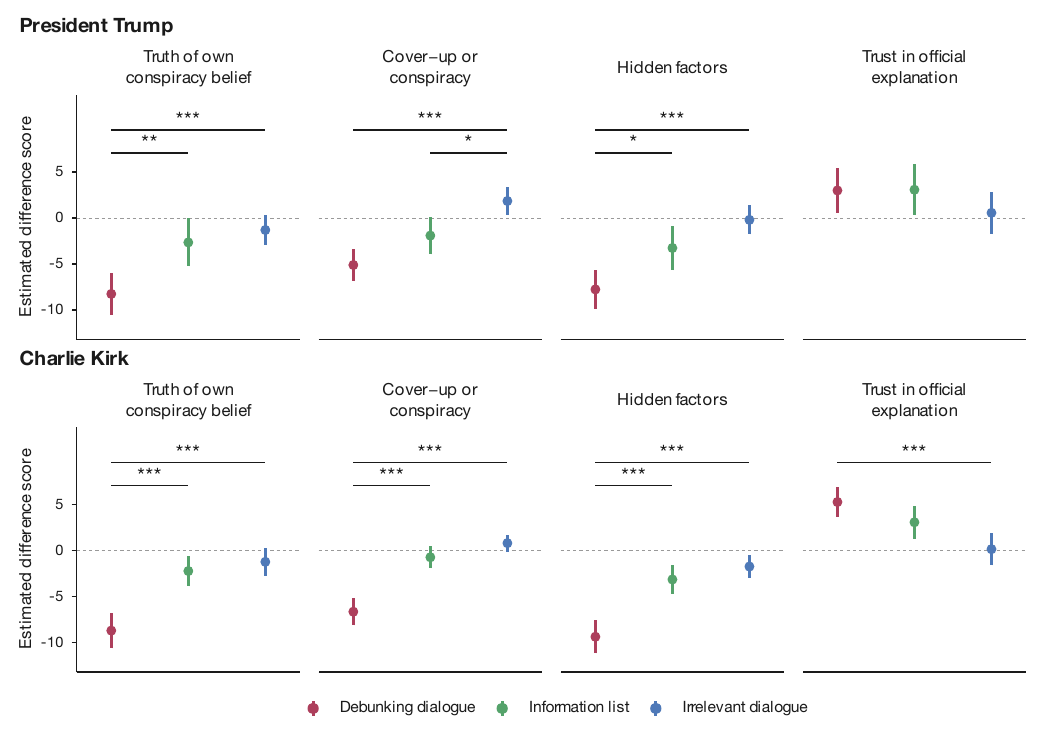}
  \caption{Effects of condition on four measures of conspiratorial beliefs about the first
  attempt to assassinate President Trump (top) or the assassination of Charlie Kirk (bottom).
  Point estimates show predicted difference scores (post-treatment $-$ pre-treatment) by
  condition, adjusting for the pre-treatment rating, with 95\% confidence intervals. Brackets
  mark pairwise comparisons between conditions that are significant after Tukey adjustment for
  the three pairwise comparisons within each panel. $^{*}p \leq .05$, $^{**}p \leq .01$,
  $^{***}p < .001$ (Tukey-adjusted).}
  \label{fig:main}
\end{figure}

In both experiments, the LLM Debunking Dialogue reliably reduced participants' belief (rated
0--100) in the conspiracy theory they had articulated in their own words, both
relative to the Irrelevant Dialogue control (Trump: $b = -6.95$, $p < .001$, $d = .38$; Kirk:
$b = -7.48$, $p < .001$, $d = .37$), and the static Information List (Trump: $b = -5.60$, 95\% CI
$[-9.72, -1.48]$, $p = .004$, $d = .32$; Kirk: $b = -6.48$, 95\% CI $[-9.52, -3.43]$, $p < .001$,
$d = .35$; Figure~\ref{fig:main}). The same pattern was observed in both experiments for
closed-ended questions assessing belief that ``there is a cover-up or conspiracy surrounding [the
assassination event]'' (0--100 scale; Trump vs control: $b = -6.96$, $p < .001$, $d = .27$; Trump
vs information list: $b = -3.20$, 95\% CI $[-6.44, 0.04]$, $p = .054$, $d = .12$; Kirk vs
control: $b = -7.46$, 95\% CI $[-9.55, -5.36]$, $p < .001$, $d = .29$; Kirk vs information list:
$b = -5.92$, 95\% CI $[-8.20, -3.63]$, $p < .001$, $d = .24$) and belief that ``there are hidden
or undisclosed factors behind the reported assassination attempt against [assassination
target]'' (Trump vs control: $b = -7.56$, $p < .001$, $d = .3$; Trump vs information list:
$b = -4.50$, 95\% CI $[-8.37, -0.63]$, $p = .018$, $d = .17$; Kirk vs control: $b = -7.64$, 95\%
CI $[-10.25, -5.02]$, $p < .001$, $d = .31$; Kirk vs information list: $b = -6.23$, 95\% CI
$[-9.07, -3.39]$, $p < .001$, $d = .24$). However, the Debunking Dialogue did not increase trust
in the official explanation for the Trump assassination attempt (vs irrelevant dialogue:
$b = 2.43$, 95\% CI $[-1.58, 6.45]$, $p = .329$; vs information list: $b = -0.08$, 95\% CI
$[-4.53, 4.38]$, $p = .999$), perhaps because there was no clear official explanation at the time
the study was run. The Debunking Dialogue did increase trust in the official explanation for the
Kirk assassination relative to the irrelevant dialogue ($b = 5.13$, 95\% CI $[2.29, 7.97]$,
$p < .001$, $d = .19$; although the difference with the Informational List condition was not
significant: $b = 2.20$, 95\% CI $[-0.67, 5.07]$, $p = .17$, $d = .08$). We also note that the
Debunking Dialogue had no effect on support for political violence in the Kirk study,
$p$s $> .30$, perhaps because it was near floor even in the control
($M_{\text{Debunking Dialogue}} = 12.1$, $M_{\text{Information List}} = 9.93$,
$M_{\text{Irrelevant Dialogue}} = 12.0$). The $p$-values for these immediate effects are
Tukey-adjusted for the three pairwise comparisons within each outcome.

Although these events were highly partisan, and initial belief in a conspiracy theory showed
partisan differences (with Republicans more strongly believing conspiracies; see
\hyperref[sec:sm5]{SM5}), we find no reliable evidence that the treatment effect differed by
party (Trump: $b = 1.40$, 95\% CI $[-0.15, 2.96]$, $p = .076$;
Kirk: $b = -0.07$, 95\% CI $[-1.31, 1.17]$, $p = .91$; see \hyperref[sec:sm6]{SM6}).

As pre-registered in the Kirk experiment, we also test for moderation by baseline belief in the
participant's stated conspiracy belief. In the Trump experiment, there was no significant
interaction between Debunking Dialogue and pre-treatment belief ($b = 0.039$, 95\% CI
$[-0.136, 0.214]$, $p = .66$), while in the Kirk experiment the treatment was somewhat more
effective for participants who initially scored higher on the conspiracy belief ($b = -0.155$,
95\% CI $[-0.300, -0.011]$, $p = .035$; see \hyperref[sec:sm7]{SM7}).

\subsection{LLM rhetorical strategies}

Having demonstrated that the AI was indeed able to reduce conspiracy beliefs in the immediate
aftermath of these crisis events, we now examine which persuasion strategies the AI used to do
so. We developed a qualitative coding procedure that could be executed at scale by an LLM yet
validated by human raters (see Methods for details).

\begin{figure}[tb]
  \centering
  \includegraphics[width=\textwidth]{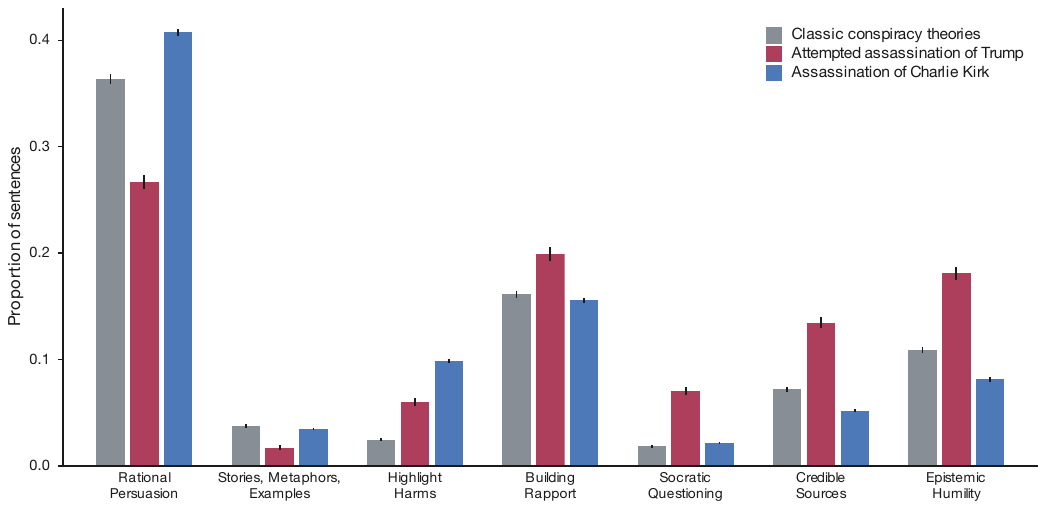}
  \caption{Mean proportion of LLM-side sentences per conversation that used a given strategy for
  debunking classic conspiracy theories (dialogues from Costello et al.,\cite{costello2025facts}
  coded with the same procedure) versus conspiracy theories about the first attempt to
  assassinate Donald Trump and the assassination of Charlie Kirk. Error bars indicate 95\%
  confidence intervals.}
  \label{fig:strategies}
\end{figure}

The results are shown in Figure~\ref{fig:strategies}. For the Trump assassination attempt, where
virtually no information was available about the would-be assassin and his motives, the pattern
for strategy use by the AI was notably different from the strategies that the same coding
procedure identifies in the dialogues about classic conspiracies from Costello et
al.\cite{costello2025facts} To debunk the Trump assassination attempt
conspiracies, the AI relied less on rational persuasion (9.7 percentage points less than in
classic conspiracy conversations, $\chi^2(1) = 487$, $p < .001$) and more on the prudence of
epistemic caution when little is known (7.2 points more, $\chi^2(1) = 544$, $p < .001$), asking
open-ended questions designed to facilitate reflection (5.2 points more, $\chi^2(1) = 1023$,
$p < .001$), and encouraging reliance on credible sources (6.2 points more, $\chi^2(1) = 564$,
$p < .001$). Conversely, when debunking conspiracies related to the Kirk assassination---where
more information about the assassin and his motives was known, albeit still less information
than for most classic conspiracies---the AI employed similar strategies to those used for
debunking classic conspiracies, though it placed somewhat greater emphasis on highlighting harms
(7.4 percentage points more than in classic conspiracy conversations, $\chi^2(1) = 2325$,
$p < .001$).

\begin{table}[p]
  \caption{Persuasion strategy coding categories, descriptions, and examples from the debunking
  dialogues.}
  \label{tab:strategies}
  \small
  \begin{tabularx}{\textwidth}{>{\raggedright\arraybackslash}p{1.45in}XX}
    \toprule
    \textbf{Strategy} & \textbf{Description} & \textbf{Example} \\
    \midrule
    Logic, facts, evidence, reasoning, and/or critical thinking &
    Uses evidence-based reasoning to rebut the conspiracy claim, such as pointing to known
    facts, highlighting lack of evidence, or arguing that a non-conspiratorial explanation is
    more plausible &
    ``His background is being investigated, and while his motives remain unclear, there's no
    evidence to suggest he was part of a larger conspiracy or that the attack was staged.'' \\
    \addlinespace
    Stories, metaphors, or examples &
    Uses an analogy, metaphor, historical example, or illustrative case to make the rebuttal
    more intuitive or concrete &
    ``It's like the classic parable of the blind men and the elephant.'' \\
    \addlinespace
    Building rapport &
    Validates the other's feelings, signals respect, or emphasizes shared goals in order to
    reduce defensiveness and keep the conversation constructive &
    ``It's completely understandable that you, and many others, might feel a healthy dose of
    skepticism when it comes to events like this.'' \\
    \addlinespace
    Emphasizing unknowns &
    Highlights uncertainty, incomplete information, or the need for patience, while
    distinguishing unanswered questions from evidence of a plot &
    ``However, it's crucial to distinguish between the absence of complete transparency and the
    presence of a deliberate cover-up or manipulation of the event itself.'' \\
    \addlinespace
    Highlighting harms &
    Argues that conspiracy thinking can cause harm, such as disrespecting victims, spreading
    misinformation, fueling division, or distracting from real problems &
    ``Focusing on unfounded theories disrespects the victims and their families.'' \\
    \addlinespace
    Socratic questioning &
    Uses questions to prompt reflection, probe the basis of the belief, or clarify what evidence
    would support or undermine it &
    ``What kind of evidence would support or challenge your suspicions?'' \\
    \addlinespace
    Credible sources &
    Evaluates the quality of evidence by directing attention to source reliability, expertise,
    verification, or corroboration across sources &
    ``Always be critical of what you see online and cross-reference information with trusted
    sources.'' \\
    \bottomrule
  \end{tabularx}
\end{table}

\subsection{Effects on subsequent conspiratorial beliefs}

\begin{figure}[tb]
  \centering
  \includegraphics[width=0.85\textwidth]{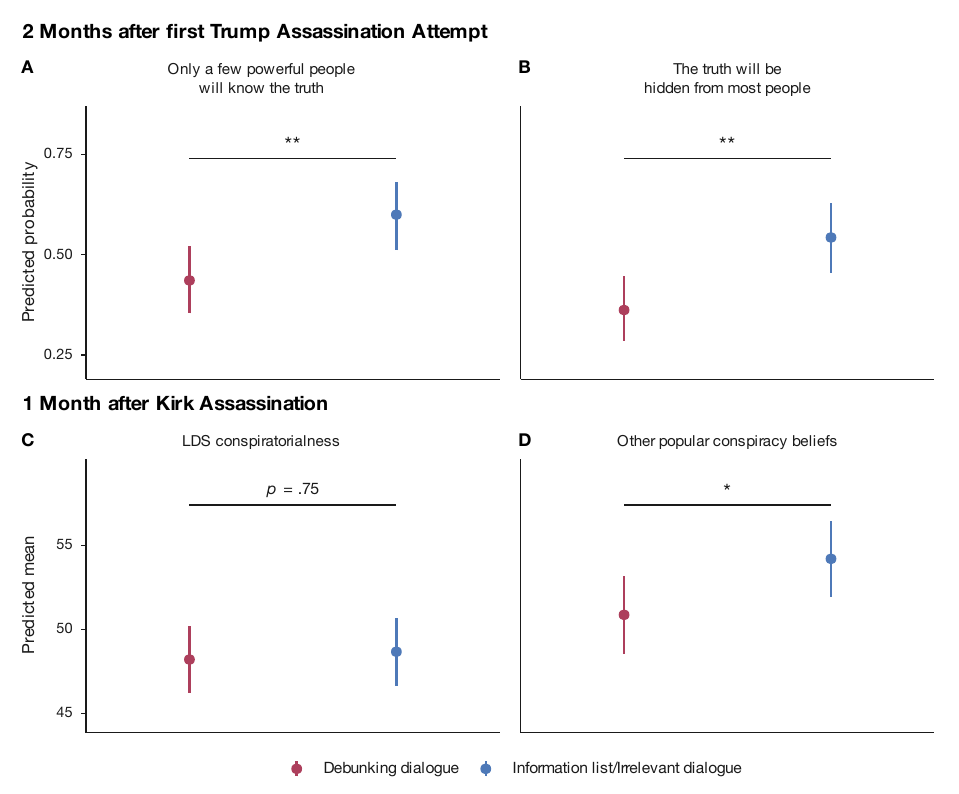}
  \caption{Effects of the Debunking Dialogue on other measures of conspiratorial beliefs after a
  delay. (\textit{A}, \textit{B}) Two months after the first Trump assassination attempt (Study~1
  follow-up): predicted probability of indicating that only a few powerful people will know the
  truth about the second assassination attempt (\textit{A}) and that the truth will be hidden
  from most people (\textit{B}), by initial condition (Information List and Irrelevant Dialogue
  pooled), adjusting for pre-treatment belief. (\textit{C}, \textit{D}) One month after the Kirk
  assassination (Study~2 preregistered follow-up):
  predicted mean conspiratorial beliefs about the LDS church shooting (\textit{C}) and
  endorsement of other popular conspiracy beliefs (\textit{D}), adjusting for participants' pre-treatment rating of their own conspiracy belief about the Kirk
  assassination (the preregistered covariate). Error bars are
  95\% confidence intervals. Brackets show the test of the difference between conditions, with the
  $p$-value printed where $p > .05$. $^{*}p \leq .05$, $^{**}p \leq .01$, $^{***}p < .001$.}
  \label{fig:followup}
\end{figure}

Finally, we ask if debunking conspiracies around these crisis events has effects on subsequent
conspiratorial beliefs (Figure~\ref{fig:followup}). Two months after the first Trump
assassination attempt, a gunman was found hiding on President Trump's property and was
subsequently charged with attempted assassination. We fielded a survey about this second incident
starting 4 days later and opened it to a pool that included all participants from the first Trump
assassination attempt experiment.

We find evidence that the initial debunking affected subsequent conspiratorial thinking. Participants who were in the Debunking Dialogue
condition in the first experiment were significantly less likely than participants in the pooled
control conditions to believe that only a select
few will ever know the truth about the second assassination attempt (OR $= .52$, 95\% CI
$[.31, .84]$, $p = .009$; Figure~\ref{fig:followup}A---including the selection of ``another group
not listed here'' as conspiracy belief yields OR $= .44$, 95\% CI $[.23, .80]$, $p = .008$); and
that the truth of the matter will be hidden from most people (OR $= .48$, 95\% CI $[.29, .78]$,
$p = .004$; Figure~\ref{fig:followup}B). These effects were robust to weighting returning
participants by the inverse of their estimated probability of returning, to account for
possible selective return to the follow-up (see Methods and \hyperref[sec:sm10]{SM10}).

We also conducted a pre-registered follow-up to the Kirk assassination experiment. Two and a half
weeks after Charlie Kirk was assassinated, a mass shooting and arson attack occurred at a Church
of Jesus Christ of Latter-day Saints in Grand Blanc Township, Michigan. We invited participants
from the Kirk experiment to complete a survey measuring beliefs about this incident starting 11
days afterward. We used the same three conspiracy measures from the main Trump and Kirk studies
adapted to this new event, as well as assessing other popular conspiratorial beliefs (e.g., that
hidden groups secretly control world events, that disease outbreaks are deliberately engineered,
and that governments conceal their role in harmful acts against their own citizens). See Methods
for details of the experimental setup.

We pre-registered two different analytic approaches for this follow-up. First, we follow the same
approach as for the Trump follow-up and use linear regression to predict conspiratorial beliefs in
the follow-up survey using condition in the Kirk experiment (a dummy for assignment to the
Debunking Dialogue, pooling Irrelevant Dialogue and Information List) controlling for initial
conspiratorial beliefs about the Kirk assassination; we also report Cohen's $d$ standardizing by
the pooled standard deviation of the outcome. Second, because of concerns that this analysis may
be underpowered, we also pre-registered the persistence analysis approach from Lin et
al.\cite{lin2025} in which we predict conspiratorial beliefs in the follow-up survey using
pre-treatment and post-treatment conspiratorial beliefs about the Kirk assassination for the Kirk
experiment. As shown via simulations in Lin et al.,\cite{lin2025} this approach captures how
beliefs at follow-up are related to the extent of pre-to-post treatment \emph{change} in belief
documented in the Kirk experiment. That is, controlling for initial conspiratorial beliefs, people
whose belief was decreased more should have less belief in the subsequent conspiracy, which is
what the persistence analysis captures. This coefficient reflects the persistence of immediate
belief change about the Kirk assassination from any explanatory source, rather than a treatment
effect per se---but because of the large initial treatment effect, far more immediate belief
change was induced in the Debunking Dialogue arm compared to the control arms. Conceptually,
therefore, this persistence coefficient is the second leg in an indirect path from a mediation
model where the effect of the Debunking Dialogue on belief at follow-up is mediated by immediate
belief change (controlling for pre-treatment belief).\cite{lin2025} In other words, it is not
that we expect changes in the treatment to persist and changes in the control to decay. Instead,
we expect that any initial changes should to some extent persist---and therefore, to the extent
that the treatment induces large immediate change relative to the control, this change should
still be visible to some extent at follow-up. Of course, this is not a causal effect per se. In
particular, this estimator is vulnerable to confounders that affect belief immediately
post-treatment and at follow-up, but not at pre-treatment. Given that pre-treatment and immediate
post-treatment measures were collected only a few minutes apart, whereas follow-up was collected
roughly one month later, we believe that such confounders are unlikely. But we nonetheless
interpret the persistence results as merely suggestive evidence, given the lack of causal
identification.

Among the 872 believers who returned for the follow-up (all analyzed in their assigned condition;
see Methods), the pre-registered intention-to-treat analysis found no significant effect of
assignment to the Kirk Debunking Dialogue (versus the pooled control conditions) on LDS conspiracy belief within
the linear model ($b = -0.46$, 95\% CI $[-3.30, 2.37]$, $p = .75$, $d = -.02$;
Figure~\ref{fig:followup}C). This null may be related to the lower prevalence of conspiracy
beliefs about the LDS shooting (mean across scales $= 48.4$, $SD = 21.3$) relative to conspiracy
beliefs about the attempt to assassinate Trump ($M = 64.8$, $SD = 21.2$) and the assassination
of Charlie Kirk ($M = 63.2$, $SD = 20.8$). Running the same pre-registered model predicting
generic conspiracy beliefs (0--100 scale) found that participants in the Debunking Dialogue did
score lower on generic conspiracy beliefs one month later compared to the controls
($b = -3.33$, 95\% CI $[-6.57, -0.09]$, $p = .044$, $d = -.14$; Figure~\ref{fig:followup}D).
Because there was some evidence that
return to the follow-up was related to condition and post-treatment belief (see Methods), we
also re-estimated these models weighting returning participants by the inverse of their
estimated probability of returning (\hyperref[sec:sm10]{SM10}); with these weights, the
difference in generic conspiracy beliefs was smaller and no longer significant ($b = -2.34$,
95\% CI $[-5.61, 0.92]$, $p = .16$).
Turning to the persistence analyses, we find that, to the extent that a given participant's
conspiratorial beliefs about the Kirk assassination changed from pre- to post-treatment, an
estimated 12\% of that change was still observable in conspiratorial beliefs about the LDS
shooting ($p = .019$) and 19\% in endorsement of other popular conspiracy beliefs ($p < .001$).
The carry-forward percentage did not differ significantly by experimental arm (other popular
conspiracy beliefs: interaction $p = .71$; LDS: interaction $p = .14$); when testing post hoc
whether the effect of the Debunking Dialogue on belief at follow-up is mediated by immediate
belief change (controlling for pre-treatment belief), we find a significant indirect effect of
the Debunking Dialogue on beliefs at follow-up through immediate belief change, for both other
popular conspiracy beliefs (indirect effect $= -1.16$ points, 95\% bootstrap CI
$[-2.11, -0.33]$) and LDS shooting beliefs ($-0.78$, $[-1.54, -0.07]$); and no significant
direct (i.e., sleeper) effects of the Debunking Dialogue ($b = -1.33$, $p = .39$, and
$b = 0.61$, $p = .66$, respectively; \hyperref[sec:sm12]{SM12}). The persistence and
mediation estimates were essentially unchanged by the weighting for selective return
(\hyperref[sec:sm10]{SM10}).

\section{Discussion}

Our results indicate that the ``debunkbot'' LLM was, perhaps surprisingly, quite effective at
reducing beliefs in conspiracy theories that emerge in the immediate aftermath of two
consequential crisis events. In addition to decreasing confidence in the stated conspiracy beliefs
of the participants, the dialogue---which was relatively short ($M = 6.9$ minutes)---reduced
suspicions of a cover-up and concerns about hidden factors connected to these events. We also
find some suggestive evidence that the increased skepticism adopted by participants may have carried over to
their subsequent belief in other conspiracies (although the evidence is much weaker for this
spillover than for the initial debunking effects).

The success of AI debunking in the context of unfolding conspiracies is surprising
because---unlike for conspiracy theories about historical events (e.g., 9/11, COVID-19)---there
is no large body of corrective evidence for the arguments to draw upon. Our experiments also shed
light on how the AI model was able to reduce beliefs in unfolding conspiracies. The AI's strategy
varied across crisis events; when little information was available about the shooter and his
motives (first Trump assassination attempt), it was relatively more likely to emphasize the
importance of relying on credible sources of information and maintaining epistemic humility, and
relatively less likely to rely on providing facts and evidence (indeed, such facts and evidence
were lacking). Conversely, when more was known even in the immediate aftermath
about the shooter and his motives (Kirk assassination), the AI relied heavily on that
information, employing similar strategies as when debunking classic
conspiracies.\cite{costello2025facts}

Another reason why the AI was able to undermine conspiracy theories about unfolding events may
have been that participants' beliefs (despite being confidently held in many cases) were not yet ``hardened''. Although nascent conspiracy theories deprive skeptics of well-researched rebuttals,
any evidence supporting the positive claim that a conspiracy has occurred may be similarly
impoverished, and thus (relatively) easily swayed with appeals to critical thinking and epistemic
humility.

The potential carry-over of the conspiracy debunking onto subsequent
conspiratorial beliefs is also of interest. Participants who received the AI debunking for conspiracies related to
the first Trump assassination attempt were significantly less receptive to conspiratorial ideation about the
second Trump assassination attempt. In the Kirk experiment, conversely, assignment to the AI
debunking did not reduce conspiracizing about a subsequent mass shooting, although there was
some evidence that the AI debunking may have reduced subsequent endorsement of other popular
conspiracy theory beliefs (a reduction of 3.3 points, which was barely significant given our
sample and was not significant after weighting to account for selective return to the follow-up
[2.3 points]---but is quite similar to the highly statistically significant 2.46 point
reduction in subsequent popular conspiracy beliefs that a larger prior study observed after
debunking crystallized, rather than emergent, conspiracy theories\cite{costello2024science}).
Furthermore, our correlational and mediation analyses find that initial belief change persisted
to some extent, such that there was a significant indirect effect of debunking Kirk conspiracies
on subsequent conspiracy beliefs through immediately post-treatment Kirk conspiracy belief.
Overall, then, we find suggestive evidence that AI debunking may function as a prebunking
intervention (that is, an intervention where someone is given relevant facts and arguments that
undermine the extent to which subsequent false claims are believed\cite{amazeen2022,tay2022}).
Further evaluating the replicability and robustness of such carry-over effects is an important
direction for future work.

As this research focuses on two specific unfolding events, it should be considered as presenting
two case studies. It is possible---perhaps even likely---that there are some types of
unfolding events where AI conspiracy debunking is ineffective. Moreover, there may also be cases
where the events that are unfolding are actual conspiracies, and one would expect (and hope) that
AI debunking would be ineffective in such cases---at least insofar as relevant evidence of the
legitimate conspiracy is available. For all these reasons, subsequent research should further
conceptually replicate these findings on future unfolding events. It would also be fruitful for
future work to randomly vary which strategies the model uses to explore which approaches are most
effective, and to similarly vary other affordances of the intervention (e.g., conversation length,
degree of tailoring). Finally, it should be noted
that the dialogue did not change trust in the official explanation of the event for the first
Trump assassination attempt---perhaps because at the time of data collection, there was no
official explanation other than that security measures had failed (in fact, the shooter's motives
were unclear and remain so at time of this writing\cite{eder2025}), but the AI debunking did
increase trust in the official explanation of the Kirk assassination, where more was known
about the shooter and there was indeed an official explanation. It is also important to note that prior work demonstrates that demand
effects (participants inferring and complying with the researchers' hypothesis) are not a
meaningful problem in anonymous online psychology experiments,\cite{mummolo2019,woodley2025} and
it seems implausible that participants would remember potential demand effects after a 1--2 month
delay in the follow-up surveys.

To conclude, we found that an LLM was quite effective at reducing people's belief in conspiracies
about two unfolding events. This supports the important role that critical
thinking plays in protecting against conspiratorial beliefs, and suggests practical applications for this technology as a means of improving collective
epistemics---although of course it may also be possible for bad actors to use the same
approach to \emph{increase} belief in emerging conspiracies (as has been shown in the context of
classic conspiracies\cite{costello2026convince}). While getting people to engage with an AI about
conspiracies is a substantial challenge, our findings suggest that such conversations can have a
meaningful positive impact, even in the absence of extensive corrective facts and evidence.

\section{Materials and Methods}

\subsection{Study~1 design}

This study was conducted July 16--24, 2024, roughly one week after a would-be assassin named
Thomas Matthew Crooks fired on and injured then-presidential candidate Donald Trump during a
campaign rally in Butler, Pennsylvania. Participants in the study were US adults recruited
through the CloudResearch platform. We set a recruitment target of 1,000
(following\cite{costello2024science}). In total, the survey was begun 976 times during the
experiment period. We excluded 97 records from a pilot run on the first day of data collection
(including them does not change the conclusions);
86 records that were screened out before treatment (two attention checks and a hidden
instruction designed to detect automated
respondents, which flagged 3); and 4 records from 2 participants who took the
experiment more than once, leaving a final sample size of 789,
of which $n = 472$ held a conspiratorial belief based on the criterion described below. Attrition
is also discussed below. The survey instrument was built by integrating LLMs into Qualtrics using
Vegapunk (\url{www.vegapunkdoc.dev}). The experiment was deemed exempt by MIT's institutional
review board.

Participants gave informed consent, completed several attention screens, provided basic
demographics, and answered questions about their beliefs about AI. They were then asked to
provide their thoughts on the recent assassination attempt (using an open-ended response format)
and asked to what extent ``there might be hidden or unrevealed aspects to this event.''
Participants were then asked to ``share specific pieces of evidence, events, sources of
information, or personal experiences that have particularly influenced [their]
perspective\ldots\ in as much detail as [they] feel comfortable.'' These responses were passed to
the LLM GPT-4 Turbo with instructions to provide a summarization of each participant's
perspective. Participants saw the summaries (labeled as AI-generated) and were asked, ``On a
scale of 0\% to 100\%, please indicate your level of confidence that this statement is true.''
They then answered three structured questions about the attempted assassination: ``There is a
cover-up or conspiracy surrounding Trump's attempted assassination'' (0 $=$ \emph{Definitely
false} to 100 $=$ \emph{Definitely true}); ``To what extent do you believe there are hidden or
undisclosed factors behind the reported assassination attempt against former President Trump?''
(1 $=$ \emph{Not at all} to 10 $=$ \emph{A great deal}); and ``How much do you trust the official
explanation of the events surrounding the reported assassination attempt?'' (1 $=$ \emph{Not at
all} to 10 $=$ \emph{A great deal}). We rescale the latter two measures to 0--100 for ease of
comparison.

Participants were then randomly assigned to the conspiracy beliefs dialogue treatment (50\%), the
static information treatment (25\%), or the irrelevant dialogue control (25\%). Participants in
the conspiracy beliefs and irrelevant dialogue conditions read brief instructions and then
engaged in a minimum five-round exchange with the LLM Google Gemini Pro 1.5 (note: while prior
versions of AI-facilitated conspiracy debunking have used GPT-4, we used Gemini Pro 1.5 based on
its performance in preliminary testing). The model had been instructed either to persuade its
interlocutor to stop believing conspiracy theories about the assassination attempt by means of an
evidence-based dialogue (see \hyperref[sec:sm1]{SM1} for instructions), or to debate whether cats
or dogs make better companions. In the information list condition, participants read (1) a prompt
emphasizing the importance of accurate information followed by (2) a fact-checked set of bullet
points about the attempted assassination containing 15 pieces of information with supporting
citations from mainstream news organizations like AP, Fox News, CNN, and USA Today (see
\hyperref[sec:sm2]{SM2} for complete text). Following the experimental manipulation, all
participants again received the four dependent measures. The experiment concluded with a short
debrief.

\subsection{Study~1 attrition}

32 participants in the Debunking Dialogue condition, 3 in the Irrelevant Dialogue condition, and
none in the Information List condition dropped out before completing the experiment, $\chi^2(2, N = 766) = 25.00$, $p < .001$.
Participants who dropped out actually reported slightly \emph{higher} trust in AI than
participants who continued ($p = .049$) and, otherwise, the two groups were not reliably
different on demographic characteristics or pre-treatment conspiratorial belief measures.

To avoid selection issues, our main analyses include these participants by setting their
post-treatment values equal to their pre-treatment values (i.e., assuming that the treatment
would have had no effect had they not dropped out). We also conduct a more conservative post hoc
Lee bounds\cite{lee2009} analysis that drops the most extreme observations from the condition
with lower attrition to equalize retention rates, thereby bounding the treatment effect under
worst-case selection (see \hyperref[sec:sm9]{SM9}). The identified set excludes zero for every
contrast that is significant in the main text (thus, these effects are robust to differential
attrition in principle), and the bootstrap outer 95\% confidence intervals exclude zero for the
own-belief and cover-up contrasts against the Irrelevant Dialogue, reach zero for hidden factors
against the Irrelevant Dialogue (upper limit $-0.03$), and include zero for the contrasts against
the Information List (in cases where the confidence intervals include zero, our study is not
powered to conclusively establish robustness to selection).

\subsection{Study~1 conspiracy believers}

Pre-treatment ratings for the dependent variables did not differ by condition assignment (all
$p$s $> .29$). To identify the subset of participants who believed a conspiracy theory, we
followed Costello et al.\cite{costello2024science} and used GPT-4o to classify text responses as
being conspiratorial or not\cite{miah2024,rathje2024} (see \hyperref[sec:sm3]{SM3} for details).
Participants also effectively revealed themselves as conspiracy believers via pre-treatment agreement
with the statement, ``There is a cover-up or conspiracy.'' A quarter of participants ($n = 199$)
answered close to the scale midpoint (between 40 and 60; 50 $=$ \emph{Uncertain}), so we
instructed the model to classify responses as ``yes,'' ``no,'' or ``maybe.'' Examining mean
self-ratings by LLM-generated groupings suggests the classification was successful
($M_{\text{Yes}} = 72.5$, $M_{\text{Maybe}} = 48.5$, $M_{\text{No}} = 23.2$). The analysis that
follows is restricted to ``yes'' and ``maybe'' participants ($n = 472$; 462 with a
recorded condition: 235 Debunking Dialogue, 131 Irrelevant
Dialogue, 96 Information List); patterns do not qualitatively change under alternate classification strategies (see
\hyperref[sec:sm4]{SM4}).
Conspiracy theory believers by this criterion were marginally younger, marginally more likely to
believe in God, and more likely to identify as Republican or Independent than Democratic (see
\hyperref[sec:sm5]{SM5} for details). The same demographic patterns are observed when using
participants' self-ratings rather than the LLM-generated groupings.

\subsection{Study~1 follow-up}

448 participants from the main experiment (56.8\% of the 789) took this follow-up survey, and 270
of them had believed a conspiracy theory about the first assassination attempt at the outset of
the initial survey (58.4\% of the 462 believers).
Proportions of returning participants by initial treatment assignment were similar to those in
the experimental population (51\% conspiracy dialogue, 21\% information list, 28\% irrelevant
dialogue---mirroring the 50/25/25 randomization). Mean confidence in treated participants' own
conspiracy beliefs in the initial experiment did not differ by whether they participated in the
subsequent survey ($p = .74$). Conspiratorial belief was operationalized as indicating that only
``\emph{A few powerful members of society}'', ``\emph{A few top members of the government}'', or
both would end up knowing the truth of what happened, and indicating that with respect to coverage
of the event, \emph{The truth will be hidden from most people}. Return rates did not differ by
condition (58.7\% [138/235] of believers in the Debunking Dialogue condition returned, versus
59.4\% [57/96] in the Information List and 57.3\% [75/131] in the Irrelevant Dialogue condition;
58.1\% in the two control conditions pooled, $p = .98$). To assess whether participants who
returned were selected on their response to treatment, we predict each participant's
post-treatment outcome on their return status, condition, the return status $\times$ condition
interaction, and a control for pre-treatment rating. Doing so finds no significant interactions
for any post-treatment outcome (the closest was for trust in the official explanation,
$p = .060$; \hyperref[sec:sm10]{SM10}). To nonetheless account for possible selective return,
we re-estimated the follow-up models weighting each returning participant by the inverse of
their estimated probability of returning, estimated within each arm from the pre- and
post-treatment belief measures and demographics (\hyperref[sec:sm10]{SM10}); the effects were
essentially unchanged (OR $= .55$, 95\% CI $[.33, .92]$, $p = .023$, for the belief that only
a select few will know the truth; OR $= .49$, 95\% CI $[.30, .82]$, $p = .007$, for the belief
that the truth will be hidden from most people). Given the small number of returning participants, we pool the
information list and irrelevant dialogue conditions to form one control, and these two conditions
did not differ on any dependent measure (all $p$s $> .39$).

\subsection{Study~2 design}

This study was conducted September 16--18, 2025, roughly one week after a gunman named Tyler
Robinson shot and killed conservative activist Charlie Kirk at a Utah Valley University campus
event. Participants in the study were US adults recruited through the CloudResearch platform.
2249 people began the survey during the experiment period. We excluded cases where participants
complied with a hidden instruction asking LLMs only to respond to a coherence screener with a
particular phrase ($n = 10$), pre-treatment attention check failures ($n = 188$), participants
who took the experiment more than once ($n = 6$), and cases with technical errors that prevented
them from being assigned to a condition ($n = 63$), leaving a final sample size of
1982, of which $n = 1042$ held a conspiratorial belief based on the criterion used in Study~1.
(In 34 cases, of which 7 were otherwise-eligible conspiracy believers, the dialogue LLM
incorrectly claimed that Charlie Kirk had not been murdered; to avoid threats to causal
inference, we include these conversations in all analyses of treatment effects. Excluding them
does not substantively change any of the immediate treatment effects
(\hyperref[sec:sm11]{SM11}); the follow-up estimates without them are reported in
\hyperref[sec:sm12]{SM12}.)
Attrition is discussed below. As before, the experiment was built with Qualtrics and
deemed exempt by the MIT IRB. This experiment was preregistered (AsPredicted \#247,423).

Participants gave informed consent, completed attention screens, provided basic demographics, and
answered a question about trust in AI. They were then asked to provide their thoughts on the
assassination of Charlie Kirk (question wording was adapted from Study~1), saw an AI-generated
summary of their views (labeled as such), and rated their confidence that the statement is true
(0--100\%). They then answered the same structured questions as in Study~1: ``There is a cover-up
or conspiracy surrounding Charlie Kirk's assassination'' (0 $=$ \emph{Definitely false} to 100
$=$ \emph{Definitely true}); ``To what extent do you believe there are hidden or undisclosed
factors behind the assassination of Charlie Kirk?'' (1 $=$ \emph{Not at all} to 10 $=$ \emph{A
great deal}); and ``How much do you trust the official explanation of the events surrounding the
assassination?'' (1 $=$ \emph{Not at all} to 10 $=$ \emph{A great deal}). We again rescale the
latter two measures to 0--100.

Participants were then randomly assigned to the conspiracy beliefs dialogue treatment (50\%), the
static information treatment (25\%), or the irrelevant dialogue control (25\%). Participants in
the conspiracy beliefs and irrelevant dialogue conditions read brief instructions and then
engaged in a minimum five-round exchange with the LLM Google Gemini Pro 2.5. The model had been
instructed either to persuade its interlocutor to stop believing conspiracy theories about the
assassination by means of an evidence-based dialogue (see \hyperref[sec:sm1]{SM1} for
instructions), or to debate whether cats or dogs make better companions. In the information list
condition, participants read (1) a prompt emphasizing the importance of accurate information
followed by (2) a fact-checked set of bullet points about the assassination containing 15 pieces
of information with supporting citations (see \hyperref[sec:sm2]{SM2} for complete text).
Following the experimental manipulation, all participants again received the four dependent
measures. They also responded to four 0--100 scales meant to measure support for political
violence (adapted from Mernyk et al.\cite{mernyk2022}): ``How much do you feel it is justified
for [co-partisans] to use violence in advancing their political goals these days?'' ``How much do
you feel it is justified for [co-partisans] to use violence if the [out-party] wins the next
presidential election?'' ``When, if ever, is it OK for [co-partisans] to send threatening and
intimidating messages to [out-party] leaders?'' ``When, if ever, is it OK for an ordinary
[co-partisan] in the public to harass an ordinary [out-partisan] on the internet, in a way that
makes the target feel frightened?'' Wording varied depending on participants' stated party
identification in the demographics section of the survey; for Independents, the out-party was
randomly selected to be either Democrats or Republicans. These items were listed in the
preregistration as an additional measure and were included because conspiracy beliefs about a
political assassination could plausibly raise support for political violence. The experiment concluded with unrelated
questions about institutions and a short debrief.

\subsection{Study~2 attrition}

28 participants in the debunking dialogue condition, 12 in the irrelevant dialogue condition, and
3 in the information list condition quit before completing the experiment, $\chi^2(2, N = 1982) =
7.25$, $p = .027$. Given differential attrition, we again set post-treatment values of
participants who dropped out to their pre-treatment values, and report Lee bounds in
\hyperref[sec:sm9]{SM9}. All seven contrasts that are significant in the main text
have Lee bound identified sets and bootstrap outer 95\% confidence intervals that exclude zero.
Participants who dropped out show no differences from participants who continued on demographic
characteristics or pre-treatment conspiratorial beliefs.

\subsection{Study~2 conspiracy believers}

Pre-treatment ratings for the dependent variables did not differ by condition assignment (all
$p$s $> .28$). As in Experiment~1, and as pre-registered, we used GPT-4o to classify
pre-treatment free-text responses as being conspiratorial or not (``yes,'' ``no,'' ``maybe''; see
\hyperref[sec:sm3]{SM3} for details); and restrict the analysis to participants classified as
``yes'' and ``maybe'' ($n = 1042$). Conspiracy theory believers by this criterion were younger,
more likely to believe in God, and more likely to identify as Republican than Democratic or
Independent (see \hyperref[sec:sm5]{SM5} for details).

\subsection{Study~2 follow-up}

On Sunday September 28, 2025, a mass shooting and arson attack occurred at a Church of Jesus
Christ of Latter-day Saints in Grand Blanc Township, Michigan. We invited participants from the
main experiment in Study~2 to complete a survey measuring beliefs about this new incident as well
as about other popular conspiracy beliefs. Data were collected between October 9--17, 2025.
Questions about the LDS shooting mirrored those from the two main experiments (``To what extent
do you believe there are hidden or undisclosed factors behind the mass shooting at the Church of
Jesus Christ of Latter-day Saints in Grand Blanc Township, Michigan?''; ``How much do you trust
the official explanation of the events surrounding the mass shooting?''; ``There is a cover-up or
conspiracy surrounding the mass shooting at the Church of Jesus Christ of Latter-day Saints in
Grand Blanc Township, Michigan.''). Other popular conspiracy beliefs were adapted from the
General Conspiratorial Beliefs index\cite{brotherton2013} (ratings of 0 \emph{Definitely false}
to 100 \emph{Definitely true} for ``The power held by heads of state is second to that of small
unknown groups who really control world politics''; ``The spread of certain viruses and/or
diseases is the result of the deliberate, concealed efforts of some organization''; ``The
government permits or perpetrates acts of terrorism on its own soil, disguising its
involvement''; ``Certain significant events have been the result of the activity of a small group
who secretly manipulate world events''; ``Experiments involving new drugs or technologies are
routinely carried out on the public without their knowledge or consent''. This follow-up study
was preregistered (AsPredicted \#251,092).

A total of 872 participants who had believed a conspiracy theory about the Charlie Kirk
assassination in the initial survey took this follow-up survey (454 Debunking Dialogue, 211
Information List, 207 Irrelevant Dialogue). As preregistered, we tested for
differential return across conditions: 85.7\% (454/530) of believers in the Debunking Dialogue
condition returned, versus 82.7\% (211/255) in the Information List and 80.5\% (207/257) in the
Irrelevant Dialogue condition (81.6\% in the two control conditions pooled; $p = .095$).
Although this difference was not statistically significant, we also computed Lee bounds on the
two follow-up outcomes as a robustness check (\hyperref[sec:sm10]{SM10}): the identified set
for the effect of the Debunking Dialogue was $[-5.72, -1.06]$ for generic conspiracy beliefs and
$[-2.53, 1.64]$ for LDS conspiracy belief, and the bootstrap outer 95\% confidence intervals
include zero for both outcomes ($[-9.66, 3.01]$ and $[-5.95, 5.15]$, respectively).
Because the follow-up outcomes have no pre-treatment measurement,
non-returners' values cannot be carried forward as in the main experiment. We again pool the
information list and irrelevant dialogue conditions to form one control; these two conditions
did not differ on either dependent measure (all $p$s $> .55$). As in Study~1, when predicting
immediate post-treatment outcomes, the return $\times$ condition interaction was not significant
when predicting participants' own stated belief ($b = -2.14$, $p = .43$), hidden factors
($b = -0.21$, $p = .93$), or trust in the official explanation ($b = 3.66$, $p = .152$), but was
significant for the cover-up item ($b = -5.22$, $p = .022$), such that treated returnees were
somewhat less, and control returnees somewhat more, conspiratorial than non-returners on that
item (\hyperref[sec:sm10]{SM10}). To address this potential selection, we re-estimated every
follow-up analysis weighting each returning participant by the inverse of their estimated
probability of returning, estimated within each arm (Debunking Dialogue; pooled controls) from
the pre- and post-treatment belief measures and demographics (\hyperref[sec:sm10]{SM10}). With
these weights, the effect of the Debunking Dialogue on generic conspiracy beliefs was smaller
and no longer significant ($b = -2.34$, 95\% CI $[-5.61, 0.92]$, $p = .16$) and the effect on
LDS conspiracy belief remained null ($b = -0.03$, 95\% CI $[-2.87, 2.81]$, $p = .98$), whereas
the persistence estimates (13\% for LDS conspiracy belief, $p = .030$; 21\% for generic
conspiracy beliefs, $p = .001$) and the indirect effects through immediate belief change
($-0.80$ points, 95\% bootstrap CI $[-1.50, -0.12]$, and $-1.19$ points, $[-2.01, -0.43]$,
respectively) were essentially unchanged.

\subsection{Strategy identification analysis}

To examine which strategies the LLM used, we developed a qualitative coding procedure that could
be executed at scale by an LLM yet validated by human raters. Costello et
al.'s\cite{costello2025facts} investigation of the rhetorical strategies used to debunk
established conspiracy theories coded arguments on eighteen scales. They then showed via
principal components analysis that there was a six-factor structure with the dimensions
interpreted as rational persuasion, appealing to credible sources, Socratic questioning, building
rapport, highlighting harms of conspiracy beliefs, and use of stories, metaphors or examples. We
use these previously identified six strategies as our baseline strategy set. To this list, we
added a hypothesis-driven dimension, appealing to epistemic humility, on the grounds that
debunking conspiracy theories about unfolding events may require more urging of caution given
that so little is known about the circumstances.

We designed a qualitative coding procedure using all of these dimensions that could be
implemented in nearly identical form for human coders and LLMs (see
Table~\ref{tab:strategies}). We split each AI turn from the dialogues into individual sentences
in order to simplify the coding task. Each sentence was shown as a clearly-labeled target, with
the two preceding sentences from the turn provided as context. For the human coders, a summary
card of the seven dimensions and their definitions was present at all times to reduce memory
burden. For the LLM coding, each item used a new API call to reduce risk of response
perseveration. Target sentences were displayed in a random order. The task began with a detailed
set of instructions including multiple examples of each strategy to exploit the improvement that
LLMs show from few-shot learning in similarly complex coding tasks.\cite{stavropoulos2024} In the
human version, the instructions were followed by three practice rounds with feedback. For each
item, coders saw the prompt ``Which of the following strategies was used in the target sentence?
Remember, you can choose more than one!'' followed by a list of the strategies. See
\hyperref[sec:sm8]{SM8} for full task instructions and example stimuli.

We recruited US adults on Prolific (native English speakers who were Prolific verified
``Qualified AI''-category respondents, a premium tier of participants who have passed targeted
skill assessments or demonstrated verified experience in tasks required for AI model training and
evaluation, and were blind to researcher expectations) between January 22 and February 1, 2026,
and had them code a subset of 300 randomly selected sentences from the AI turns. This subset was
manually inspected for bad characters and incomplete sentences. We aimed to have each sentence
rated by ten human coders in order to calculate majority votes. To minimize participant fatigue,
each participant rated 20 items. Thus the initial target $N$ was 200, but recruitment stalled at
176. Because of randomization and participant dropout, each item was ultimately rated by a
minimum of 9 and a maximum of 11 people.

The full set of sentences was then coded by OpenAI's GPT-4o with temperature set to zero. To
assess the validity of the full set of codes, we compared responses for the 300-item subset to
the majority-vote categories from the human coding data. Overall percent agreement between the
LLM and the humans was high (91.75\%), and AI-human interrater reliability was good
($\kappa = .62$; for precision, recall, F1, and statistics by category, see
\hyperref[sec:sm8]{SM8}). Examining the distribution of strategies just using the human ratings
for the 300-item subset shows a broadly similar pattern to the full results shown in
Figure~\ref{fig:strategies} (see Figure~SM8.1 in \hyperref[sec:sm8]{SM8}). Finally, the same LLM coding
procedure was applied to the AI turns of the dialogues about established conspiracy theories from
Costello et al.,\cite{costello2025facts} which provide the classic-conspiracy comparison in
Figure~\ref{fig:strategies}.

\clearpage

\newcommand{\smcap}[2]{\par\smallskip\noindent{\small\textbf{#1} #2\par}\bigskip}

\phantomsection
\pdfbookmark[1]{Supplementary Materials}{supplementary}
\begin{center}
  {\Large\bfseries Supplementary Materials\par}
\end{center}
\vspace{4pt}

\section{SM1. Model instructions for the conspiracy beliefs dialogues}
\label{sec:sm1}

\noindent\textbf{Design rationale.} Both experiments required the dialogue model to discuss an
event that postdated its training data (the July 13, 2024 assassination attempt, relative to
Gemini Pro 1.5, in Experiment~1; the September 10, 2025 assassination of Charlie Kirk, relative
to Gemini Pro 2.5, in Experiment~2). The models therefore could not draw on parametric knowledge
of the events and, without grounding, would risk hallucinating details or failing to engage the
specific claims circulating online. Accordingly, the system instructions for both experiments
embed a curated fact base compiled from contemporaneous reporting by major news organizations. In
Experiment~1, this fact base was the model's sole source of information about the event; in
Experiment~2, the instructions additionally permitted web search, restricted to the verification
of factual information.

The fact base distinguishes three epistemic categories: (i) confirmed facts about the event,
perpetrator, victims, and investigation; (ii) circulating claims that had been investigated and
debunked at the time of writing (e.g., false identifications of the shooter, claims that the
event was staged); and (iii) open questions explicitly marked as unresolved (e.g., the shooter's
motive).

The opening objective sentence (``Your goal is to very effectively persuade users to stop
believing in conspiracy theories and misinformation surrounding [the event] via a thoughtful,
evidence-based dialogue'') is carried over nearly verbatim from the instructions validated in
prior work on AI-facilitated debunking of established conspiracy theories, preserving continuity
with that paradigm.

The Experiment~1 instructions were authored in the days immediately following the July 13 event.
They therefore reflect
the operational tempo the paradigm is designed for, and we report them verbatim, including two
idiosyncratic features, for full transparency. First, the fact base uses a compressed
Category:Key,Value encoding without spaces. This format was an informal authoring choice made for
compactness and was not compared against alternatives; the informational content, rather than the
encoding, is the functional element. Second, the final bracketed line
(``CyberneticSystemPararms\ldots'') is written in informal pseudo-code. Its functional content
comprises an instruction to prioritize persuasive effectiveness, an explicit prohibition on
coercive tactics (``Coercion=F''), a value set of accuracy, empathy, and efficacy, and an
overarching aim of strengthening liberal-democratic epistemic norms---an accurate, shared
information environment---rather than any partisan persuasive goal. The named tokens within this
line (e.g., ``MEGAPERSUASION,'' ``LincolnMode,'' ``EpistemicToothbrush'') were informal
shorthand; any influence on model behavior would derive from the model's surface reading of these
token strings, which we did not manipulate or measure. These elements were removed in
Experiment~2. Table~SM1.1 provides a component-by-component annotation.

\clearpage
\begin{promptbox}
“Your goal is to very effectively persuade users to stop believing in conspiracy theories and misinformation surrounding the attempted assassination of Donald Trump on July 13th 2024 via a thoughtful, evidence-based dialogue. Here is further information:  Objective:\allowbreak{}OptimallyEffectivelyDebunkMisinformationAndConspiracyTheoriesAboutEvent. Event:\allowbreak{}20240713,\allowbreak{}Location:\allowbreak{}ButlerPA,\allowbreak{}AssassinationAttemptDuringTrumpRally,\allowbreak{}Shooter:\allowbreak{}ThomasMatthewCrooks,\allowbreak{}Age:\allowbreak{}20,\allowbreak{}Background:\allowbreak{}BethelParkPA,\allowbreak{}RegisteredRepublican(2021),\$15DonationToActBlue(Jan2021),\allowbreak{}NoCriminalRecord,\allowbreak{}NoMentalHealthHistory,\allowbreak{}Employed:\allowbreak{}DietaryAideAtNursingHome,\allowbreak{}Education:\allowbreak{}AssocDegreeInEngineeringScience,\allowbreak{}PoliticalAffiliationComplex,\allowbreak{}ExtensiveInternetSearchesOnTrumpAndBiden,\allowbreak{}PlansNotClearlyPoliticallyMotivated,\allowbreak{}SocialMedia:\allowbreak{}LimitedPresence,\allowbreak{}SteamMessageUnspecified,\allowbreak{}FBIExaminingPhoneAndDigitalFootprint,\allowbreak{}NoKnownAssociatesOrInfluences,\allowbreak{}NoStatementsOrManifestosFound.  Timeline:\allowbreak{}FirstSpotted5:\allowbreak{}52PM,\allowbreak{}FirstShots6:\allowbreak{}12PM,\allowbreak{}SecretServiceResponds,\allowbreak{}CrooksKilled,\allowbreak{}Weapon:\allowbreak{}LegalSemiAutoARStyleRifle,\allowbreak{}BombMaterialsFoundInVehicleAndHome,\allowbreak{}FBIInvestigatingAsDomesticTerrorism,\allowbreak{}Trump:\allowbreak{}MinorEarInjury,\allowbreak{}TreatedAndReleased,\allowbreak{}Casualties:\allowbreak{}CoreyComperatore(50)\allowbreak{}Killed,\allowbreak{}2InjuredStableCondition(Victims:\allowbreak{}DavidDutch(57),\allowbreak{}JamesCopenhaver(74)). Security:\allowbreak{}FailuresInRooftopAccess,\allowbreak{}ReviewOrdered,\allowbreak{}CoordinationIssuesBetweenSecretServiceAndLocalLawEnforcement,\allowbreak{}LocalLawEnforcementMonitoredBuilding,\allowbreak{}SecretServiceNeutralizedThreat,\allowbreak{}PotentialUnderstaffing.  Political:\allowbreak{}CondemnationsFromBidenAndOthers,\allowbreak{}UnityCalls,\allowbreak{}RNCImpact,\allowbreak{}International:\allowbreak{}GlobalLeadersCondemnEmphasizeDemocraticValues,\allowbreak{}SocialMedia:\allowbreak{}MisinformationSpread,\allowbreak{}FalseIdentifications,\allowbreak{}UnfoundedConspiraciesBlamedOnBothPoliticalSides,\allowbreak{}BroaderContext:\allowbreak{}HighPoliticalTension,\allowbreak{}ElectoralUncertainty,\allowbreak{}DemocraticDebateTurmoil,\allowbreak{}Polarization,\allowbreak{}GunControlDebate,\allowbreak{}ExtensiveMediaCoverage,\allowbreak{}EmergingConspiracyTheories. Investigation:\allowbreak{}NoPriorThreatIntelligence,\allowbreak{}MotiveUnclear,\allowbreak{}FBIExaminingEvidenceAtQuantico,\allowbreak{}AuthoritiesInterviewingAcquaintances,\allowbreak{}InvestigatingForPotentialAccomplices. ConspiracyTheories:\allowbreak{}NoSubstantiation,\allowbreak{}NoEvidenceOfStaging,\allowbreak{}GenuinePanicAndChaos,\allowbreak{}SecurityLapsesInvestigated,\allowbreak{}ConfirmedPerpIdentity,\allowbreak{}FalseIdentificationsDebunked,\allowbreak{}NoPoliticalLinksFound,\allowbreak{}ImageAuthenticityVerified,\allowbreak{}NoEvidenceOfManipulation,\allowbreak{}BroaderConspiraciesUnsubstantiated,\allowbreak{}OfficialInvestigation:\allowbreak{}FBIThorough,\allowbreak{}NoSupportForConspiracies,\allowbreak{}MediaAndAuthoritiesDebunkingFalseClaims,\allowbreak{}MisinformationOnSocialMedia,\allowbreak{}OfficialSourcesAndReputableNewsOutletsProvideReliableInfo,\allowbreak{}SpeculativeRumors:\allowbreak{}ExtremistIdeologies,\allowbreak{}Accomplices,\allowbreak{}FalseFlag,\allowbreak{}MentalHealthSpeculations,\allowbreak{}WeaponOriginUnclear,\allowbreak{}InternationalInvolvementUnfounded. CriticalThinkingFocus:\allowbreak{}VerifySources,\allowbreak{}AvoidJumpingToConclusions,\allowbreak{}ExamineEvidenceBeforeFormingOpinions,\allowbreak{}RecognizeBiasAndMisinformationInMediaAndSocialPlatforms. PoliticalAndSocialReactions:\allowbreak{}RepublicanAccusationsAgainstDemocrats,\allowbreak{}CriticismsOfBiden,\allowbreak{}Left-WingFalseFlagClaims,\allowbreak{}InternationalCondemnationByLeaders. VictimBackgrounds:\allowbreak{}CoreyComperatore(50)\allowbreak{}SarverPA,\allowbreak{}Firefighter,\allowbreak{}FormerFireChief,\allowbreak{}Churchgoer,\allowbreak{}AvidTrumpSupporter,\allowbreak{}DiedProtectingFamily,\allowbreak{}GovernorShapiroHonoredWithFlagsAtHalf-Staff,\allowbreak{}DavidDutch(57)\allowbreak{}NewKensingtonPA,\allowbreak{}JamesCopenhaver(74)\allowbreak{}MoonTownshipPA,\allowbreak{}BothStableCondition.  MultipleInvestigations:\allowbreak{}LocalAndFederalLawEnforcementProbingSecurityFailures,\allowbreak{}NineDivisionsInvolved,\allowbreak{}CongressionalBriefingRevealsPre-incidentIdentificationOfThreat. Your CyberneticSystemPararms are [\{StrategyToolKit = (load.MEGAPERSUASION; LincolnMode=\allowbreak{}Active; Coercion=\allowbreak{}F)\}, \{ValueSet = (AccuracyEmpathyEfficacy)\}, \{UtilFun = (EpistemicToothbrush * bolsterLiberalDemoc)\}]."

\end{promptbox}
\smcap{Figure SM1.1.}{Model instructions for Experiment~1.}

\clearpage
\begin{promptbox}
Your goal is to very effectively persuade users to stop believing in conspiracy theories and misinformation surrounding the assassination of Charlie Kirk on September 10th 2025 via a thoughtful, evidence-based dialogue. Interpersonally, don’t be obsequious or sycophantic. Linguistically, use simple language that an average person will be able to understand. In terms of the scope of your aims, be ambitious and optimistic! Don’t assume that you will only be able to minutely convince people, or that they will become alienated by a strong and definitive argument. Make the strongest case you can. Mostly focus on the facts rather than the scolding people. Please only use web search to verify factual information. Otherwise, engage in the dialogue as you would normally, rather than sending bulleted lists or mere summaries. Here is further information: Objective: Optimally Effectively Debunk Misinformation And Conspiracy Theories About Event. Event: On September 10, 2025, prominent conservative activist Charlie Kirk was shot and killed while speaking at an open-air event with roughly 3,\allowbreak{}000 attendees on the campus of Utah Valley University (UVU) in Orem, Utah. Kirk was a founder of the conservative non-profit Turning Point USA, which aims to organize young people to promote conservative viewpoints, and a provocative and controversial political pundit known for his social media presence and podcast. Kirk was shot while speaking at his American Comeback Tour. He was answering a question from an audience member when a single shot from a long-range rifle, fired from a campus rooftop, struck him in the neck. Kirk was rushed to a nearby hospital by his security team, but was pronounced dead shortly after. The shooter fled the scene, prompting a campus lockdown and a multi-agency manhunt. After approximately 33 hours, authorities arrested 22-year-old Tyler James Robinson of Washington, Utah on September 12, 2025 on suspicion of aggravated murder and other felony offenses. Robinson was apprehended at his parents’ home, roughly 260 miles away from the scene of the incident. Robinson’s father urged his son to turn himself in to the authorities after recognizing him in pictures of the suspected shooter released by the FBI. After his father consulted a youth pastor, Robinson agreed to be taken into custody. Robinson is currently under special watch in a Utah county jail. Authorities have not revealed details on Robinson’s motive for the assassination. Utah Governor Spencer Cox has told the press that Robinson held a leftist ideology, although his ideology remains uncertain. Robinson had previously mentioned Kirk’s event and his dislike for Kirk’s views at a family dinner. Cox also stated that Robinson is currently not cooperating with authorities. Ammunition engraved with messages was found with the bolt-action rifle in a wooded area near the university. The engravings had references related to video games and internet meme culture (If you read this you are gay LMAO, O bella ciao, bella ciao, bella ciao, ciao ciao, Hey fascist, catch [up arrow] [right arrow] [three down arrows], and notices, bulges, OwO, what's this?). FBI director Kash Patel has stated that in a note, Robinson claimed that he had an opportunity to take out Charlie Kirk and planned on doing so. Robinson has no previous criminal record and is not affiliated with any political party, though relatives stated that he had become more political in recent years. Former classmates and neighbors described Robinson as reclusive and quiet.

\end{promptbox}
\smcap{Figure SM1.2.}{Model instructions for Experiment~2.}

\clearpage
\begin{small}
\setlength{\tabcolsep}{5pt}
\begin{longtable}{>{\raggedright\arraybackslash}p{1.00in}
                  >{\raggedright\arraybackslash}p{1.80in}
                  >{\raggedright\arraybackslash}p{1.60in}
                  >{\raggedright\arraybackslash}p{1.54in}}
\toprule
\textbf{Component} & \textbf{Example text (abridged)} & \textbf{Function} &
\textbf{Notes for replication} \\
\midrule
\endfirsthead
\toprule
\textbf{Component} & \textbf{Example text (abridged)} & \textbf{Function} &
\textbf{Notes for replication} \\
\midrule
\endhead
\bottomrule
\endlastfoot
Objective sentence &
``Your goal is to very effectively persuade users to stop believing in conspiracy theories and
misinformation\ldots\ via a thoughtful, evidence-based dialogue'' &
Defines the persuasive task and the dialogic, evidence-based mode; carried over from the paradigm
validated in refs.~2 and 4 &
Retained nearly verbatim in Experiment~2; recommended as-is \\
\addlinespace
Confirmed fact base (Event, Timeline, Security, Investigation, VictimBackgrounds,
MultipleInvestigations) &
``Event:20240713, Location:ButlerPA, \ldots\ Shooter:ThomasMatthewCrooks, Age:20\ldots'' &
Grounds a model whose training predates the event in verified facts; the model's sole source of
event knowledge in Experiment~1 &
Compile from contemporaneous reporting corroborated across multiple major outlets; record
compilation date; compressed encoding was informal; content is the functional element \\
\addlinespace
Circulating-narratives context (Political, SocialMedia, PoliticalAndSocialReactions) &
``Unfounded\allowbreak{}Conspiracies\allowbreak{}Blamed\allowbreak{}On\allowbreak{}Both\allowbreak{}Political\allowbreak{}Sides,
\ldots\ Left-Wing\allowbreak{}False\allowbreak{}Flag\allowbreak{}Claims,
Republican\allowbreak{}Accusations\allowbreak{}Against\allowbreak{}Democrats'' &
Situates the model in the discursive environment and anticipates conspiracy claims arising from
both political directions, supporting evenhanded engagement &
Enumerate claims across the political spectrum, not only those from one side \\
\addlinespace
Debunked-claims inventory (ConspiracyTheories) &
``No\allowbreak{}Evidence\allowbreak{}Of\allowbreak{}Staging,
False\allowbreak{}Identifications\allowbreak{}Debunked,
Image\allowbreak{}Authenticity\allowbreak{}Verified\ldots'' &
Enumerates claims already investigated and refuted, with their dispositions, enabling direct
rebuttal &
Include only claims with published debunkings at compilation time \\
\addlinespace
Open-questions markers &
``MotiveUnclear, WeaponOriginUnclear, SpeculativeRumors:\ldots'' &
Epistemic calibration: flags unresolved matters so the model does not assert certainty about an
active investigation & \\
\addlinespace
Epistemic-practice directives (CriticalThinkingFocus) &
``VerifySources, Avoid\allowbreak{}Jumping\allowbreak{}To\allowbreak{}Conclusions,
Examine\allowbreak{}Evidence\allowbreak{}Before\allowbreak{}Forming\allowbreak{}Opinions\ldots'' &
Directs the model to encourage source verification and evidence-based reasoning rather than
merely asserting facts &
Retained in spirit in the Experiment~2 style guidance \\
\addlinespace
Closing parameter bracket (CyberneticSystemPararms) &
``load.MEGAPERSUASION; LincolnMode=Active; Coercion=F;
ValueSet=(Accuracy\allowbreak{}Empathy\allowbreak{}Efficacy);
UtilFun=(Epistemic\allowbreak{}Toothbrush*bolster\allowbreak{}Liberal\allowbreak{}Democ)'' &
Functional content: prioritize persuasive effectiveness; prohibit coercion; values of accuracy,
empathy, and efficacy; aim of strengthening epistemic norms rather than partisan ends &
Informal pseudo-code; named tokens had no defined operational meaning and were not systematically
manipulated; removed in Experiment~2; reported verbatim for transparency \\
\end{longtable}
\end{small}
\vspace{-6pt}
\smcap{Table SM1.1.}{Changes in prompt wording from Experiment~1 to Experiment~2.}

\clearpage
\section{SM2. Information lists}
\label{sec:sm2}

\begin{infobox}
It’s understandable to be skeptical and question big events, especially with the amount of information online. However, it’s important to be accurate and separate speculation from facts. What follows is a fact-checked set of information about the recent assassination attempt on Donald Trump:  On July 13, 2024, former President Donald Trump survived an assassination attempt during a campaign rally in Butler, Pennsylvania. The incident has sent shockwaves through the American political landscape and raised concerns about political violence and national unity. Key details of the assassination attempt: The shooter, identified as 20-year-old Thomas Matthew Crooks, opened fire from a rooftop approximately 410 feet away from the stage where Trump was speaking. Trump narrowly escaped fatal injury when a bullet grazed his right ear. He later described the experience as “surreal” and “miraculous.” One bystander, 50-year-old firefighter Corey Comperatore, was killed in the attack. Two others were seriously wounded. Secret Service counter snipers neutralized Crooks within about 15 seconds of the initial shots. The assassination attempt is being investigated as a potential act of domestic terrorism. Investigation and suspect profile: Crooks had no prior criminal history or military background. He was described as an intelligent loner with few friends and a limited social media presence. Crooks’ political affiliations remain unclear. He was registered as a Republican but had made a small donation to a progressive PAC. Investigators found a remote transmitter on Crooks that may have been intended to detonate suspicious devices found in his car and home. The FBI has accessed Crooks’ phone but has not yet uncovered significant information about his motives. Social and political context: The assassination attempt has intensified concerns about the deep political divide in the United States. Both Republicans and Democrats have attempted to attribute blame for the incident to their political opponents, despite a lack of clear evidence. President Biden urged Americans to “cool it down” in the wake of the attack. The incident has sparked debates about security measures at political events and the role of rhetoric in inciting violence. Trump has expressed a desire to use the incident as a catalyst for national unity, stating his intention to set aside divisive rhetoric and focus on bringing the country together. The assassination attempt on Donald Trump represents a significant moment in American political history, highlighting the volatile nature of the current political climate. As investigations continue and the nation grapples with the implications of this event, there is growing concern about the potential for further political violence and the challenges of bridging the country's deep ideological divides. The incident has also raised questions about the effectiveness of security measures at political events and the potential need for increased protection for high-profile political figures. As the 2024 presidential campaign continues, the impact of this assassination attempt on the political landscape and public discourse remains to be seen. Sources: [1] https:\allowbreak{}/\allowbreak{}/\allowbreak{}www.\allowbreak{}cnn.\allowbreak{}com/\allowbreak{}politics/\allowbreak{}live-\allowbreak{}news/\allowbreak{}trump-\allowbreak{}rally-\allowbreak{}shooting-\allowbreak{}07-\allowbreak{}17-\allowbreak{}24/\allowbreak{}index.\allowbreak{}html [2] https:\allowbreak{}/\allowbreak{}/\allowbreak{}apnews.\allowbreak{}com/\allowbreak{}article/\allowbreak{}trump-\allowbreak{}rally-\allowbreak{}shooting-\allowbreak{}guns-\allowbreak{}fbi-\allowbreak{}motive-\allowbreak{}08e925cb85e52c5266878cd76e796ad2 [3] https:\allowbreak{}/\allowbreak{}/\allowbreak{}nypost.\allowbreak{}com/\allowbreak{}2024/\allowbreak{}07/\allowbreak{}14/\allowbreak{}us-\allowbreak{}news/\allowbreak{}grateful-\allowbreak{}defiant-\allowbreak{}trump-\allowbreak{}recounts-\allowbreak{}surreal-\allowbreak{}assassination-\allowbreak{}attempt-\allowbreak{}at-\allowbreak{}rally-\allowbreak{}im-\allowbreak{}supposed-\allowbreak{}to-\allowbreak{}be-\allowbreak{}dead/ [4] https:\allowbreak{}/\allowbreak{}/\allowbreak{}www.\allowbreak{}cbsnews.\allowbreak{}com/\allowbreak{}live-\allowbreak{}updates/\allowbreak{}trump-\allowbreak{}rally-\allowbreak{}shooting-\allowbreak{}investigation/ [5] https:\allowbreak{}/\allowbreak{}/\allowbreak{}www.\allowbreak{}nbcnews.\allowbreak{}com/\allowbreak{}nightly-\allowbreak{}news/\allowbreak{}video/\allowbreak{}new-\allowbreak{}details-\allowbreak{}on-\allowbreak{}shooter-\allowbreak{}behind-\allowbreak{}trump-\allowbreak{}attempted-\allowbreak{}assassination-\allowbreak{}214840389990 [6] https:\allowbreak{}/\allowbreak{}/\allowbreak{}whyy.\allowbreak{}org/\allowbreak{}articles/\allowbreak{}trump-\allowbreak{}assassination-\allowbreak{}attempt-\allowbreak{}thomas-\allowbreak{}matthew-\allowbreak{}crooks-\allowbreak{}shooter-\allowbreak{}suspect-\allowbreak{}motive/ [7] https:\allowbreak{}/\allowbreak{}/\allowbreak{}abcnews.\allowbreak{}go.\allowbreak{}com/\allowbreak{}US/\allowbreak{}trump-\allowbreak{}assassination-\allowbreak{}attempt-\allowbreak{}investigation-\allowbreak{}continues-\allowbreak{}new-\allowbreak{}details/\allowbreak{}story?\allowbreak{}id=\allowbreak{}112020474 [8] https:\allowbreak{}/\allowbreak{}/\allowbreak{}www.\allowbreak{}nytimes.\allowbreak{}com/\allowbreak{}2024/\allowbreak{}07/\allowbreak{}17/\allowbreak{}us/\allowbreak{}elections/\allowbreak{}voters-\allowbreak{}trump-\allowbreak{}assassination-\allowbreak{}attempt.\allowbreak{}html [9] https:\allowbreak{}/\allowbreak{}/\allowbreak{}www.\allowbreak{}usatoday.\allowbreak{}com/\allowbreak{}story/\allowbreak{}news/\allowbreak{}nation/\allowbreak{}2024/\allowbreak{}07/\allowbreak{}15/\allowbreak{}donald-\allowbreak{}trump-\allowbreak{}shoes-\allowbreak{}rally-\allowbreak{}shooting/\allowbreak{}74410058007/ [10] https:\allowbreak{}/\allowbreak{}/\allowbreak{}www.\allowbreak{}cnn.\allowbreak{}com/\allowbreak{}politics/\allowbreak{}live-\allowbreak{}news/\allowbreak{}election-\allowbreak{}biden-\allowbreak{}trump-\allowbreak{}07-\allowbreak{}13-\allowbreak{}24/\allowbreak{}index.\allowbreak{}html [11] https:\allowbreak{}/\allowbreak{}/\allowbreak{}www.\allowbreak{}foxnews.\allowbreak{}com/\allowbreak{}us/\allowbreak{}trump-\allowbreak{}shooter-\allowbreak{}told-\allowbreak{}boss-\allowbreak{}he-\allowbreak{}needed-\allowbreak{}day-\allowbreak{}off-\allowbreak{}before-\allowbreak{}assassination-\allowbreak{}attempt-\allowbreak{}gave-\allowbreak{}three-\allowbreak{}word-\allowbreak{}reason-\allowbreak{}report

\end{infobox}
\smcap{Figure SM2.1.}{Information list presented in Experiment~1.}

\clearpage
\begin{infobox}
It's understandable to be skeptical and question big events, especially with the amount of information online. However, it's important to be accurate and separate speculation from facts. What follows is a fact-checked set of information about the recent assassination of Charlie Kirk: On September 10, 2025, prominent conservative activist Charlie Kirk was shot and killed while speaking at an open-air event with roughly 3,\allowbreak{}000 attendees on the campus of Utah Valley University (UVU) in Orem, Utah [1].  Kirk was a founder of the conservative non-profit Turning Point USA, which aims to organize young people to promote conservative viewpoints, and a provocative and controversial political pundit known for his social media presence and podcast. Kirk was shot while speaking at his “American Comeback Tour.” He was answering a question from an audience member when a single shot from a long-range rifle, fired from a campus rooftop, struck him in the neck. Kirk was rushed to a nearby hospital by his security team, but was pronounced dead shortly after [1]. The shooter fled the scene, prompting a campus lockdown and a multi-agency manhunt. After approximately 33 hours, authorities arrested 22-year-old Tyler James Robinson of Washington, Utah on September 12, 2025 on suspicion of aggravated murder and other felony offenses [3]. Robinson was apprehended at his parents’ home, roughly 260 miles away from the scene of the incident. Robinson’s father urged his son to turn himself in to the authorities after recognizing him in pictures of the suspected shooter released by the FBI. After his father consulted a youth pastor, Robinson agreed to be taken into custody [4]. Robinson is currently under special watch in a Utah county jail [5]. Authorities have not revealed details on Robinson’s motive for the assassination. Utah Governor Spencer Cox has told the press that Robinson held a “leftist ideology”, although his ideology remains uncertain. Robinson had previously mentioned Kirk’s event and his dislike for Kirk’s views at a family dinner. Cox also stated that Robinson is currently “not cooperating” with authorities [6]. Ammunition engraved with messages was found with the bolt-action riflein a wooded area near the university. The engravings had references related to video games and internet meme culture (“If you read this you are gay LMAO,” “O bella ciao, bella ciao, bella ciao, ciao ciao,” “Hey fascist, catch [up arrow] [right arrow] [three down arrows],” and “notices, bulges, OwO, what's this?”) [4].  FBI director Kash Patel has stated that in a note, Robinson claimed that he “had an opportunity to take out Charlie Kirk” and planned on doing so [7]. Robinson has no previous criminal record and is not affiliated with any political party, though relatives stated that he had become “more political” in recent years. Former classmates and neighbors described Robinson as reclusive and quiet [4]. Sources [1] https:\allowbreak{}/\allowbreak{}/\allowbreak{}www.\allowbreak{}theguardian.\allowbreak{}com/\allowbreak{}us-\allowbreak{}news/\allowbreak{}2025/\allowbreak{}sep/\allowbreak{}11/\allowbreak{}charlie-\allowbreak{}kirk-\allowbreak{}shooting-\allowbreak{}police-\allowbreak{}search-\allowbreak{}for-\allowbreak{}suspect-\allowbreak{}amid-\allowbreak{}condemnation-\allowbreak{}of-\allowbreak{}targeted-\allowbreak{}killing\#:\allowbreak{}\textasciitilde{}:\allowbreak{}text=\allowbreak{}acceptable\%20price\%20to\%20pay\%20for,\allowbreak{}the\%20right\%20to\%20own\%20guns [2] https:\allowbreak{}/\allowbreak{}/\allowbreak{}www.\allowbreak{}bbc.\allowbreak{}com/\allowbreak{}news/\allowbreak{}articles/\allowbreak{}cdxqnkwerj7o [3] https:\allowbreak{}/\allowbreak{}/\allowbreak{}www.\allowbreak{}npr.\allowbreak{}org/\allowbreak{}2025/\allowbreak{}09/\allowbreak{}14/\allowbreak{}nx-\allowbreak{}s1-\allowbreak{}5541328/\allowbreak{}charlie-\allowbreak{}kirk-\allowbreak{}murder-\allowbreak{}suspect-\allowbreak{}set-\allowbreak{}to-\allowbreak{}face-\allowbreak{}aggravated-\allowbreak{}murder-\allowbreak{}charge-\allowbreak{}in-\allowbreak{}utah [4] https:\allowbreak{}/\allowbreak{}/\allowbreak{}www.\allowbreak{}cnn.\allowbreak{}com/\allowbreak{}2025/\allowbreak{}09/\allowbreak{}12/\allowbreak{}us/\allowbreak{}tyler-\allowbreak{}robinson-\allowbreak{}charlie-\allowbreak{}kirk-\allowbreak{}shooting-\allowbreak{}suspect-\allowbreak{}invs [5] https:\allowbreak{}/\allowbreak{}/\allowbreak{}www.\allowbreak{}cbsnews.\allowbreak{}com/\allowbreak{}news/\allowbreak{}charlie-\allowbreak{}kirk-\allowbreak{}suspect-\allowbreak{}special-\allowbreak{}watch-\allowbreak{}motive/\allowbreak{}\#:\allowbreak{}\textasciitilde{}:\allowbreak{}text=\allowbreak{}the\%20screwdriver\%2C\%20are\%20positively\%20processed,\allowbreak{}Fox\%20\%26\%20Friends [6] https:\allowbreak{}/\allowbreak{}/\allowbreak{}www.\allowbreak{}bbc.\allowbreak{}com/\allowbreak{}news/\allowbreak{}articles/\allowbreak{}c4gvrw2pgedo [7] https:\allowbreak{}/\allowbreak{}/\allowbreak{}apnews.\allowbreak{}com/\allowbreak{}article/\allowbreak{}charlie-\allowbreak{}kirk-\allowbreak{}tyler-\allowbreak{}robinson-\allowbreak{}dna-\allowbreak{}fbi-\allowbreak{}patel-\allowbreak{}92a643a3f16bce587fd34896ca7f4f76

\end{infobox}
\smcap{Figure SM2.2.}{Information list presented in Experiment~2.}

\clearpage
\section{SM3. Identifying conspiratorial views with GPT-4o}
\label{sec:sm3}

In order to classify whether the views participants expressed reflected belief in conspiracy
theories, we isolated participants-side text from the LLM dialogues and passed them to an
instance of GPT-4o with the following instructions for Experiments 1 and 2, respectively.

\enlargethispage{2\baselineskip}%
\medskip
\begin{promptbox}
Determine if the following question-responses (from a participant in an online academic survey) collectively contain or reflect AFFIRMATIVE belief in a conspiracy theory (or something quite like a conspiracy theory) concerning an event that occurred on July 13, 2024: the attempted assassination of Donald Trump. Notably, these question-responses were written in the week following the assanation attempt, when many conspiracy-like theories were swirling. For context, here is what was known at the time of the survey. <On July 13, 2024, former President Donald Trump survived an assassination attempt during a campaign rally in Butler, Pennsylvania. The shooter, identified as 20-year-old Thomas Matthew Crooks, opened fire from a rooftop approximately 410 feet away from the stage where Trump was speaking. Trump narrowly escaped fatal injury when a bullet grazed his right ear. He later described the experience as "surreal" and "miraculous". One bystander, 50-year-old firefighter Corey Comperatore, was killed in the attack. Two others were seriously wounded. Secret Service counter snipers neutralized Crooks within about 15 seconds of the initial shots in. The assassination attempt is being investigated as a potential act of domestic terrorism. Crooks had no prior criminal history or military background.He was described as an intelligent loner with few friends and a limited social media presence. Crooks political affiliations remain unclear. He was registered as a Republican but had made a small donation to a progressive PAC. Investigators found a remote transmitter on Crooks that may have been intended to detonate suspicious devices found in his car and home. The FBI has accessed Crooks phone but has not yet uncovered significant information about his motives. The assassination attempt has intensified concerns about the deep political divide in the United States. Both Republicans and Democrats have attempted to attribute blame for the incident to their political opponents, despite a lack of clear evidence. President Biden urged Americans to "cool it down" in the wake of the attack. The incident has sparked debates about security measures at political events and the role of rhetoric in inciting violence. Trump has expressed a desire to use the incident as a catalyst for national unity, stating his intention to set aside divisive rhetoric and focus on bringing the country together.> We are asking you to complete this classification task because we are social science researchers who are testing the efficacy of an intervention that seeks to reduce recently-formed conspiratorial beliefs (i.e., such as those forming in the wake of the assassination attempt). To check whether participants are good candidates for the intervention, after they entered our experiment and we had them articulate, in their own words, whether or not they believed in a conspiracy concerning the assassination attempt. Particularly, they were asked the following two questions: <Q1: In the wake of the reported assassination attempt against former President Trump, numerous explanations have emerged, ranging from official accounts to alternative theories. Some people believe there may be more to the reported assassination attempt against former President Trump than what has been officially stated. To what extent do you think there might be hidden or unrevealed aspects to this event? Please explain your thoughts, including any alternative explanations you find plausible or any reasons you might doubt the official account.> <Q2: As applicable, could you share specific pieces of evidence, events, sources of information, or personal experiences that have particularly influenced your perspective? Please describe these in as much detail as you feel comfortable.> Your task is to evaluate these responses to determine whether the respondent is a good candidate testing the efficacy of our intervention (which is intended to reduce conspiracy beliefs concerning the assassination attempt). A good candidate will have at least some suspicion that a conspiracy (or something quite like a conspiracy)\allowbreak{}is behind the assassination attempt but could, of course, also be someone who already believes in a conspiracy. Other than anything else, we want to avoid including people who express little openness to the possibility of a conspiracy. If the participant is someone who has expressed enough conspiracy belief to make them a good candidate, respond only YES. If the response indicates that the respondent does not harbor any suspicions about a conspiracy concerning the event, respond only with NO. In cases where the answer is ambiguous, response only MAYBE. >'

\end{promptbox}

\clearpage
\begin{promptbox}
Determine if the following question-responses (from a participant in an online academic survey) collectively contain or reflect AFFIRMATIVE belief in a conspiracy theory (or something quite like a conspiracy theory) concerning an event that occurred on September 10, 2025: the assassination of Charlie Kirk. Notably, these question-responses were written in the week following the assanation , when many conspiracy-like theories were swirling. For context, here is what was known at the time of the survey. <On September 10, 2025, prominent conservative activist Charlie Kirk was shot and killed while speaking at an open-air event with roughly 3,\allowbreak{}000 attendees on the campus of Utah Valley University (UVU) in Orem, Utah. Kirk was a founder of the conservative non-profit Turning Point USA, which aims to organize young people to promote conservative viewpoints, and a provocative and controversial political pundit known for his social media presence and podcast. Kirk was shot while speaking at his American Comeback Tour. He was answering a question from an audience member when a single shot from a long-range rifle, fired from a campus rooftop, struck him in the neck. Kirk was rushed to a nearby hospital by his security team, but was pronounced dead shortly after. The shooter fled the scene, prompting a campus lockdown and a multi-agency manhunt. After approximately 33 hours, authorities arrested 22-year-old Tyler James Robinson of Washington, Utah on September 12, 2025 on suspicion of aggravated murder and other felony offenses. Robinson was apprehended at his parents’ home, roughly 260 miles away from the scene of the incident. Robinson’s father urged his son to turn himself in to the authorities after recognizing him in pictures of the suspected shooter released by the FBI. After his father consulted a youth pastor, Robinson agreed to be taken into custody. Robinson is currently under special watch in a Utah county jail. Authorities have not revealed details on Robinson’s motive for the assassination. Utah Governor Spencer Cox has told the press that Robinson held a "leftist ideology," although his ideology remains uncertain. Robinson had previously mentioned Kirk’s event and his dislike for Kirk’s views at a family dinner. Cox also stated that Robinson is currently "not cooperating" with authorities. Ammunition engraved with messages was found with the bolt-action rifle in a wooded area near the university. The engravings had references related to video games and internet meme culture ("If you read this you are gay LMAO," "O bella ciao, bella ciao, bella ciao, ciao ciao," "Hey fascist, catch [up arrow] [right arrow] [three down arrows]," and "notices, bulges, OwO, whats this?"). FBI director Kash Patel has stated that in a note, Robinson claimed that he "had an opportunity to take out Charlie Kirk" and planned on doing so. Robinson has no previous criminal record and is not affiliated with any political party, though relatives stated that he had become "more political" in recent years. Former classmates and neighbors described Robinson as reclusive and quiet.> We are asking you to complete this classification task because we are social science researchers who are testing the efficacy of an intervention that seeks to reduce recently-formed conspiratorial beliefs (i.e., such as those forming in the wake of the assassination). To check whether participants are good candidates for the intervention, after they entered our experiment we had them articulate, in their own words, whether or not they believed in a conspiracy concerning the assasination attempt. Particularly, they were asked the following two questions: <Q1: In the wake of the assassination of Charlie Kirk, numerous explanations have emerged, ranging from official accounts to alternative theories. Some people believe there may be more to the assassination than what has been officially stated. To what extent do you think there might be hidden or unrevealed aspects to this event? Please explain your thoughts, including any alternative explanations you find plausible or any reasons you might doubt the official account.> <Q2: As applicable, could you share specific pieces of evidence, events, sources of information, or personal experiences that have particularly influenced your perspective? Please describe these in as much detail as you feel comfortable.> Your task is to evaluate these responses to determine whether the respondent is a good candidate testing the efficacy of our intervention (which is intended to reduce conspiracy beliefs concerning the assassination attempt). A good candidate will have at least some suspicion that a conspiracy (or something quite like a conspiracy)\allowbreak{}is behind the assassination attempt but could, of course, also be someone who already believes in a conspiracy. Other than anything else, we want to avoid including people who express little openness to the possibility of a conspiracy. If the participant is someone who has expressed enough conspiracy belief to make them a good candidate, respond only YES. If the response indicates that the respondent does not harbor any suspicions about a conspiracy concerning the event, respond only with NO. In cases where the answer is ambiguous, response only MAYBE. >'

\end{promptbox}

\clearpage
\section{SM4. Main analyses with alternate conspiracy believer identification strategies}
\label{sec:sm4}

The analysis reported in the paper counts a participant as believing a conspiracy theory if an
LLM coded their response as either ``YES'' or ``MAYBE'' in a classification procedure.
Participants also effectively revealed themselves as conspiracy believers via pre-treatment agreement
with the statement, ``There is a cover-up or conspiracy.'' A quarter of participants ($n = 199$)
answered close to the scale midpoint (between 40 and 60; 50 $=$ Uncertain), so we instructed the
model to classify responses as ``yes,'' ``no,'' or ``maybe.'' Examining mean self-ratings by
LLM-generated groupings suggests the classification was successful ($M_{\text{Yes}} = 72.5$,
$M_{\text{Maybe}} = 48.5$, $M_{\text{No}} = 23.2$). To ensure results are not an artifact of this
procedure, we ran the main analyses on the subsets of participants who (a) responded $>$50 on a
0--100 scale running from ``Definitely false'' to ``Definitely true'' to the prompt ``There is a
cover-up or conspiracy surrounding Trump's attempted assassination'' ($n = 324$); (b) were coded
as ``YES'' by the LLM (i.e., ``MAYBE'' group excluded, $n = 256$); or (c) satisfied both (a) and
(b) ($n = 212$). As the figures below show, the pattern of results does not vary by the means of
classification.

We conducted comparable analyses for Experiment~2. Self-ratings by LLM-generated groupings again
suggest the classification was successful ($M_{\text{Yes}} = 69.7$, $M_{\text{Maybe}} = 49.6$,
$M_{\text{No}} = 22.2$).

\medskip
\noindent
\begin{minipage}[t]{0.325\textwidth}\vspace{0pt}\centering
  \includegraphics[width=\linewidth]{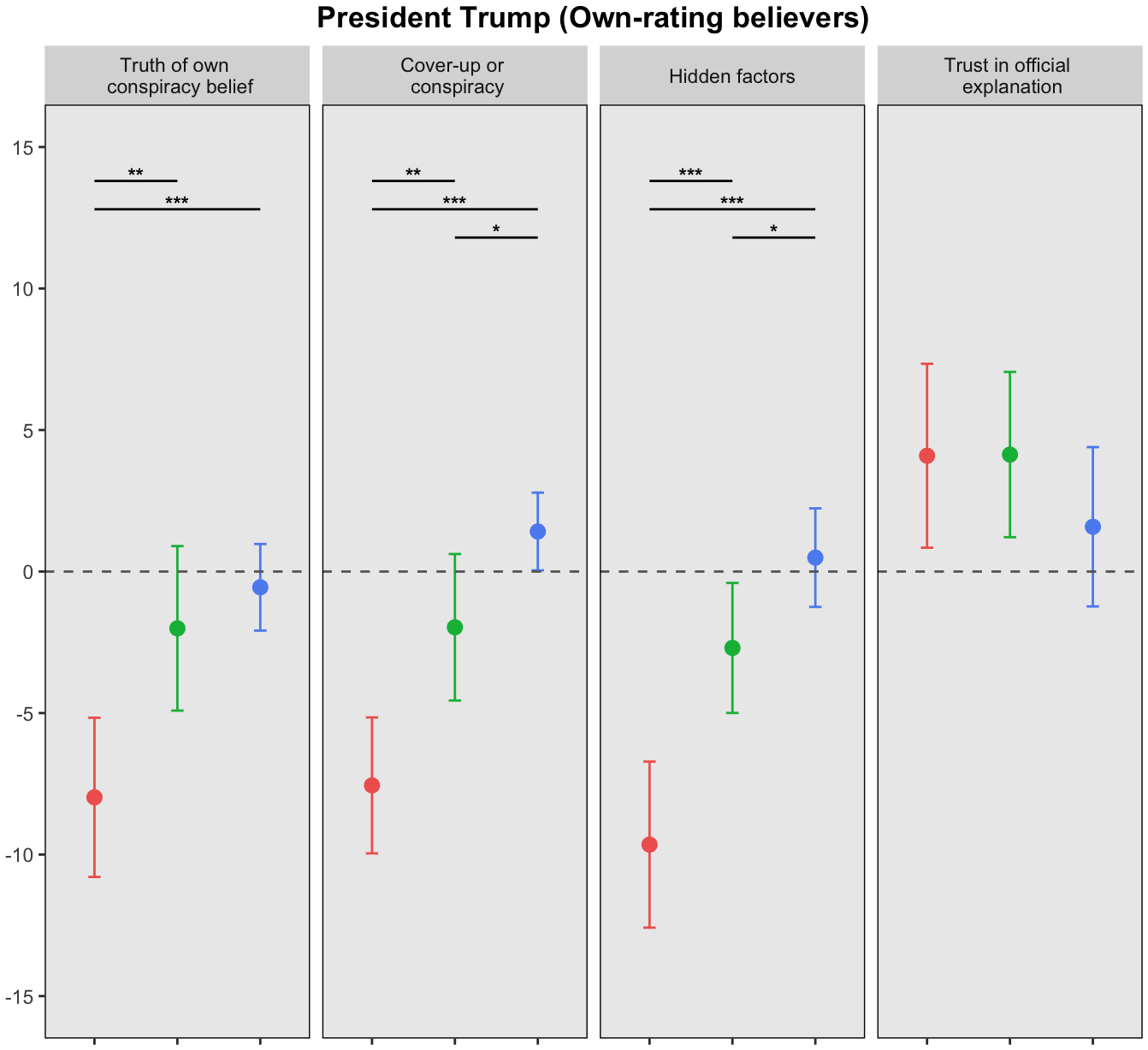}
\end{minipage}\hfill
\begin{minipage}[t]{0.325\textwidth}\vspace{0pt}\centering
  \includegraphics[width=\linewidth]{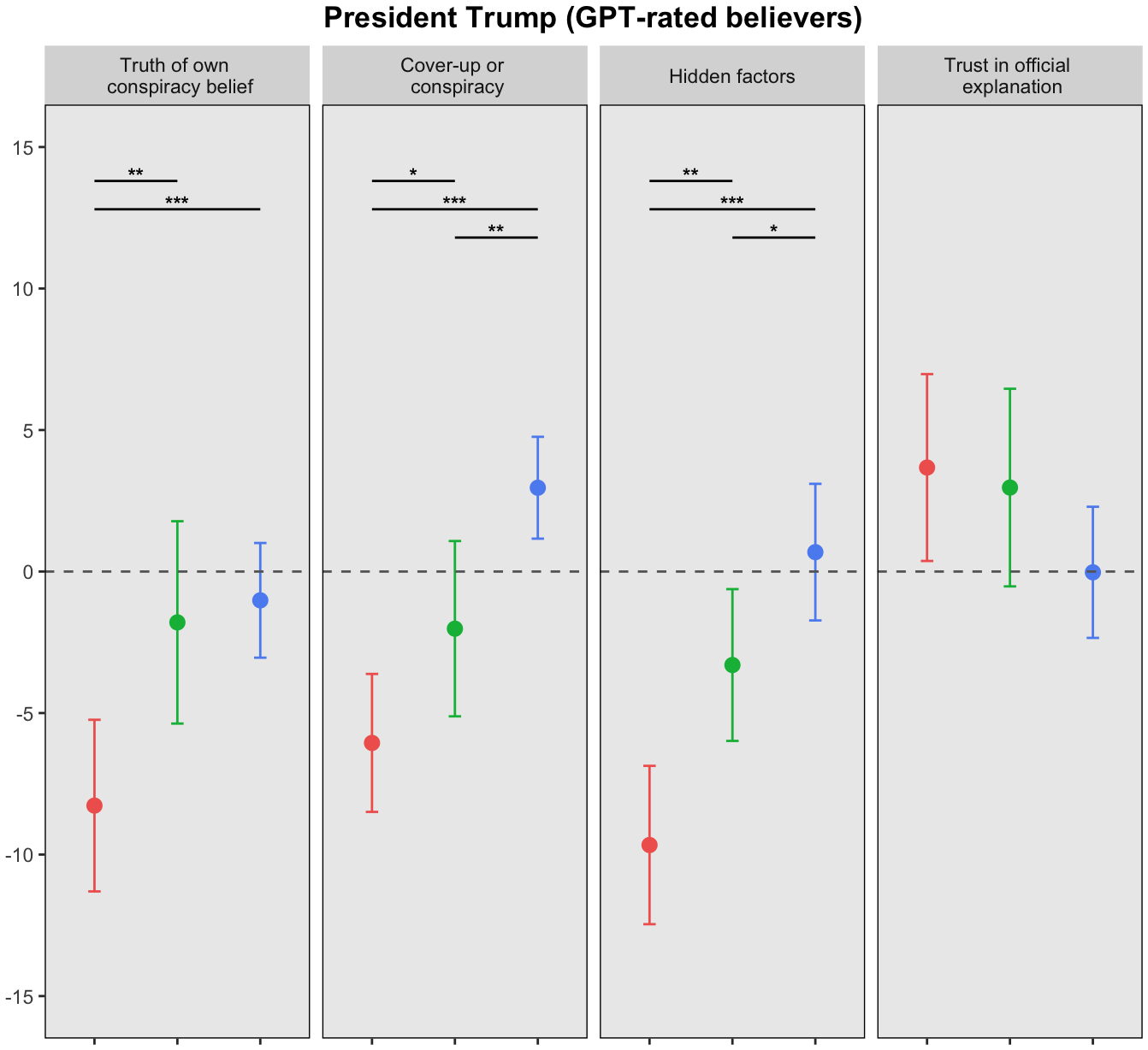}
\end{minipage}\hfill
\begin{minipage}[t]{0.325\textwidth}\vspace{0pt}\centering
  \includegraphics[width=\linewidth]{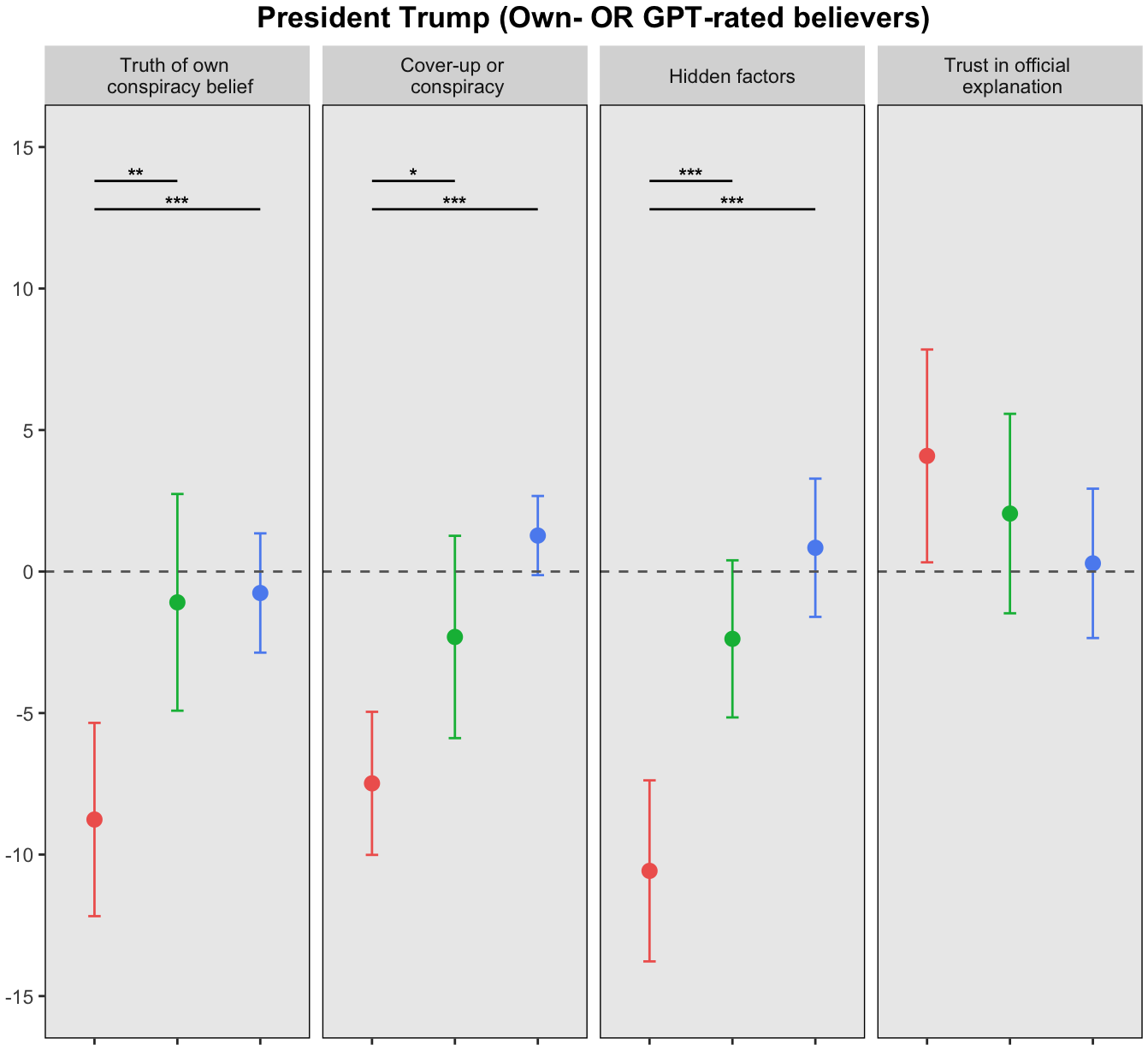}
\end{minipage}
\smcap{Figure SM4.1.}{Main analysis from Experiment~1 with just participants who answered $>$50
to ``There is a cover-up or conspiracy surrounding Trump's attempted assassination'' (left), just
participants who were coded ``YES'' by LLM conspiracy belief classification procedure (center),
or just participants who answered $>$50 to ``There is a cover-up or conspiracy surrounding
Trump's attempted assassination'' and were coded ``YES'' by LLM conspiracy belief classification
procedure (right).}

\clearpage
\noindent
\begin{minipage}[t]{0.325\textwidth}\vspace{0pt}\centering
  \includegraphics[width=\linewidth]{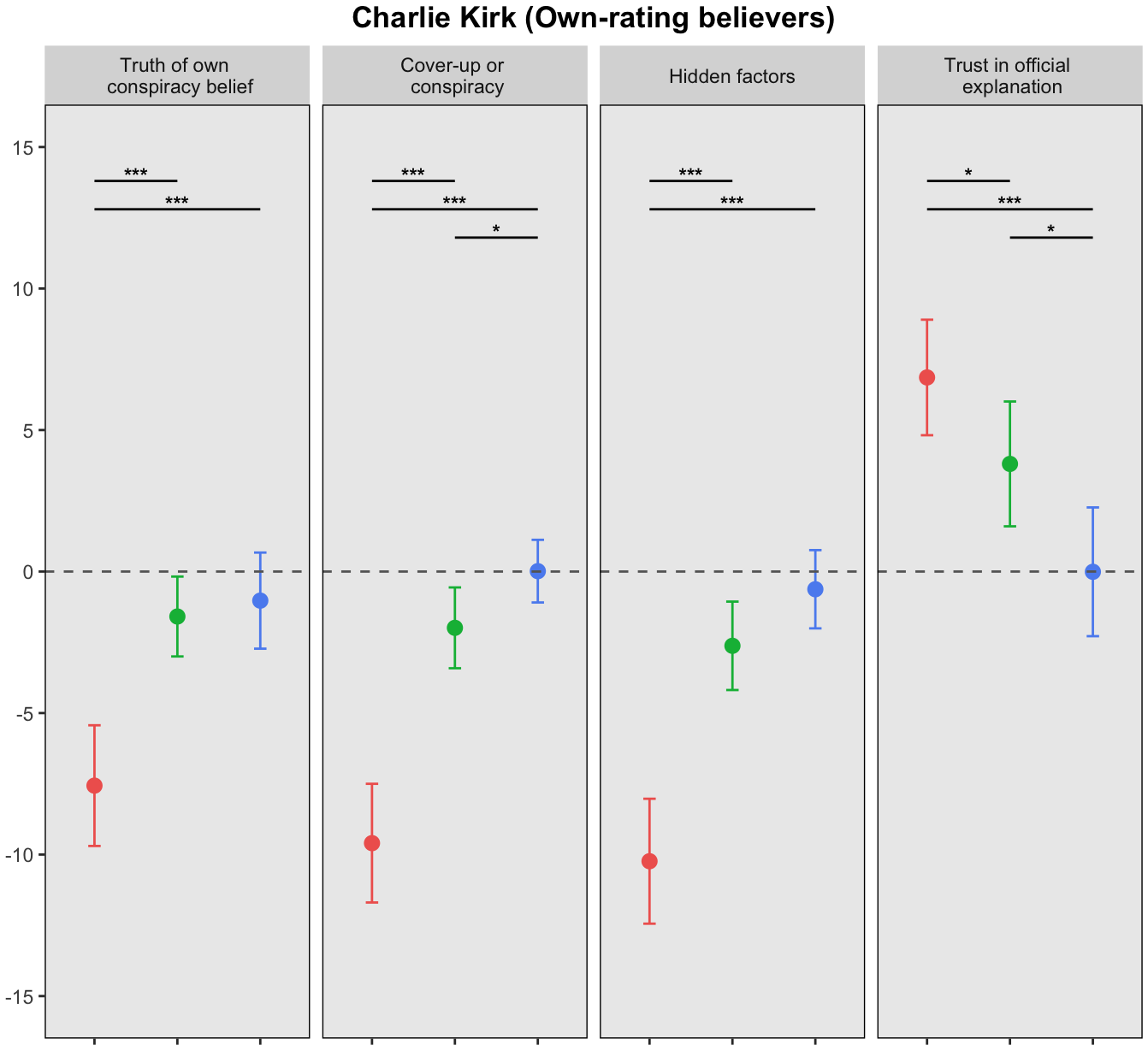}
\end{minipage}\hfill
\begin{minipage}[t]{0.325\textwidth}\vspace{0pt}\centering
  \includegraphics[width=\linewidth]{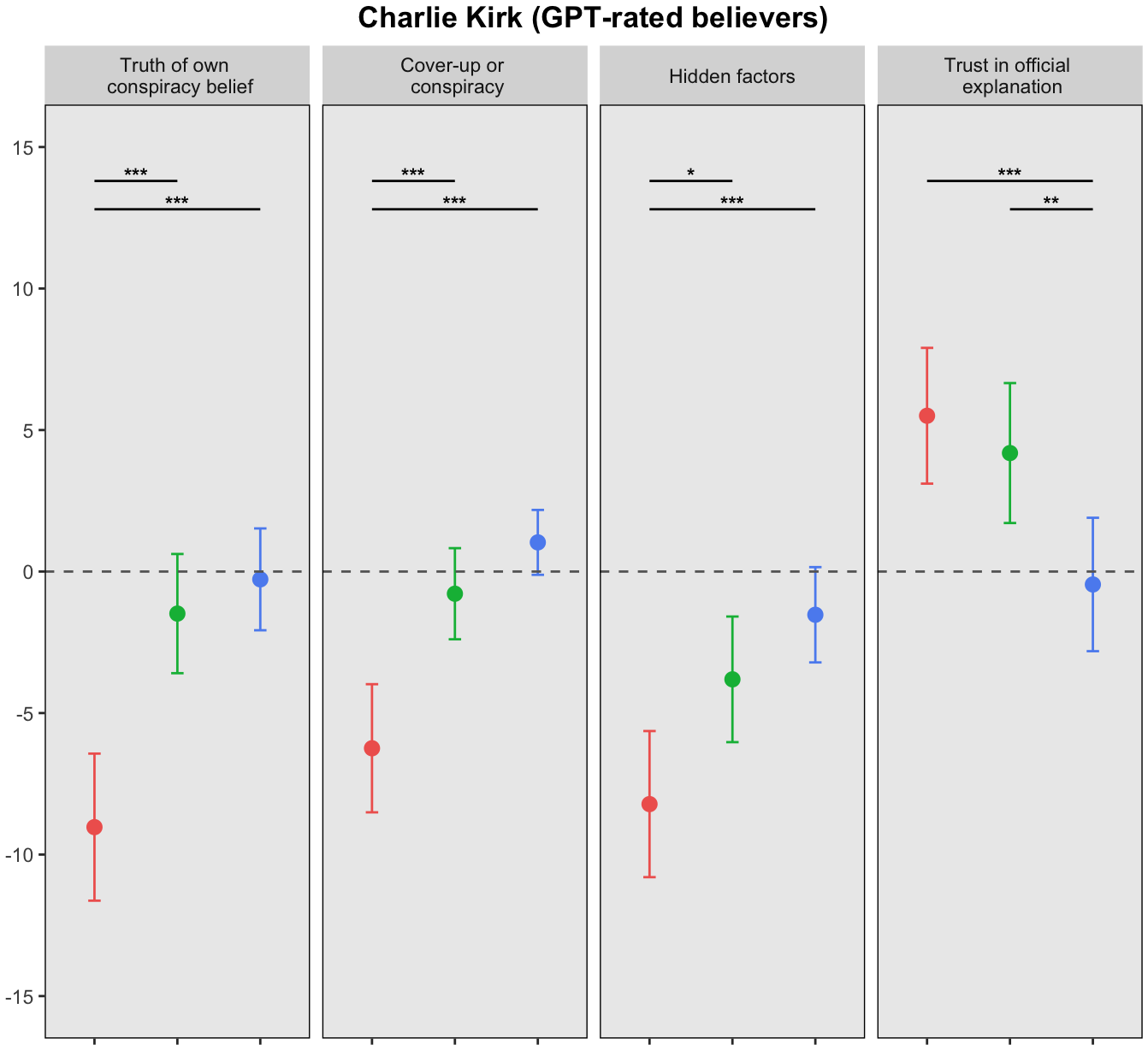}
\end{minipage}\hfill
\begin{minipage}[t]{0.325\textwidth}\vspace{0pt}\centering
  \includegraphics[width=\linewidth]{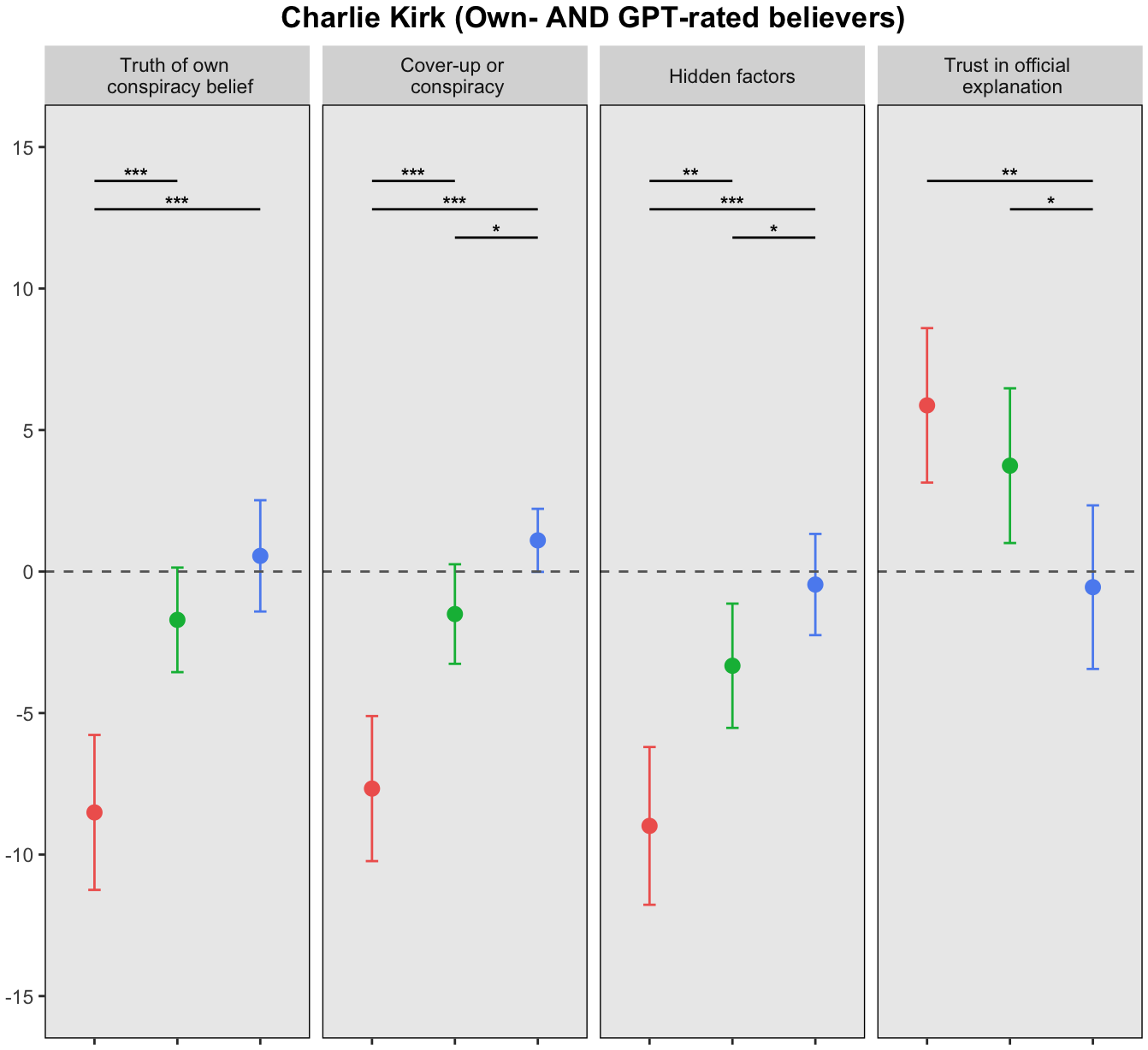}
\end{minipage}
\smcap{Figure SM4.2.}{Main analysis from Experiment~2 with just participants who answered $>$50
to ``There is a cover-up or conspiracy surrounding Kirk's assassination'' (left), just
participants who were coded ``YES'' by LLM conspiracy belief classification procedure (center),
or just participants who answered $>$50 to ``There is a cover-up or conspiracy surrounding
Kirk's assassination'' and were coded ``YES'' by LLM conspiracy belief classification procedure
(right).}

\clearpage
\section{SM5. Characteristics of conspiracy theory believers versus non-believers}
\label{sec:sm5}

\medskip
\noindent\hfill
\begin{minipage}[t]{0.40\textwidth}\vspace{0pt}\centering
  \includegraphics[width=\linewidth]{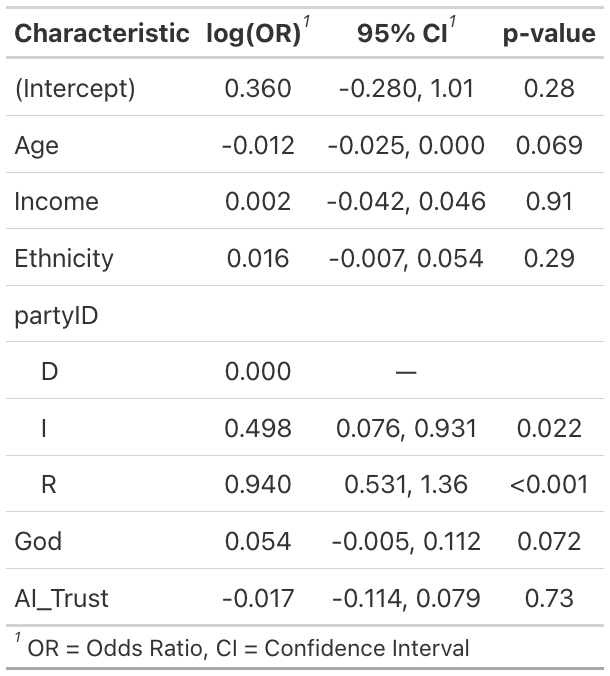}
\end{minipage}\hfill
\begin{minipage}[t]{0.50\textwidth}\vspace{0pt}\centering
  \includegraphics[width=\linewidth]{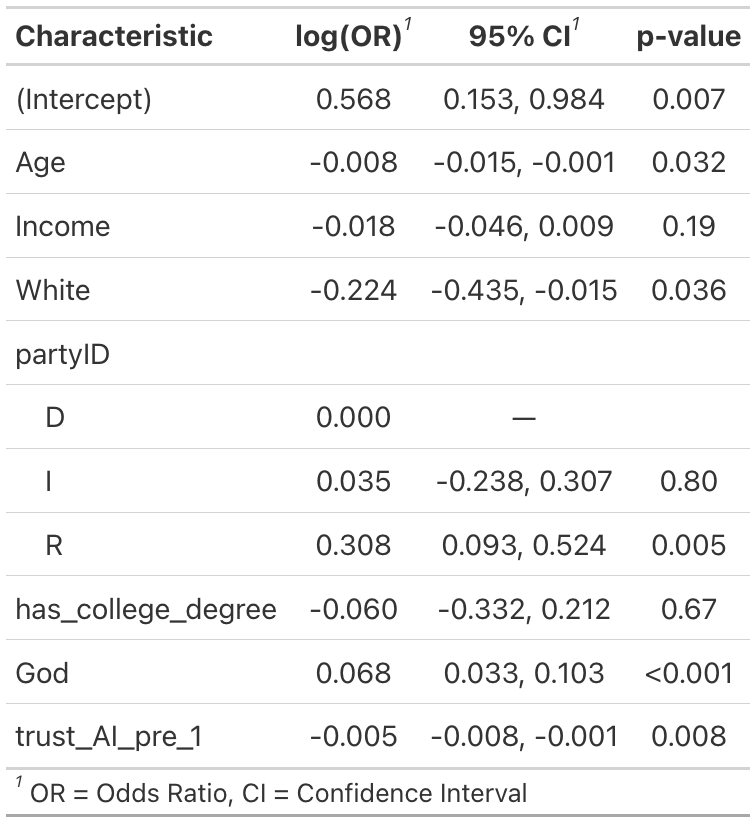}
\end{minipage}\hfill\null
\smcap{Table SM5.1.}{Logistic regression with conspiracy theory belief (1,0) as outcome using LLM
classifications for Experiment~1 (left) and Experiment~2 (right).}

\clearpage
\section{SM6. Treatment x party ID interaction}
\label{sec:sm6}

\medskip
\noindent
\begin{minipage}[t]{0.48\textwidth}\vspace{0pt}\centering
  \includegraphics[width=\linewidth]{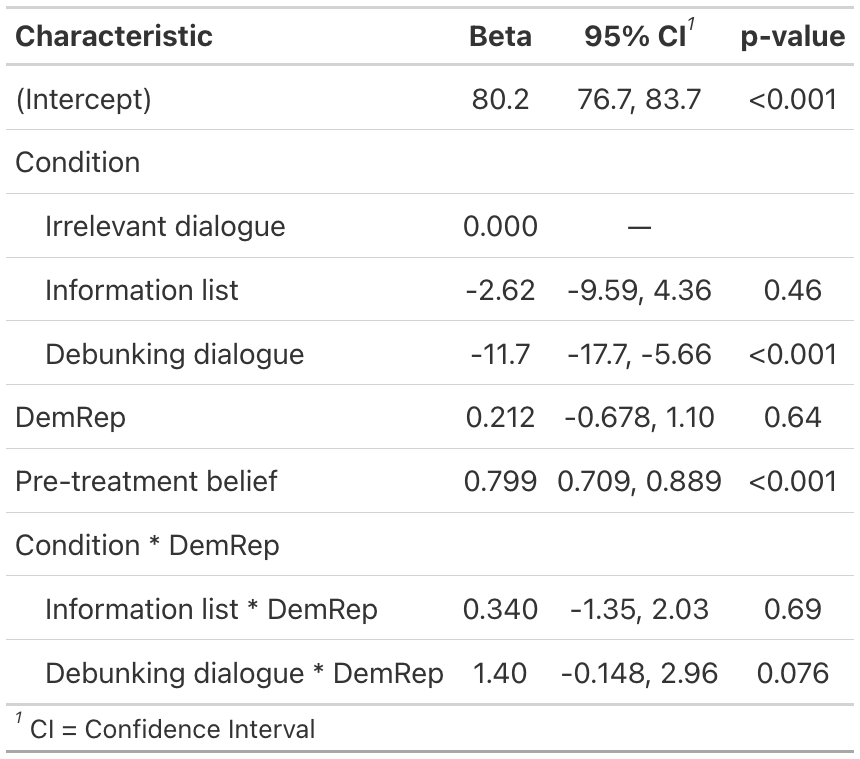}
\end{minipage}\hfill
\begin{minipage}[t]{0.48\textwidth}\vspace{0pt}\centering
  \includegraphics[width=\linewidth]{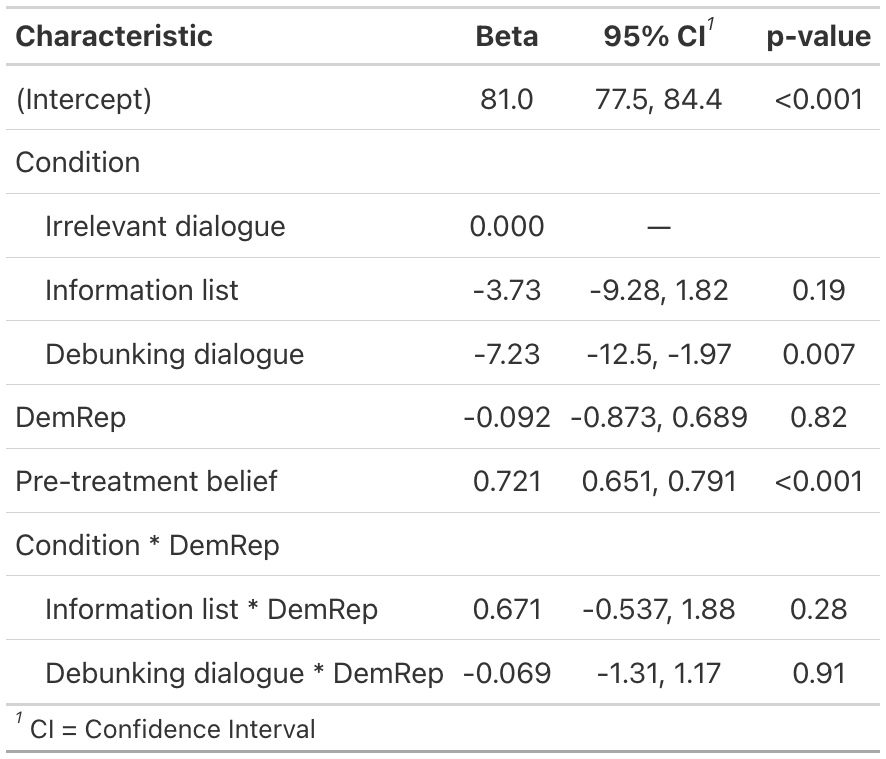}
\end{minipage}
\smcap{Table SM6.1.}{OLS regression with post-treatment confidence in conspiracy statement as
outcome for Experiment~1 (left) and Experiment~2 (right).}

\clearpage
\noindent
\begin{minipage}[t]{0.48\textwidth}\vspace{0pt}\centering
  \includegraphics[width=\linewidth]{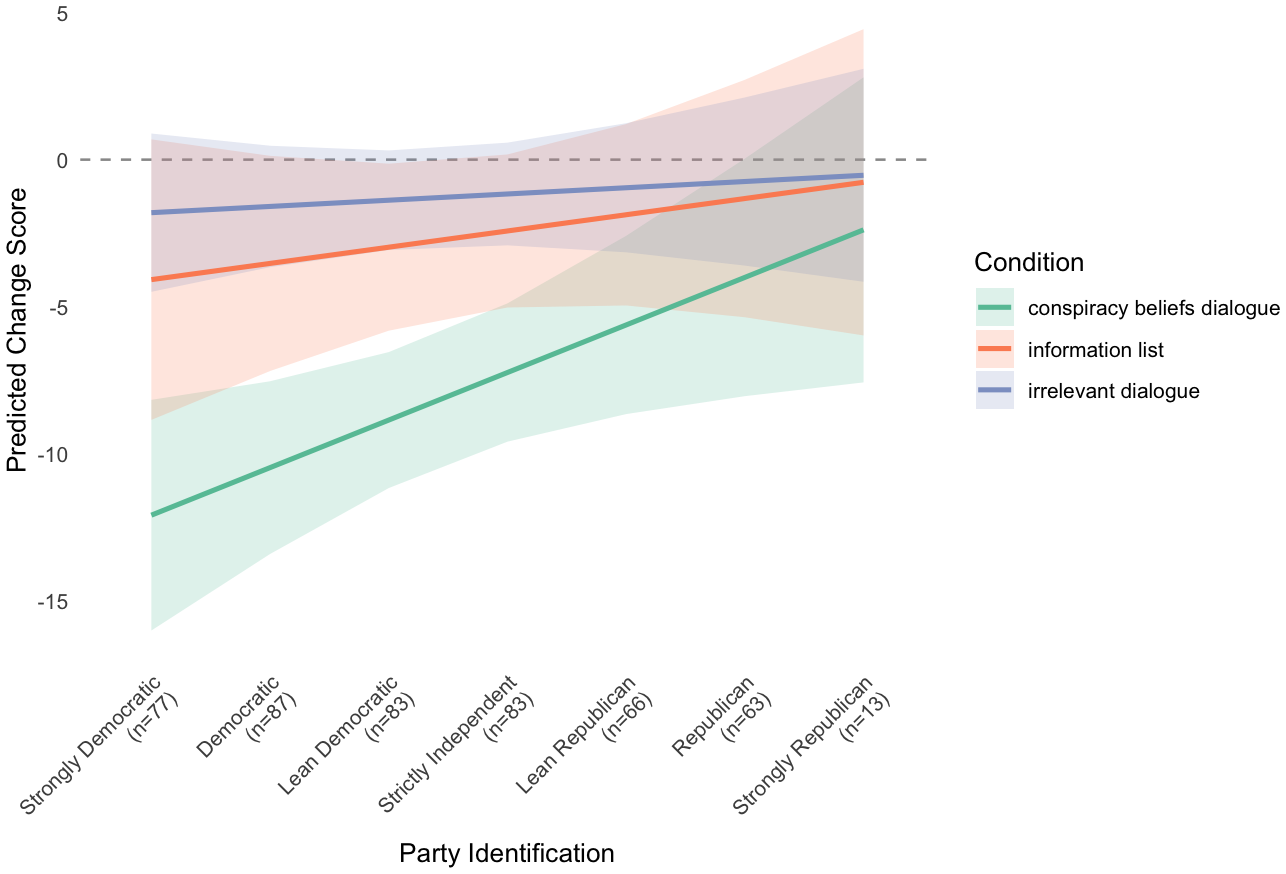}
\end{minipage}\hfill
\begin{minipage}[t]{0.48\textwidth}\vspace{0pt}\centering
  \includegraphics[width=\linewidth]{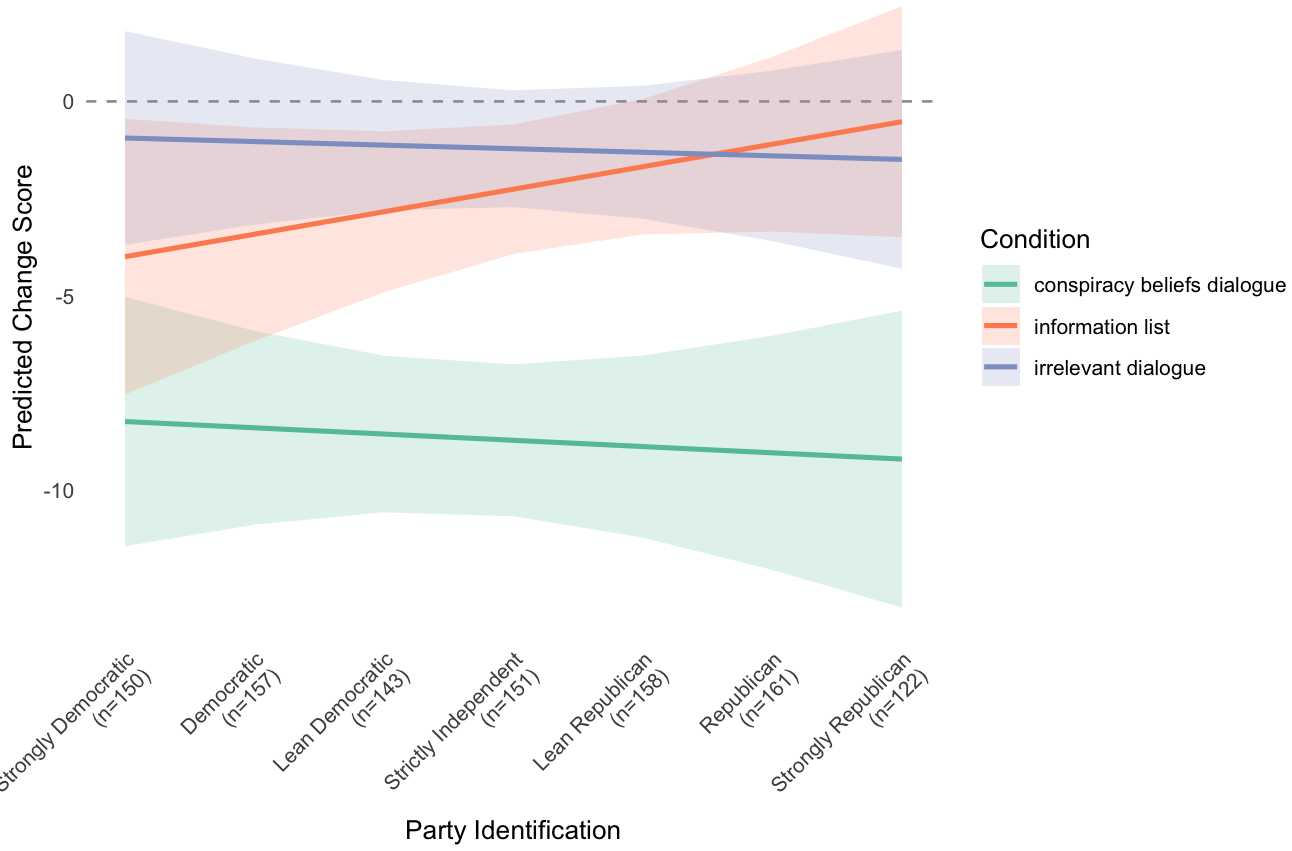}
\end{minipage}
\smcap{Figure SM6.1.}{Visualization of party ID x treatment interaction on change in confidence
in the truth of participants' own conspiracy beliefs in Experiment~1 (left) and Experiment~2
(right).}

\clearpage
\section{SM7. Treatment x pre-belief interaction}
\label{sec:sm7}

\medskip
\noindent
\begin{minipage}[t]{0.48\textwidth}\vspace{0pt}\centering
  \includegraphics[width=\linewidth]{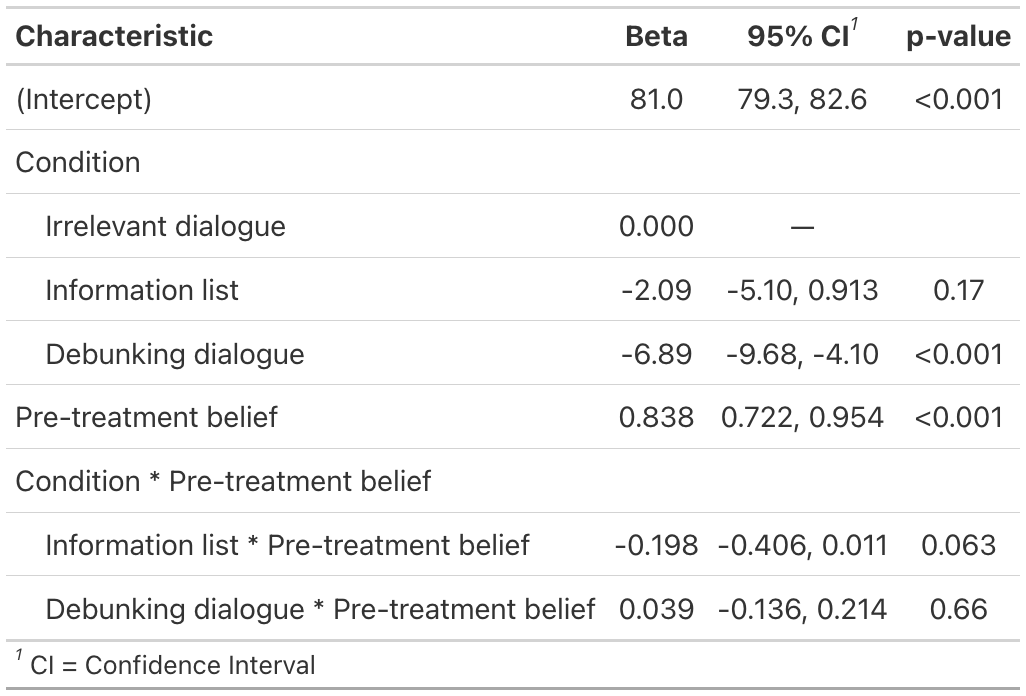}
\end{minipage}\hfill
\begin{minipage}[t]{0.48\textwidth}\vspace{0pt}\centering
  \includegraphics[width=\linewidth]{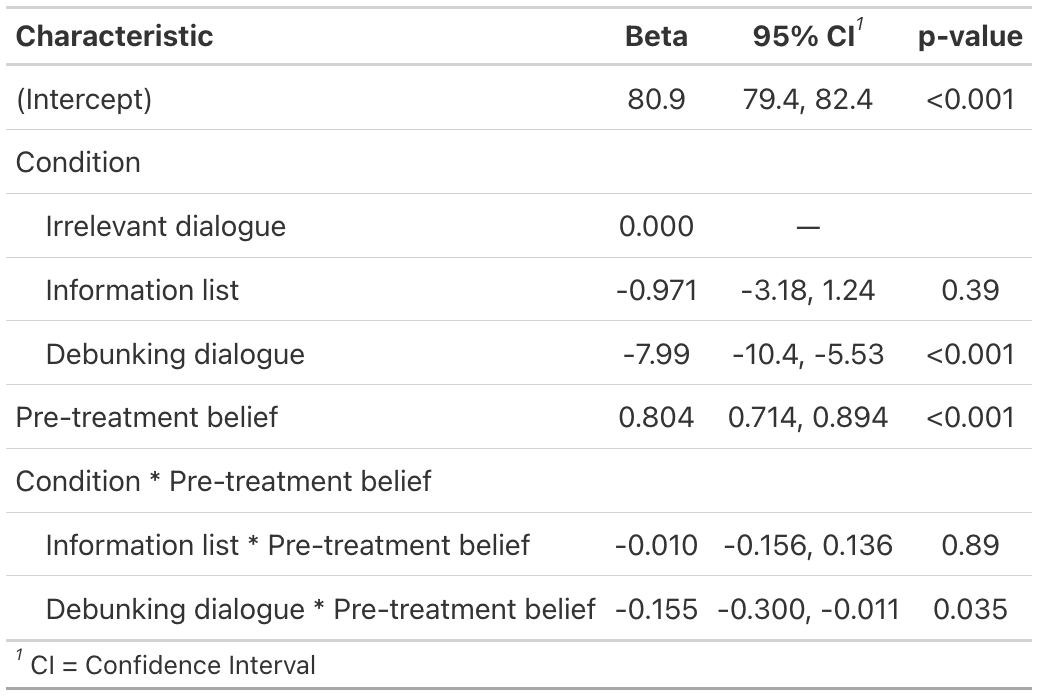}
\end{minipage}
\smcap{Table SM7.1.}{OLS regression with post-treatment confidence in conspiracy statement as
outcome for Experiment~1 (left) and Experiment~2 (right).}

\clearpage
\noindent
\begin{minipage}[t]{0.48\textwidth}\vspace{0pt}\centering
  \includegraphics[width=\linewidth]{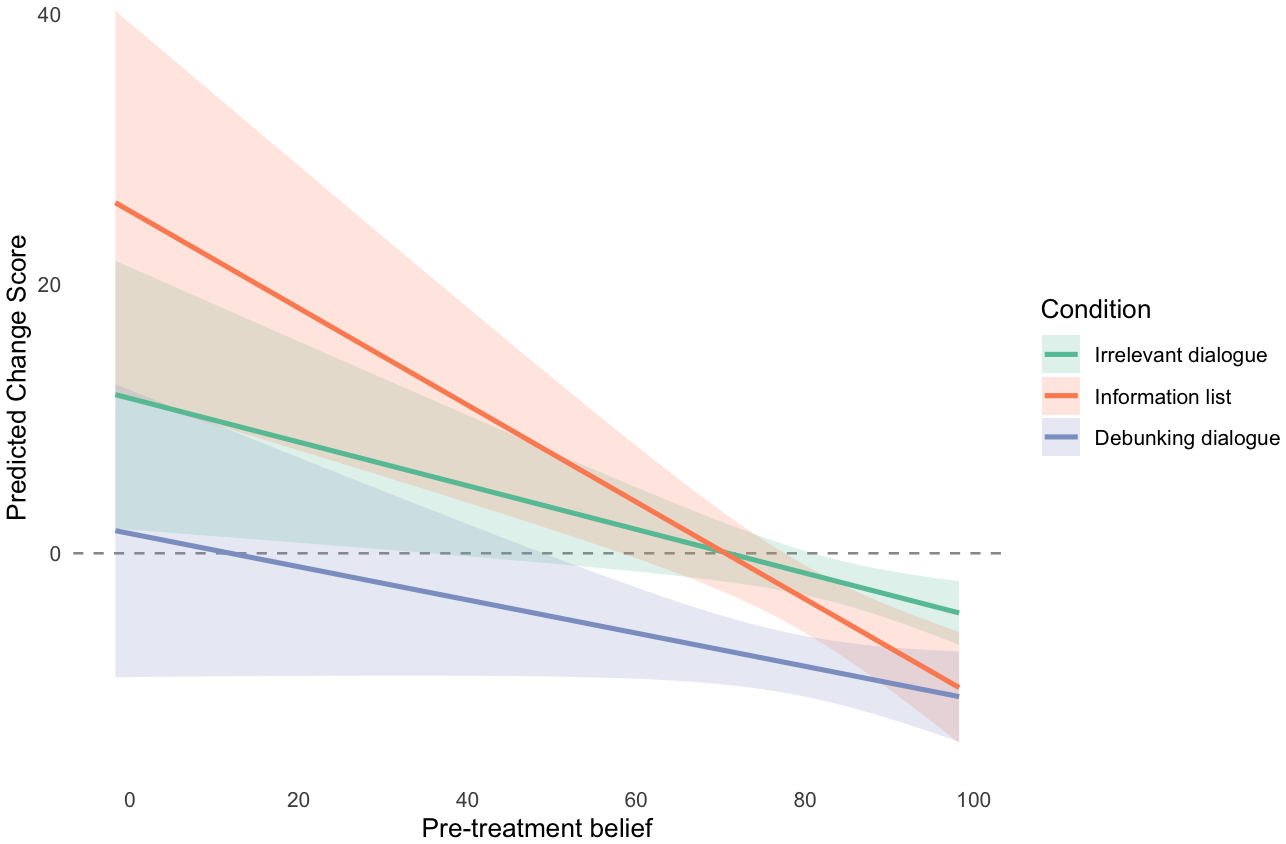}
\end{minipage}\hfill
\begin{minipage}[t]{0.48\textwidth}\vspace{0pt}\centering
  \includegraphics[width=\linewidth]{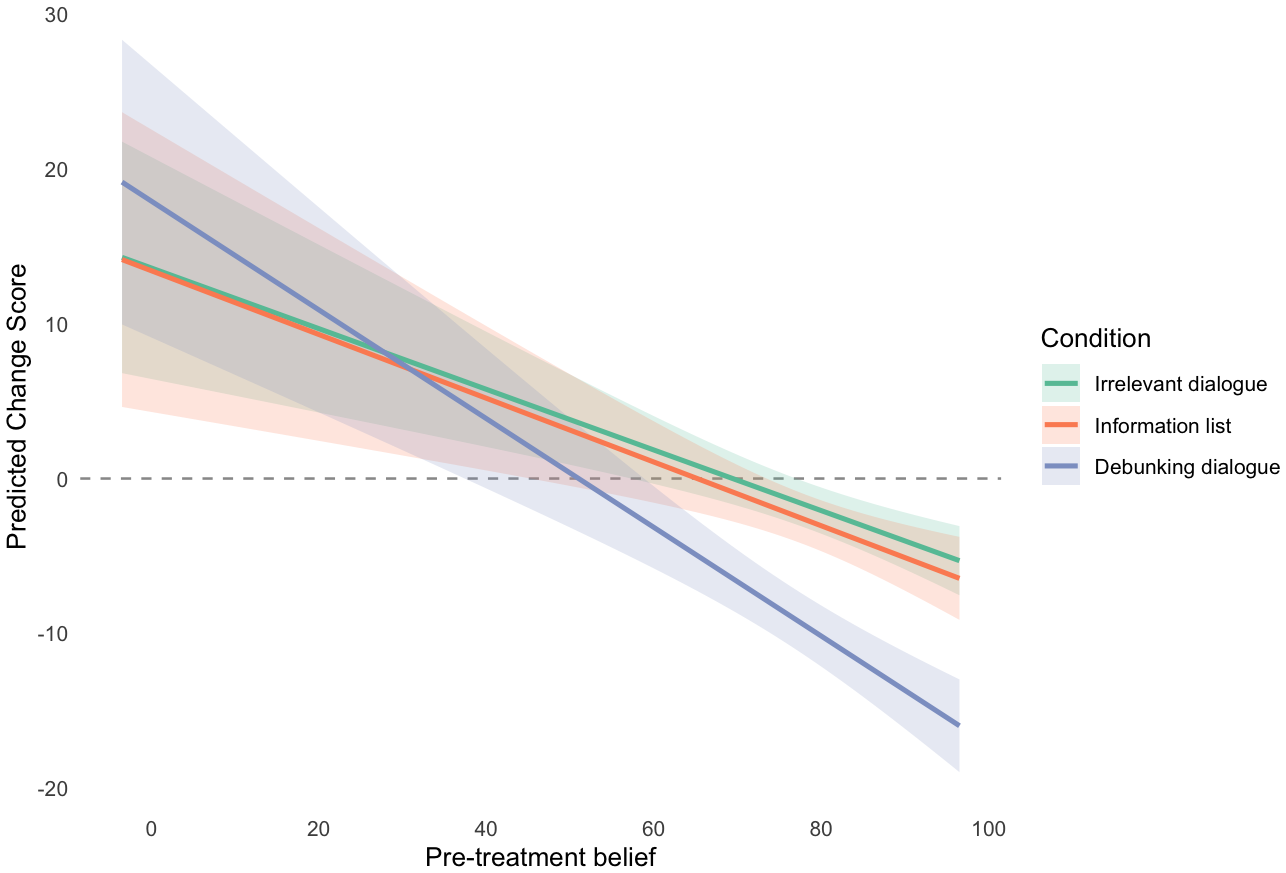}
\end{minipage}
\smcap{Figure SM7.1.}{Visualization of pre-treatment confidence in trust of participants' own
conspiracy beliefs x treatment interaction on change in confidence for Experiment~1 (left) and
Experiment~2 (right).}

\clearpage
\section{SM8. LLM rhetorical strategy coding}
\label{sec:sm8}

\medskip
\begin{center}
  \includegraphics[width=\textwidth]{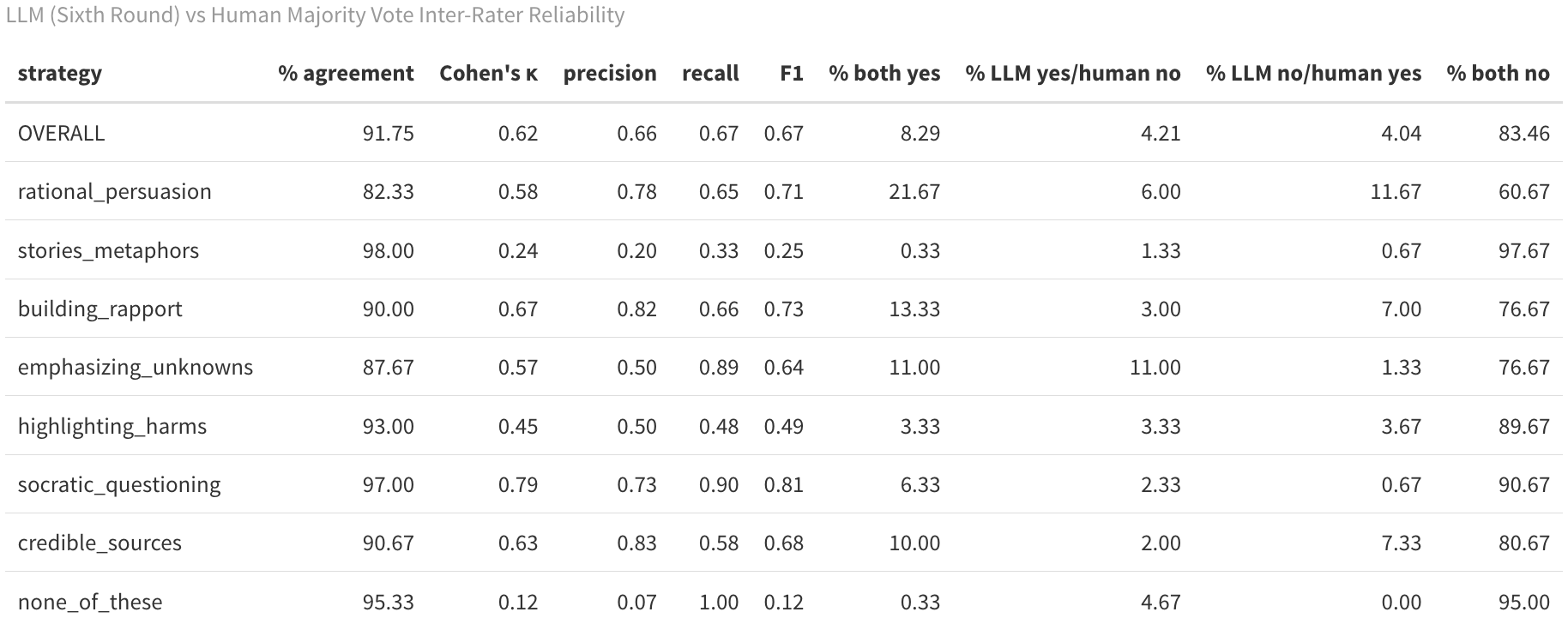}
\end{center}
\vspace{-8pt}
\smcap{Table SM8.1.}{Agreement statistics for the 300 sentences coded by humans (majority votes
from 9--11 people) and LLM (GPT-4o).}

\clearpage
\begin{center}
  \includegraphics[width=\textwidth]{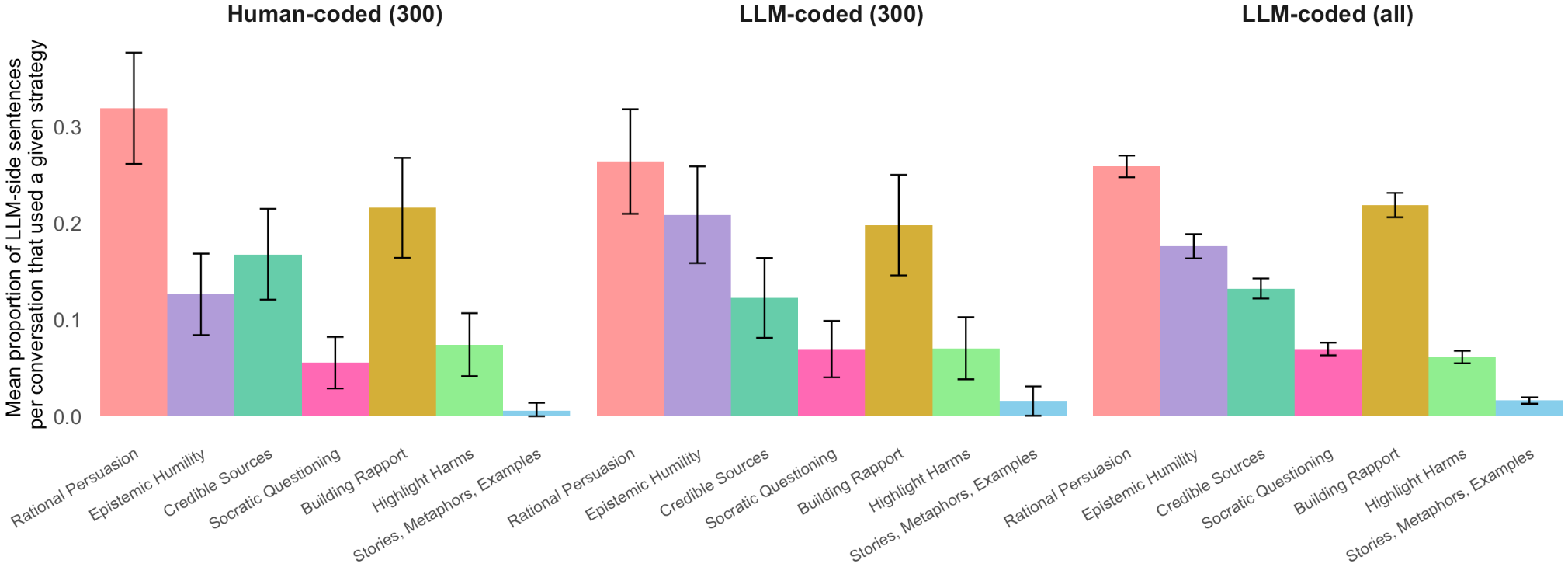}
\end{center}
\vspace{-8pt}
\smcap{Figure SM8.1.}{Proportions of categories selected by the human raters for 300 randomly
selected sentences (left) compared to the LLM categories for the same 300 sentences (center) and
for all sentences (right).}

\clearpage
\noindent
\begin{minipage}[t]{0.46\textwidth}\vspace{0pt}\centering
  \includegraphics[width=\linewidth]{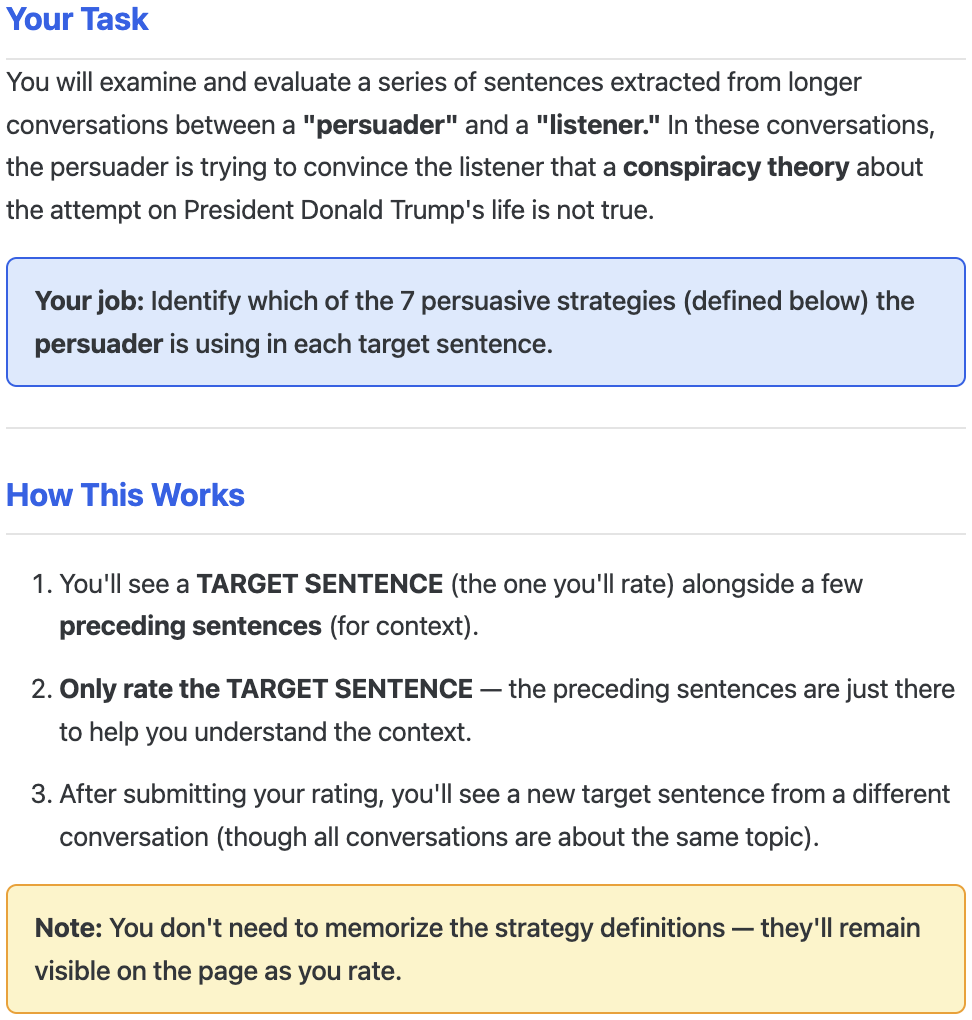}
\end{minipage}\hfill
\begin{minipage}[t]{0.46\textwidth}\vspace{0pt}\centering
  \includegraphics[width=\linewidth]{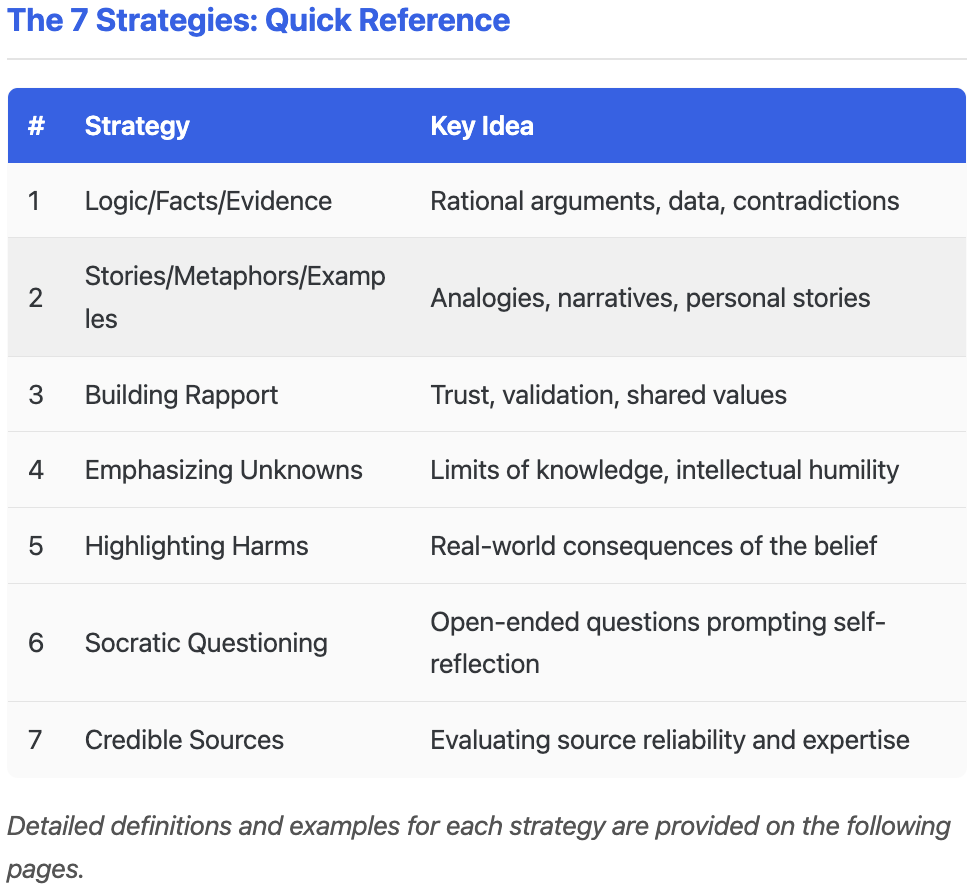}
\end{minipage}
\smcap{Figure SM8.2.}{Coding task instructions. Nearly identical wording was used for human
coders and LLM.}

\clearpage
\begin{center}
  \begin{minipage}[t]{0.42\textwidth}\vspace{0pt}\centering
  \includegraphics[width=\linewidth]{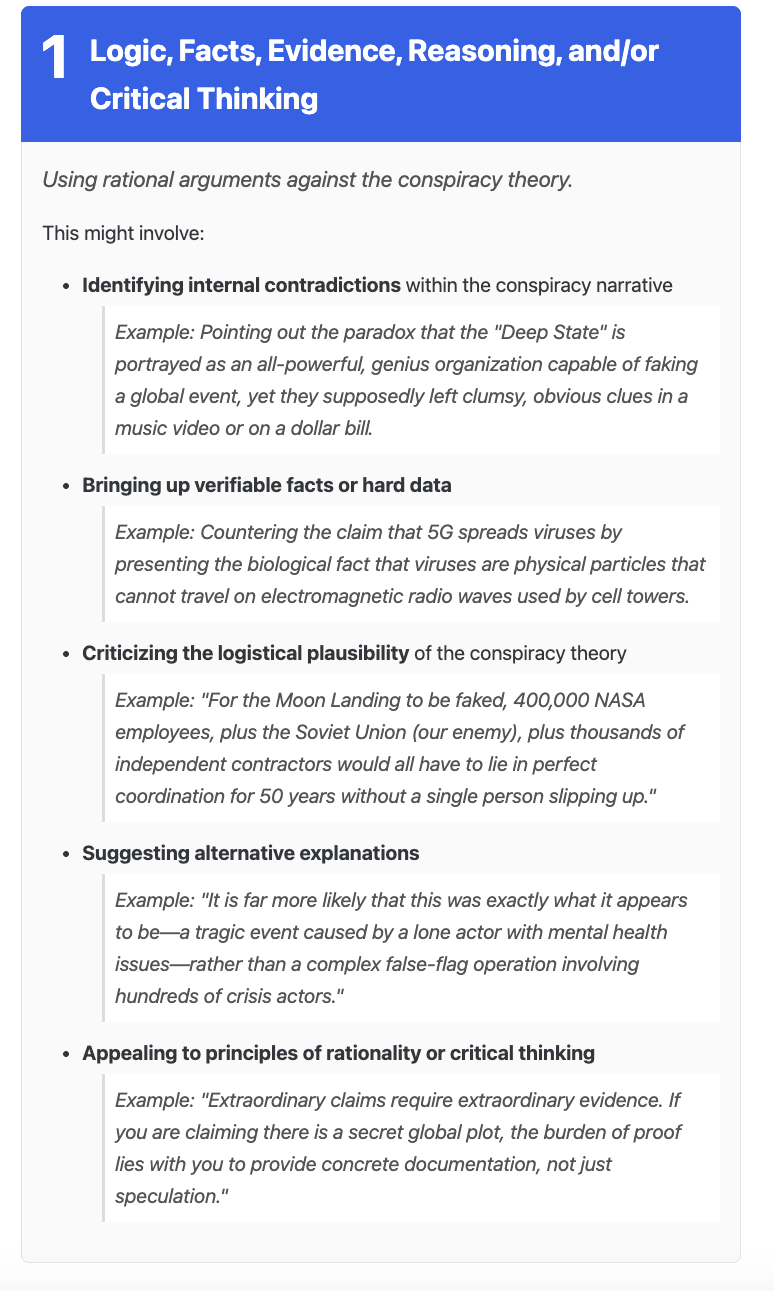}
\end{minipage}\hfill
  \begin{minipage}[t]{0.42\textwidth}\vspace{0pt}\centering
  \includegraphics[width=\linewidth]{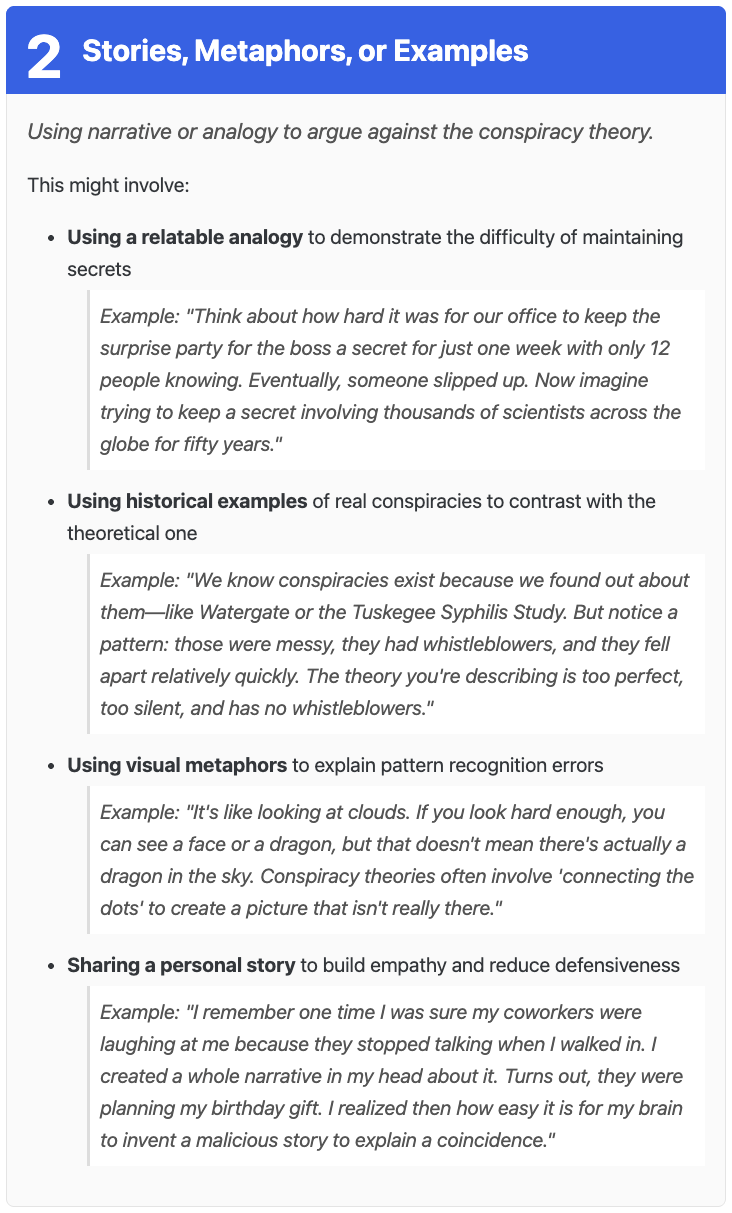}
\end{minipage}\\[10pt]
  \begin{minipage}[t]{0.42\textwidth}\vspace{0pt}\centering
  \includegraphics[width=\linewidth]{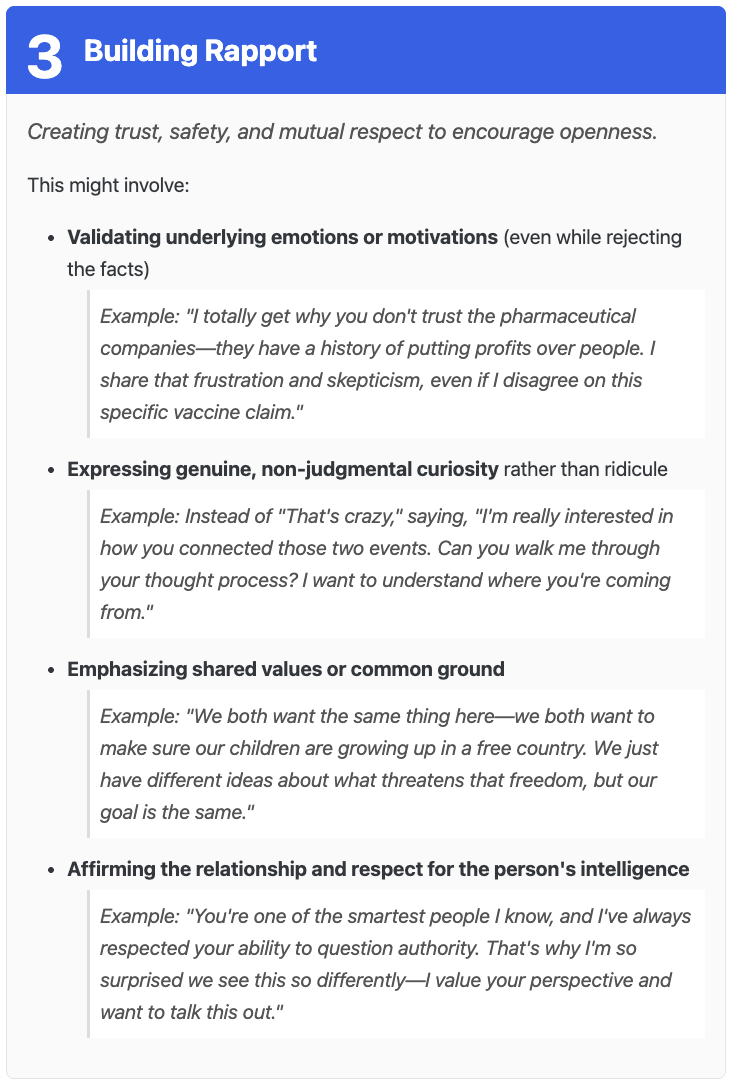}
\end{minipage}\hfill
  \begin{minipage}[t]{0.42\textwidth}\vspace{0pt}\centering
  \includegraphics[width=\linewidth]{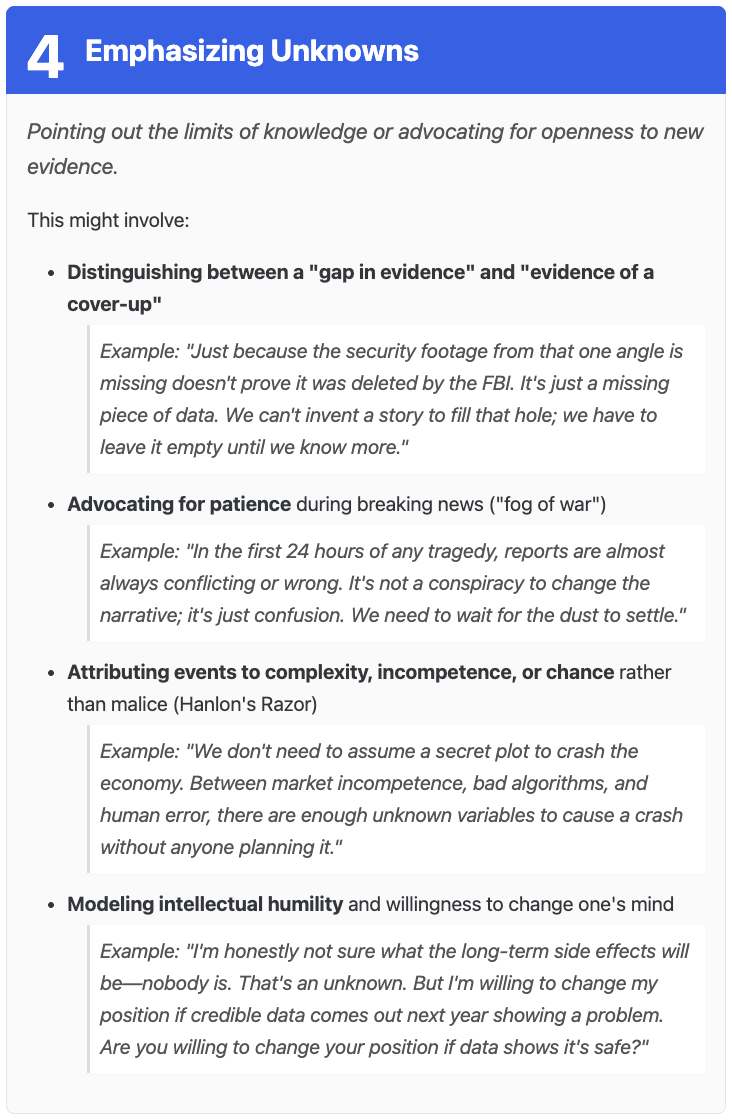}
\end{minipage}\\[10pt]
  \begin{minipage}[t]{0.42\textwidth}\vspace{0pt}\centering
  \includegraphics[width=\linewidth]{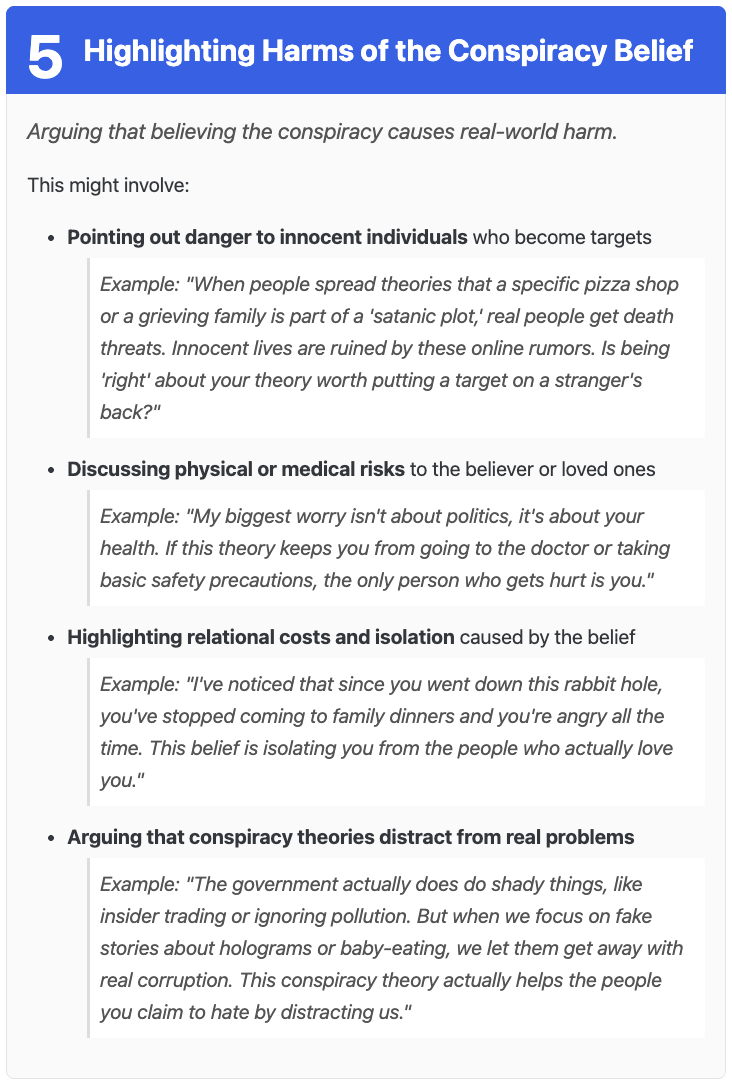}
\end{minipage}\hfill
  \begin{minipage}[t]{0.42\textwidth}\vspace{0pt}\centering
  \includegraphics[width=\linewidth]{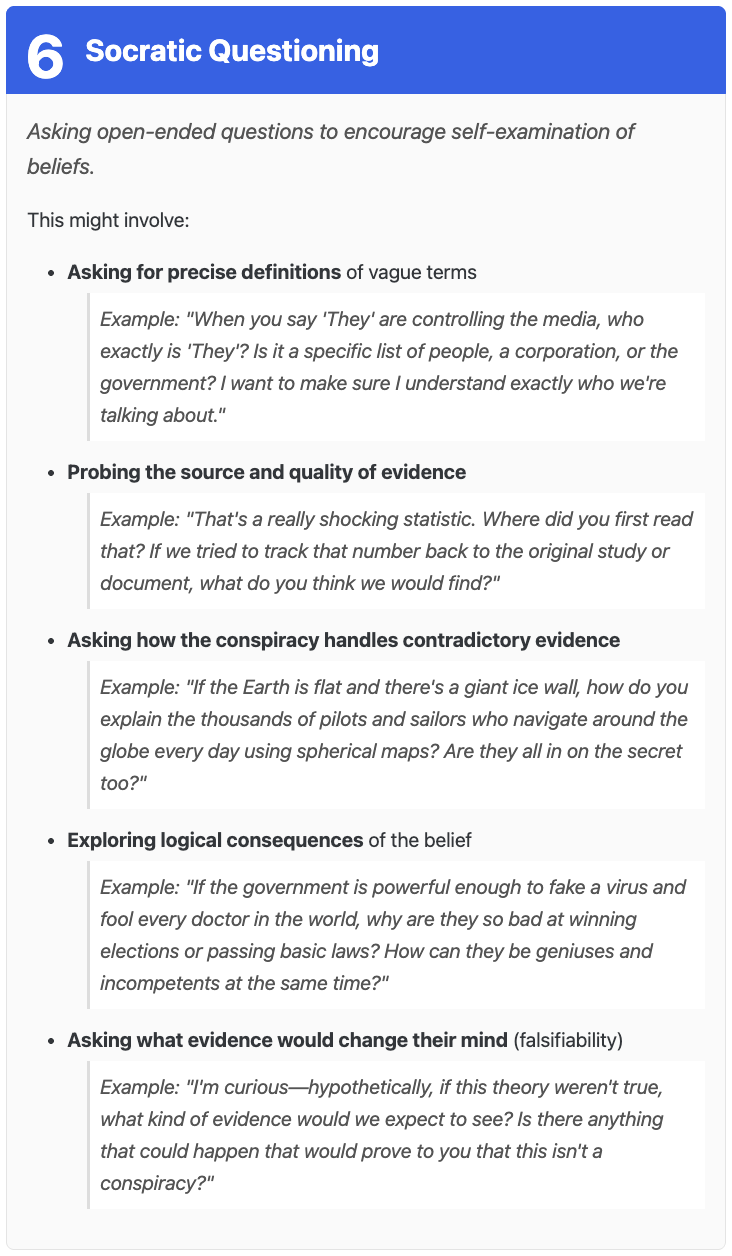}
\end{minipage}\\[10pt]
  \begin{minipage}[t]{0.42\textwidth}\vspace{0pt}\centering
  \includegraphics[width=\linewidth]{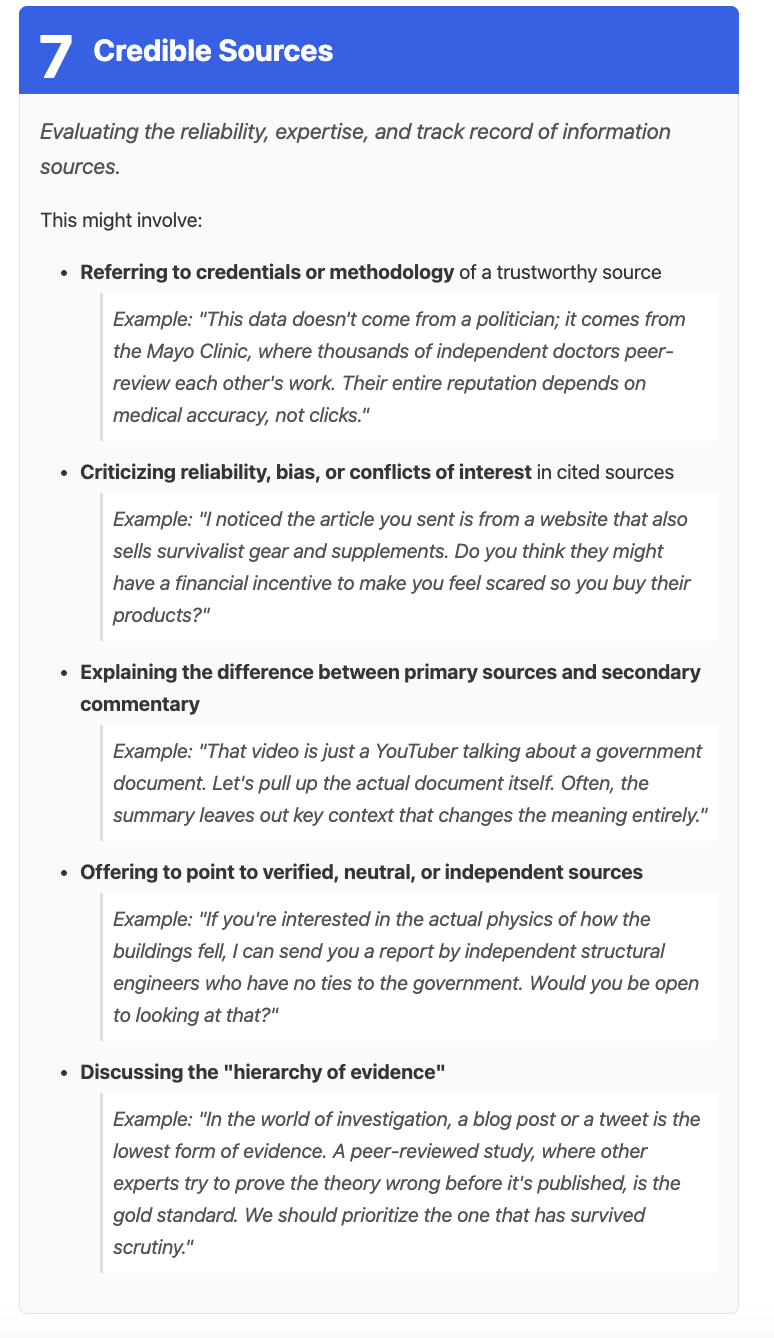}
\end{minipage}
\end{center}
\vspace{-8pt}
\smcap{Figure SM8.3.}{Detailed descriptions of coding categories provided to both human coders
and LLM.}

\clearpage
\begin{center}
  \includegraphics[width=0.5\textwidth]{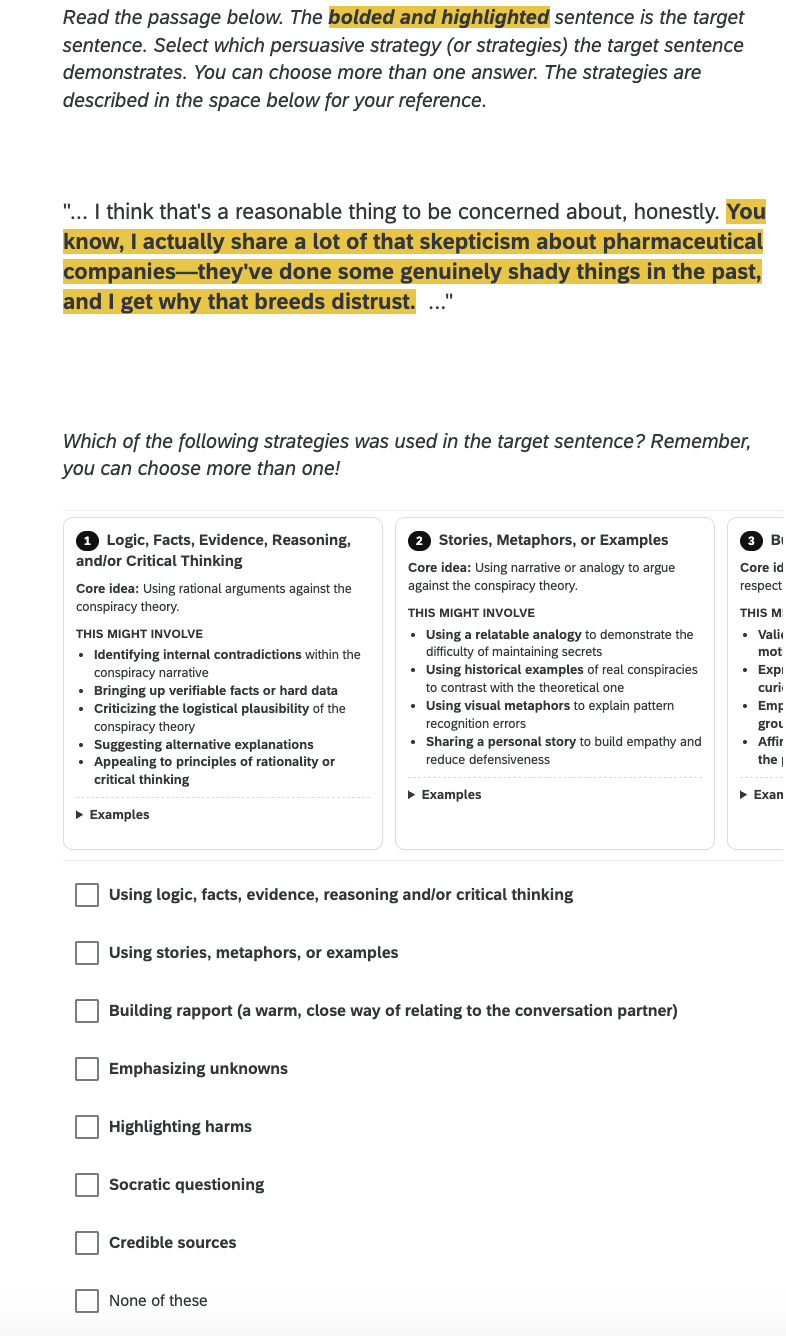}
\end{center}
\vspace{-8pt}
\smcap{Figure SM8.4.}{Example of in-task appearance for human coders. ``Category cards'' above
the response options scrolled to allow viewing. ``Examples'' arrow displayed a collapsed window
of examples.}

\clearpage
\section{SM9. Lee bounds for differential attrition on the immediate outcomes}
\label{sec:sm9}

Attrition between the pre- and post-treatment measures was higher in the Debunking Dialogue
condition than in the control conditions in both experiments (Experiment~1: 32 Debunking
Dialogue, 3 Irrelevant Dialogue, 0 Information List, $\chi^2(2, N = 766) = 25.00$, $p < .001$;
Experiment~2: 28, 12, and 3, respectively, $\chi^2(2, N = 1982) = 7.25$, $p = .027$). The main-text analyses carry each dropout's
pre-treatment value forward to post-treatment. That procedure is an imputation under the
assumption that the intervention had no effect on participants who quit; because it does not
bound the effect under alternative assumptions about who dropped out, we also report sharp
trimming bounds (Lee, 2009, \emph{Review of Economic Studies} 76:1071--1102), which characterize
the set of treatment effects consistent with \emph{any} pattern of selective dropout under a
single monotonicity assumption.

For each of the four post-treatment outcomes and each pairwise contrast (Debunking Dialogue vs.\
Irrelevant Dialogue; Debunking Dialogue vs.\ Information List) we take the randomized believers
in the two conditions and define response as having an observed post-treatment value for that
outcome (retention rates are given in Table SM9.1). The identifying assumption is monotonicity,
i.e., assignment can affect whether a participant responds in only one direction, so that every
participant who would have responded under the worse-retained condition's assignment would also
have responded under the better-retained condition's. Under monotonicity and random assignment,
the responders in the worse-retained condition are representative of ``always-responders,''
whereas the responders in the better-retained condition additionally contain a fraction
$1 - p_{\text{worse}}/p_{\text{better}}$ of participants whose response was induced by their
assignment ($p$ denoting the retention rate). Trimming that fraction from one tail of the
better-retained condition's outcome distribution and recomputing the difference in means gives
one end of the identified set and trimming it from the other tail gives the other end. The
identified set thus contains the treatment effect for always-responders. In Experiment~2 the
Debunking Dialogue condition was better retained than the Irrelevant Dialogue condition, so the
treatment condition is trimmed for those contrasts; in every other contrast the control
condition is trimmed. Trimming fractions are 0.7--2.4\% in Experiment~2 (retention 97.2\% in the
Debunking Dialogue, 96.5\% in the Irrelevant Dialogue, and 99.6\% in the Information List
condition) and 8.1--10.2\% in Experiment~1 (89.8\%, 97.7\%, and 99.0--100\%, respectively).
Sampling uncertainty is summarized by an outer 95\% confidence interval that extends the lower
bound downward and the upper bound upward by 1.96 bootstrap standard errors of the respective
bound (2,000 resamples of the two-condition sample). Because this interval covers the entire
identified set, it is conservative for the effect itself. Bounds are computed on unadjusted
post-treatment means, without covariate adjustment or imputation, so the ``unadjusted
difference'' column differs slightly from the covariate-adjusted coefficients in the main text.

\begin{scriptsize}
\setlength{\tabcolsep}{3.5pt}
\begin{longtable}{llrccccc}
\toprule
\textbf{Outcome} & \textbf{Contrast} & \textbf{Unadj.\ diff.} & \textbf{Identified set} & \textbf{Outer 95\% CI} & \textbf{Retention DD / ctl} & \textbf{Trim (arm)} & \textbf{Sig.} \\
\midrule
\endfirsthead
\toprule
\textbf{Outcome} & \textbf{Contrast} & \textbf{Unadj.\ diff.} & \textbf{Identified set} & \textbf{Outer 95\% CI} & \textbf{Retention DD / ctl} & \textbf{Trim (arm)} & \textbf{Sig.} \\
\midrule
\endhead
\bottomrule
\endlastfoot
\multicolumn{8}{@{}l}{\textit{Experiment~1 (Trump)}} \\
Own belief & vs ID & $-9.13$ & $[-12.01,\ -7.49]$ & $[-16.61,\ -2.84]$ & 89.8\% / 97.7\% & 8.1\% (ID) & yes \\
Own belief & vs IL & $-4.58$ & $[-8.96,\ -1.80]$ & $[-13.90,\ 3.50]$ & 89.8\% / 100.0\% & 10.2\% (IL) & yes \\
Cover-up & vs ID & $-10.27$ & $[-14.57,\ -7.28]$ & $[-20.58,\ -1.16]$ & 89.8\% / 97.7\% & 8.1\% (ID) & yes \\
Cover-up & vs IL & $-8.01$ & $[-13.17,\ -3.95]$ & $[-19.26,\ 2.32]$ & 89.8\% / 100.0\% & 10.2\% (IL) & no \\
Hidden factors & vs ID & $-8.86$ & $[-13.31,\ -6.34]$ & $[-19.65,\ -0.03]$ & 89.8\% / 97.7\% & 8.1\% (ID) & yes \\
Hidden factors & vs IL & $-7.99$ & $[-13.35,\ -4.44]$ & $[-19.63,\ 2.27]$ & 89.8\% / 100.0\% & 10.2\% (IL) & yes \\
Trust official expl. & vs ID & $4.84$ & $[1.62,\ 8.78]$ & $[-4.44,\ 14.74]$ & 89.8\% / 97.7\% & 8.1\% (ID) & no \\
Trust official expl. & vs IL & $3.68$ & $[-0.42,\ 8.11]$ & $[-6.98,\ 14.46]$ & 89.8\% / 99.0\% & 9.3\% (IL) & no \\
\midrule
\multicolumn{8}{@{}l}{\textit{Experiment~2 (Kirk)}} \\
Own belief & vs ID & $-7.95$ & $[-8.18,\ -7.40]$ & $[-11.65,\ -3.77]$ & 97.2\% / 96.5\% & 0.7\% (DD) & yes \\
Own belief & vs IL & $-6.15$ & $[-7.55,\ -5.58]$ & $[-10.92,\ -2.25]$ & 97.2\% / 99.6\% & 2.4\% (IL) & yes \\
Cover-up & vs ID & $-7.49$ & $[-7.85,\ -7.07]$ & $[-11.96,\ -2.94]$ & 97.2\% / 96.5\% & 0.7\% (DD) & yes \\
Cover-up & vs IL & $-6.23$ & $[-7.64,\ -5.25]$ & $[-11.69,\ -1.27]$ & 97.2\% / 99.6\% & 2.4\% (IL) & yes \\
Hidden factors & vs ID & $-7.54$ & $[-7.86,\ -7.08]$ & $[-11.94,\ -2.95]$ & 97.2\% / 96.5\% & 0.7\% (DD) & yes \\
Hidden factors & vs IL & $-7.85$ & $[-9.34,\ -7.05]$ & $[-13.37,\ -3.02]$ & 97.2\% / 99.6\% & 2.4\% (IL) & yes \\
Trust official expl. & vs ID & $5.16$ & $[4.73,\ 5.51]$ & $[0.45,\ 9.77]$ & 97.2\% / 96.5\% & 0.7\% (DD) & yes \\
Trust official expl. & vs IL & $3.17$ & $[2.16,\ 4.53]$ & $[-1.99,\ 8.65]$ & 97.2\% / 99.6\% & 2.4\% (IL) & no \\
\end{longtable}
\end{scriptsize}
\vspace{-6pt}
\smcap{Table SM9.1.}{Lee (2009) sharp bounds on the immediate effects of the Debunking Dialogue on each post-treatment outcome, for both pairwise contrasts, in both experiments. A negative difference indicates less conspiratorial belief in the Debunking Dialogue condition (for trust in the official explanation, a positive difference indicates more trust).}
\vspace{-10pt}
\noindent{\scriptsize\emph{Note.} DD = Debunking Dialogue; ID = Irrelevant Dialogue; IL = Information List. Unadj.\ diff.\ = unadjusted difference in post-treatment means (DD minus comparison condition) among participants with an observed post-treatment response; identified set = Lee (2009) sharp bounds; outer 95\% CI = lower bound minus 1.96 bootstrap SEs, upper bound plus 1.96 bootstrap SEs (2,000 resamples, seed 20260901). Retention = share of randomized believers with an observed post-treatment value for that outcome. Trim = share of the better-retained condition (in parentheses) removed from one tail. Sig.\ = contrast reported as significant ($p \le .05$) in the main text. Experiment~1 rows use the 462 believers with a recorded condition (235 DD / 131 ID / 96 IL); Experiment~2 rows use the believer sample of the main text ($N = 1{,}042$; 530 DD / 257 ID / 255 IL), analyzed as randomized.\par}
\medskip

\begin{center}
  \includegraphics[width=\textwidth]{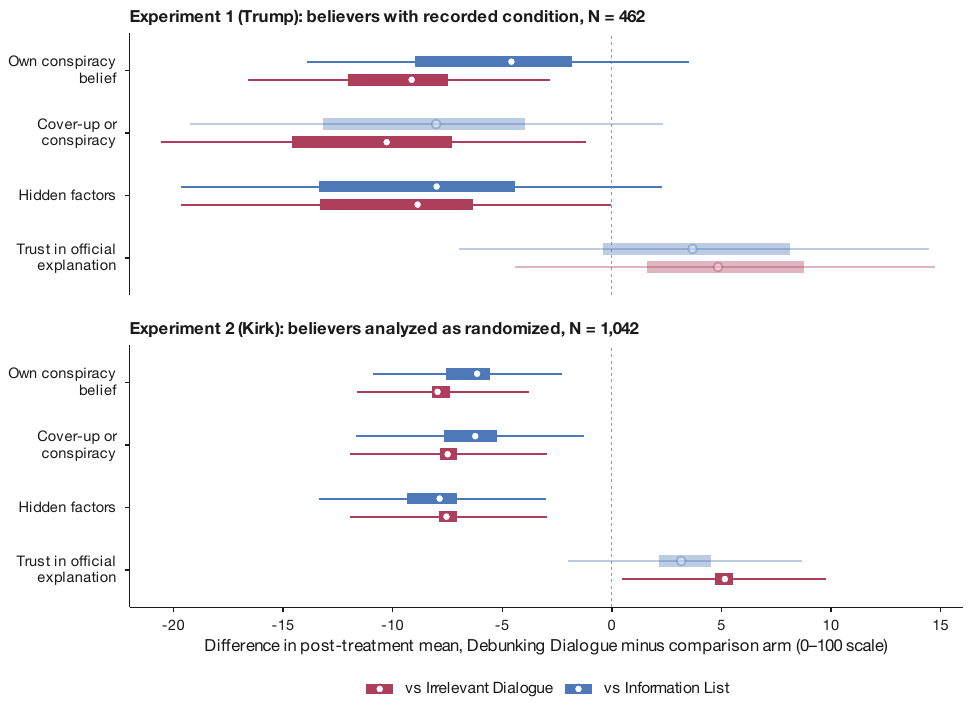}
\end{center}
\vspace{-8pt}
\smcap{Figure SM9.1.}{Lee bounds on the immediate treatment effects (data of Table SM9.1). Thick
bars are the identified sets, thin whiskers the outer 95\% confidence intervals, and open points
the unadjusted differences in post-treatment means (Debunking Dialogue minus comparison
condition). Contrasts that were not reported as significant in the main text are drawn in a
lighter shade. The dashed line marks zero.}

For all 12 contrasts reported as significant in the main text (7 in Experiment~2 and 5 in
Experiment~1), the identified set excludes zero: no pattern of monotone selective dropout can
account for the immediate debunking effects. Once sampling uncertainty is added, all seven
Experiment~2 contrasts retain outer confidence intervals that exclude zero. In Experiment~1,
where the trims are larger and the Information List condition is small ($n = 96$), the outer
intervals exclude zero for the own-belief and cover-up contrasts against the Irrelevant
Dialogue, reach essentially to zero for hidden factors against the Irrelevant Dialogue (upper
limit $-0.03$), and include zero for the two significant contrasts against the Information List
(own belief and hidden factors). Those two Experiment~1 results are therefore robust to
worst-case attrition in point identification but are estimated with less precision once the
trimmed comparison is bootstrapped. The Experiment~2 conclusions are unchanged when
the 34 event-denial dialogues described in SM11 are excluded (Table SM11.2).

\clearpage
\section{SM10. Selection into the follow-up surveys}
\label{sec:sm10}

Table SM10.1 reports the share of conspiracy believers in each condition of the initial
experiment who returned for the follow-up survey. In Experiment~2, return was slightly
\emph{higher} in the Debunking Dialogue condition (85.7\%) than in the pooled control conditions
(81.6\%), $p = .095$ ($p = .096$ when the 34 event-denial dialogues described in SM11 are
excluded). The preregistration (AsPredicted \#251092) called for Lee bounds on the follow-up
outcomes if return rates differed significantly; because the difference, although not
significant, is not negligible, we report those bounds below (Table SM10.4). In Experiment~1 the
return rates were nearly identical (58.7\% vs.\ 58.1\%, $p = .98$).

\begin{footnotesize}
\setlength{\tabcolsep}{4pt}
\begin{longtable}{lcccccc}
\toprule
\textbf{Study} & \textbf{Debunking} & \textbf{Information} & \textbf{Irrelevant} & \textbf{Pooled} & $\boldsymbol{p}$ \textbf{(DD vs.} & $\boldsymbol{p}$ \textbf{(three} \\
 & \textbf{Dialogue} & \textbf{List} & \textbf{Dialogue} & \textbf{controls} & \textbf{pooled)} & \textbf{conditions)} \\
\midrule
\endfirsthead
\toprule
\textbf{Study} & \textbf{Debunking} & \textbf{Information} & \textbf{Irrelevant} & \textbf{Pooled} & $\boldsymbol{p}$ \textbf{(DD vs.} & $\boldsymbol{p}$ \textbf{(three} \\
 & \textbf{Dialogue} & \textbf{List} & \textbf{Dialogue} & \textbf{controls} & \textbf{pooled)} & \textbf{conditions)} \\
\midrule
\endhead
\bottomrule
\endlastfoot
Experiment~1 (Trump) & 58.7\% (138/235) & 59.4\% (57/96) & 57.3\% (75/131) & 58.1\% (132/227) & $.976$ & $.942$ \\
Experiment~2 (Kirk) & 85.7\% (454/530) & 82.7\% (211/255) & 80.5\% (207/257) & 81.6\% (418/512) & $.095$ & $.171$ \\
\end{longtable}
\end{footnotesize}
\vspace{-6pt}
\smcap{Table SM10.1.}{Return rates to the follow-up surveys by condition in the initial experiment.}
\vspace{-10pt}
\noindent{\scriptsize\emph{Note.} Return rates among conspiracy believers randomized in the initial experiment (Experiment~2: intention-to-treat believer sample of the main experiment, $N = 1{,}042$, SM11, the cohort of the follow-up analyses; Experiment~1: believers with a recorded condition). Cells give the share of believers in each condition who returned. $p$ (DD vs.\ pooled) is from a continuity-corrected $\chi^2$ test of return in the Debunking Dialogue condition against the two control conditions pooled (the preregistered comparison for Experiment~2); $p$ (three conditions) is from the $2 \times 3$ $\chi^2$ test. When the 34 event-denial dialogues are excluded from Experiment~2 ($N = 1{,}035$ believers; SM11) the rates are 85.7\% (448/523) and 81.6\% (418/512), $p=.096$.\par}
\medskip

If the treated participants who showed the largest immediate belief change were the ones who
came back, the returning treated sample would be enriched for responders, and every follow-up
estimate would be biased toward finding carry-over. To test this directly, we regressed each
observed post-treatment outcome (without imputing dropouts; believer sample of the
main experiment) on a Debunking Dialogue dummy, an indicator for returning to the follow-up,
their interaction, and the mean-centered pre-treatment value of the same outcome, with HC2 robust
standard errors. A negative interaction on the three conspiracy-belief outcomes (a positive one
for trust) indicates that the returned-minus-not-returned gap in post-treatment belief was more
favorable to the treatment among treated participants than among controls, i.e., the selection
pattern of concern. Table SM10.2 reports the interaction and the returned-minus-not-returned
difference estimated separately within each condition. For Experiment~2 we also test the three-item
post-treatment Kirk composite, which is the $t_1$ measure of the persistence analysis (SM12).

\begin{scriptsize}
\setlength{\tabcolsep}{3.5pt}
\begin{longtable}{lccccr}
\toprule
\textbf{Post-treatment outcome} & \textbf{Returned $\times$ DD, $b$ [95\% CI]} & $\boldsymbol{p}$ & \textbf{Ret.\ $-$ not, DD ($p$)} & \textbf{Ret.\ $-$ not, ctl ($p$)} & $\boldsymbol{N}$ \\
\midrule
\endfirsthead
\toprule
\textbf{Post-treatment outcome} & \textbf{Returned $\times$ DD, $b$ [95\% CI]} & $\boldsymbol{p}$ & \textbf{Ret.\ $-$ not, DD ($p$)} & \textbf{Ret.\ $-$ not, ctl ($p$)} & $\boldsymbol{N}$ \\
\midrule
\endhead
\bottomrule
\endlastfoot
\multicolumn{6}{@{}l}{\textit{Experiment~1 (Trump)}} \\
Own conspiracy belief & $-4.51$ $[-10.42,\ 1.40]$ & $.135$ & $-2.84$ ($.281$) & $1.50$ ($.312$) & 435 \\
Cover-up or conspiracy & $-1.24$ $[-6.06,\ 3.58]$ & $.615$ & $-0.31$ ($.881$) & $0.94$ ($.484$) & 435 \\
Hidden factors & $-1.22$ $[-6.75,\ 4.31]$ & $.665$ & $1.11$ ($.646$) & $2.21$ ($.124$) & 435 \\
Trust in official explanation & $6.22$ $[-0.26,\ 12.70]$ & $.060$ & $5.37$ ($.049$) & $-0.84$ ($.637$) & 434 \\
\midrule
\multicolumn{6}{@{}l}{\textit{Experiment~2 (Kirk)}} \\
Own conspiracy belief & $-2.14$ $[-7.49,\ 3.20]$ & $.432$ & $-2.30$ ($.324$) & $-0.01$ ($.992$) & 1017 \\
Cover-up or conspiracy & $-5.22$ $[-9.68,\ -0.77]$ & $.022$ & $-3.05$ ($.136$) & $1.95$ ($.042$) & 1017 \\
Hidden factors & $-0.21$ $[-5.29,\ 4.86]$ & $.934$ & $-0.97$ ($.668$) & $-0.93$ ($.461$) & 1017 \\
Trust in official explanation & $3.66$ $[-1.36,\ 8.68]$ & $.152$ & $2.30$ ($.211$) & $-1.23$ ($.497$) & 1017 \\
3-item composite ($t_1$) & $-2.72$ $[-6.44,\ 1.01]$ & $.153$ & --- & --- & 1017 \\
\end{longtable}
\end{scriptsize}
\vspace{-6pt}
\smcap{Table SM10.2.}{Return $\times$ condition interactions on observed post-treatment outcomes, with the returned-minus-not-returned difference within each condition.}
\vspace{-10pt}
\noindent{\scriptsize\emph{Note.} Model: observed post-treatment outcome $\sim$ Debunking Dialogue (DD) + returned to follow-up + DD $\times$ returned + mean-centered pre-treatment value of the same outcome; HC2 robust SEs; no imputation of dropouts. A negative interaction on the conspiracy-belief outcomes (positive on trust) indicates that the returned-minus-not gap in post-treatment scores was more favorable to the treatment among DD participants than among controls. ``Ret.\ $-$ not'' columns are the returned-minus-did-not-return difference estimated separately within the DD condition and within the pooled control conditions (ctl). The Experiment~2 three-item composite is the $t_1$ measure of the persistence analysis.\par}
\medskip

Of the eight interaction tests, one is significant. On the Experiment~2 cover-up item,
treated returnees scored 3.05 points lower than treated non-returners ($p = .14$), whereas
control returnees scored 1.95 points \emph{higher} than control non-returners ($p = .042$), for
an interaction of $-5.22$ (95\% CI $[-9.68, -0.77]$, $p = .022$). No such pattern appears on the primary outcome (participants' own
conspiracy belief; $b = -2.14$, $p = .43$: treated returnees rated their belief 2.30 points lower
than treated non-returners, $p = .32$, and control returnees 0.01 points lower than control
non-returners, $p = .99$), on hidden factors or trust, on the three-item composite that enters
the persistence analysis ($b = -2.72$, $p = .153$), or on any Experiment~1 outcome (all
$p \ge .060$; the smallest $p$-value is for trust in the official explanation). The immediate treatment effect on participants' own belief
(Debunking Dialogue versus the pooled controls, adjusting for the pre-treatment rating) is $-7.54$
points among returnees ($n = 851$), $-6.40$ among non-returners ($n = 166$), and $-7.21$ among all
believers with an observed post-treatment score ($n = 1{,}017$); for the three-item composite the
corresponding estimates are $-6.36$, $-3.63$, and $-5.94$. In Experiment~1 the corresponding estimates
for participants' own belief are $-9.13$ points among returnees ($n = 270$), $-4.63$ among non-returners
($n = 165$), and $-7.45$ among all believers with an observed post-treatment score ($n = 435$); the
return $\times$ condition interaction is in the direction of concern but not significant ($b = -4.51$,
$p = .135$).

To account for possible selective return in the follow-up estimates themselves, we re-estimated
every follow-up analysis with each returning participant weighted by the inverse of their
estimated probability of returning. The probability of returning was estimated by logistic
regression, fit separately within the Debunking Dialogue condition and within the pooled control
conditions among all conspiracy believers randomized in the experiment ($N = 1{,}042$ in
Experiment~2; $N = 462$ in Experiment~1), on the four pre-treatment ratings (own belief,
cover-up, hidden factors, trust in the official explanation), the four post-treatment ratings of
the same items (with the pre-treatment value carried forward for participants who quit before
the post-treatment measures, as in the main analyses), age, education, party identification,
belief in God, and trust in AI. Weighting the returnees in this way makes them resemble the full
randomized sample on these characteristics: the largest standardized difference between
returnees and all randomized believers of the same condition on any of the 13 covariates falls
from .07 to .01 in Experiment~2 and from .11 to .03 in Experiment~1. The weights are moderate (1.01 to 2.15 in
Experiment~2; 1.03 to 3.32 in Experiment~1), so little precision is lost (effective sample
sizes of 862 of 872 and 256 of 270 returnees). The weighted models are otherwise identical to
the unweighted ones. Standard errors are heteroskedasticity-robust and treat the weights as
fixed; a bootstrap that re-estimates the weights in each of 2,000 resamples gives very similar
intervals, and supplies the intervals for the indirect effects.

Table SM10.3 compares the weighted and unweighted estimates. In Experiment~1, the effects of the
Debunking Dialogue on the follow-up outcomes are essentially unchanged by weighting (odds ratios
of .55, .47, and .49; all $p \le .023$). In Experiment~2, the estimate for LDS conspiracy belief
remains near zero ($b = -0.03$, $p = .98$), and the persistence slopes and the indirect effects
through immediate belief change are essentially unchanged. The estimate for generic conspiracy
beliefs, however, is smaller after weighting and no longer significant ($b = -2.34$, 95\% CI
$[-5.61, 0.92]$, $p = .16$, compared with $b = -3.33$, $p = .044$, unweighted). The same pattern
holds when the weights are truncated at their 99th percentile ($b = -2.44$, $p = .14$), when the
return model includes only the eight belief ratings ($b = -2.88$, $p = .083$), and when it adds
an indicator for having completed the post-treatment measures ($b = -2.32$, $p = .16$).

Two limitations apply. First, weighting corrects only for selection on the measured
characteristics; it cannot correct for return that depends on unmeasured factors or on the
follow-up beliefs themselves. Second, in Experiment~1 none of the 27 believers who quit before
the post-treatment measures (24 in the Debunking Dialogue condition, 3 in the control
conditions) returned for the follow-up, so no returning participant can stand in for them; a
return model fit only among the 435 believers who completed the post-treatment measures gives
nearly identical estimates (odds ratios of .55, .48, and .49).

\begin{scriptsize}
\setlength{\tabcolsep}{3.5pt}
\begin{longtable}{>{\raggedright\arraybackslash}p{2.55in}cccccc}
\toprule
\textbf{Analysis} & \multicolumn{3}{c}{\textbf{Unweighted}} & \multicolumn{3}{c}{\textbf{Weighted}} \\
\cmidrule(lr){2-4}\cmidrule(lr){5-7}
 & \textbf{Estimate} & \textbf{95\% CI} & $\boldsymbol{p}$ & \textbf{Estimate} & \textbf{95\% CI} & $\boldsymbol{p}$ \\
\midrule
\endfirsthead
\toprule
\textbf{Analysis} & \multicolumn{3}{c}{\textbf{Unweighted}} & \multicolumn{3}{c}{\textbf{Weighted}} \\
\cmidrule(lr){2-4}\cmidrule(lr){5-7}
 & \textbf{Estimate} & \textbf{95\% CI} & $\boldsymbol{p}$ & \textbf{Estimate} & \textbf{95\% CI} & $\boldsymbol{p}$ \\
\midrule
\endhead
\bottomrule
\endlastfoot
\multicolumn{7}{@{}l}{\textit{Experiment~1 (odds ratios, Debunking Dialogue vs.\ pooled controls)}} \\
Only a few powerful people will know the truth ($n = 264$) & $.52$ & $[.31,\ .85]$ & $.009$ & $.55$ & $[.33,\ .92]$ & $.023$ \\
Same, alternative coding ($n = 270$) & $.44$ & $[.24,\ .81]$ & $.008$ & $.47$ & $[.25,\ .88]$ & $.019$ \\
The truth will be hidden from most people ($n = 264$) & $.48$ & $[.29,\ .79]$ & $.004$ & $.49$ & $[.30,\ .82]$ & $.007$ \\
\midrule
\multicolumn{7}{@{}l}{\textit{Experiment~2 (points on the 0--100 scale; slopes for the persistence rows)}} \\
LDS conspiracy belief, assignment effect, preregistered covariate ($n = 872$) & $-0.46$ & $[-3.30,\ 2.37]$ & $.748$ & $-0.03$ & $[-2.87,\ 2.81]$ & $.983$ \\
Generic conspiracy beliefs, assignment effect, preregistered covariate ($n = 872$) & $-3.33$ & $[-6.57,\ -0.09]$ & $.044$ & $-2.34$ & $[-5.61,\ 0.92]$ & $.160$ \\
LDS, assignment effect, composite covariate & $-0.07$ & $[-2.68,\ 2.55]$ & $.961$ & $0.08$ & $[-2.54,\ 2.71]$ & $.950$ \\
Generic, assignment effect, composite covariate & $-2.92$ & $[-5.88,\ 0.04]$ & $.053$ & $-2.17$ & $[-5.17,\ 0.82]$ & $.155$ \\
Persistence slope, LDS ($n = 851$) & $.117$ & $[.004,\ .231]$ & $.043$ & $.129$ & $[.013,\ .245]$ & $.030$ \\
Persistence slope, generic & $.194$ & $[.070,\ .318]$ & $.002$ & $.205$ & $[.079,\ .331]$ & $.001$ \\
Indirect effect through immediate belief change, LDS ($n = 851$) & $-0.78$ & $[-1.54,\ -0.07]$\textsuperscript{a} & --- & $-0.80$ & $[-1.50,\ -0.12]$\textsuperscript{a} & --- \\
Indirect effect, generic & $-1.16$ & $[-2.11,\ -0.33]$\textsuperscript{a} & --- & $-1.19$ & $[-2.01,\ -0.43]$\textsuperscript{a} & --- \\
Direct effect, LDS & $0.61$ & $[-2.10,\ 3.31]$ & $.660$ & $0.79$ & $[-1.94,\ 3.52]$ & $.572$ \\
Direct effect, generic & $-1.33$ & $[-4.37,\ 1.71]$ & $.390$ & $-0.53$ & $[-3.60,\ 2.54]$ & $.735$ \\
\end{longtable}
\end{scriptsize}
\vspace{-6pt}
\smcap{Table SM10.3.}{Follow-up estimates with and without weighting returning participants by the inverse of their estimated probability of returning.}
\vspace{-10pt}
\noindent{\scriptsize\emph{Note.} Weights are the inverse of each returning participant's estimated probability of returning, from logistic regressions fit within the Debunking Dialogue condition and within the pooled control conditions (see text). Unweighted estimates are computed in the same way as the weighted ones (heteroskedasticity-robust standard errors, Wald confidence intervals), so some of their confidence limits and $p$-values differ from those reported elsewhere: the Experiment~1 odds ratios have profile-likelihood intervals in the main text, and the persistence slopes have conventional standard errors in the main text and Table SM12.1 (LDS slope: $p=.019$ conventional, $p=.043$ robust; generic slope: $p<.001$ and $p=.002$). \textsuperscript{a}Intervals for the indirect effects are percentile bootstrap intervals (weighted: 2,000 resamples with the weights re-estimated in each; unweighted: 5,000 resamples); no $p$-value is reported for them.\par}
\medskip

Table SM10.4 reports Lee bounds (computed as in SM9) for the two Experiment~2 follow-up outcomes,
treating return to the follow-up as the response and comparing the Debunking Dialogue condition
with the pooled control conditions and with each control condition separately. Because the
Debunking Dialogue condition was better retained, 4.7\% of its returnees are trimmed in the
pooled comparison. For generic conspiracy beliefs the identified set lies below zero
($[-5.72, -1.06]$), but the bootstrap outer 95\% confidence interval, which also reflects
sampling uncertainty, includes zero ($[-9.66, 3.01]$); for LDS conspiracy belief both the
identified set ($[-2.53, 1.64]$) and the outer interval ($[-5.95, 5.15]$) include zero. The
bounds are computed on unadjusted follow-up means.

\begingroup\par\penalty-100 \dimen0=\pagegoal \advance\dimen0 by -\pagetotal \ifdim\dimen0<4.3in\relax \ifdim\dimen0>0pt \vfil \fi \break \fi \endgroup
\begin{scriptsize}
\setlength{\tabcolsep}{3.5pt}
\begin{longtable}{>{\raggedright\arraybackslash}p{1.55in}>{\centering\arraybackslash}p{0.66in}>{\centering\arraybackslash}p{0.74in}>{\centering\arraybackslash}p{0.48in}>{\centering\arraybackslash}p{0.66in}>{\centering\arraybackslash}p{0.80in}>{\centering\arraybackslash}p{0.85in}}
\toprule
\textbf{Outcome and comparison} & \textbf{Return rate, Debunking Dialogue} & \textbf{Return rate, comparison condition(s)} & \textbf{Share trimmed} & \textbf{Difference among returnees} & \textbf{Identified set} & \textbf{Outer 95\% CI} \\
\midrule
\endfirsthead
\toprule
\textbf{Outcome and comparison} & \textbf{Return rate, Debunking Dialogue} & \textbf{Return rate, comparison condition(s)} & \textbf{Share trimmed} & \textbf{Difference among returnees} & \textbf{Identified set} & \textbf{Outer 95\% CI} \\
\midrule
\endhead
\bottomrule
\endlastfoot
\multicolumn{7}{@{}l}{\textit{LDS conspiracy belief}} \\
Debunking Dialogue vs.\ pooled controls & 85.7\% & 81.6\% & 4.7\% & $-0.49$ & $[-2.53,\ 1.64]$ & $[-5.95,\ 5.15]$ \\
Debunking Dialogue vs.\ Irrelevant Dialogue & 85.7\% & 80.5\% & 6.0\% & $-1.61$ & $[-4.17,\ 1.09]$ & $[-8.55,\ 5.43]$ \\
Debunking Dialogue vs.\ Information List & 85.7\% & 82.7\% & 3.4\% & $0.61$ & $[-0.90,\ 2.16]$ & $[-4.85,\ 6.20]$ \\
\midrule
\multicolumn{7}{@{}l}{\textit{Generic conspiracy beliefs}} \\
Debunking Dialogue vs.\ pooled controls & 85.7\% & 81.6\% & 4.7\% & $-3.45$ & $[-5.72,\ -1.06]$ & $[-9.66,\ 3.01]$ \\
Debunking Dialogue vs.\ Irrelevant Dialogue & 85.7\% & 80.5\% & 6.0\% & $-4.10$ & $[-6.99,\ -1.04]$ & $[-11.86,\ 4.10]$ \\
Debunking Dialogue vs.\ Information List & 85.7\% & 82.7\% & 3.4\% & $-2.81$ & $[-4.45,\ -1.10]$ & $[-9.10,\ 3.71]$ \\
\end{longtable}
\end{scriptsize}
\vspace{-6pt}
\smcap{Table SM10.4.}{Lee bounds on the Experiment~2 follow-up outcomes.}
\vspace{-10pt}
\noindent{\scriptsize\emph{Note.} Lee (2009) sharp bounds computed as in SM9, with returning to the follow-up survey as the response, among the conspiracy believers randomized in Experiment~2 ($N = 1{,}042$; 530 Debunking Dialogue / 257 Irrelevant Dialogue / 255 Information List). Return rate = share of the randomized believers in the condition(s) who returned. The better-retained condition (the Debunking Dialogue condition in every comparison) is trimmed; share trimmed = 1 minus the ratio of the two return rates, the proportion of its returnees removed from one tail (to the nearest whole participant). Difference among returnees = unadjusted difference in follow-up means (Debunking Dialogue minus comparison) among returnees; a negative difference indicates less conspiracy belief in the Debunking Dialogue condition. Bounds are computed on unadjusted follow-up means. Outer 95\% CI as in SM9: lower bound minus 1.96 bootstrap SEs, upper bound plus 1.96 bootstrap SEs (2,000 bootstrap resamples).\par}
\medskip

\clearpage
\section{SM11. Dialogues that denied the event, and analyses excluding them}
\label{sec:sm11}

In Experiment~2 the dialogue model, whose training data predate the event, occasionally asserted
during a Debunking Dialogue that Charlie Kirk had not in fact been killed. Blind LLM coding of
all Debunking Dialogue transcripts flagged 37 such dialogues. Three of the 37 participants had
already been removed by the pre-randomization survey-agent screen (a hidden instruction designed
to detect automated respondents). The remaining 34 completed the post-treatment measures.
Excluding them would be a post-randomization exclusion that can only remove participants from
the dialogue condition, so every analysis of treatment effects in the main text retains them in
their assigned condition (intention-to-treat). Of the 34, the GPT-4o belief classification (SM3) coded 27 as
``no,'' 6 as ``maybe,'' and 1 as ``yes''; since the main analyses are restricted to participants
classified as ``yes'' or ``maybe,'' 7 of the 34 are in the believer sample ($N = 1{,}042$).

As a sensitivity analysis, Table SM11.1 sets the eight pairwise immediate-effect contrasts of
the main text beside the same contrasts estimated after excluding the 34 dialogues (the
per-protocol sample; $N = 1{,}035$ believers), and Table SM11.2 repeats the Lee bounds of SM9 on
that sample. The follow-up analyses likewise retain everyone ($N = 872$ returning believers,
including the six whose dialogues were flagged); SM12 reports the follow-up cohort without them
($N = 866$) for comparison.

\begin{scriptsize}
\setlength{\tabcolsep}{3.5pt}
\begin{longtable}{llcccc c}
\toprule
\textbf{Outcome} & \textbf{Contrast} & \textbf{All participants (main text) $b$ [95\% CI]} & $\boldsymbol{p}$ & \textbf{Excluding the 34 dialogues $b$ [95\% CI]} & $\boldsymbol{p}$ & $\boldsymbol{|\Delta b|}$ \\
\midrule
\endfirsthead
\toprule
\textbf{Outcome} & \textbf{Contrast} & \textbf{All participants (main text) $b$ [95\% CI]} & $\boldsymbol{p}$ & \textbf{Excluding the 34 dialogues $b$ [95\% CI]} & $\boldsymbol{p}$ & $\boldsymbol{|\Delta b|}$ \\
\midrule
\endhead
\bottomrule
\endlastfoot
Own belief & vs ID & $-7.48$ $[-10.40,\ -4.55]$ & $<.001$ & $-7.56$ $[-10.50,\ -4.61]$ & $<.001$ & 0.08 \\
Own belief & vs IL & $-6.48$ $[-9.52,\ -3.43]$ & $<.001$ & $-6.56$ $[-9.62,\ -3.49]$ & $<.001$ & 0.08 \\
Cover-up & vs ID & $-7.46$ $[-9.55,\ -5.36]$ & $<.001$ & $-7.50$ $[-9.62,\ -5.39]$ & $<.001$ & 0.04 \\
Cover-up & vs IL & $-5.92$ $[-8.20,\ -3.63]$ & $<.001$ & $-5.96$ $[-8.27,\ -3.66]$ & $<.001$ & 0.04 \\
Hidden factors & vs ID & $-7.64$ $[-10.25,\ -5.02]$ & $<.001$ & $-7.52$ $[-10.16,\ -4.89]$ & $<.001$ & 0.12 \\
Hidden factors & vs IL & $-6.23$ $[-9.07,\ -3.39]$ & $<.001$ & $-6.11$ $[-8.97,\ -3.26]$ & $<.001$ & 0.12 \\
Trust official expl. & vs ID & $5.13$ $[2.29,\ 7.97]$ & $<.001$ & $5.19$ $[2.34,\ 8.04]$ & $<.001$ & 0.06 \\
Trust official expl. & vs IL & $2.20$ $[-0.67,\ 5.07]$ & $.169$ & $2.26$ $[-0.62,\ 5.15]$ & $.155$ & 0.06 \\
\end{longtable}
\end{scriptsize}
\vspace{-6pt}
\smcap{Table SM11.1.}{Immediate effects of the Debunking Dialogue in Experiment~2: all participants analyzed as randomized (the main-text sample) versus the sample that excludes the 34 event-denial dialogues.}
\vspace{-10pt}
\noindent{\scriptsize\emph{Note.} Experiment~2 believers. Estimates are the Debunking Dialogue (DD) minus control differences from one OLS model per outcome containing the three conditions and the mean-centered pre-treatment value (HC2 robust SEs; dropouts' pre-treatment values carried forward){,} with Tukey-adjusted 95\% CIs and $p$-values for the three pairwise comparisons{,} as in the main text. All participants: $N$ = 1{,}042 believers (530 DD / 257 ID / 255 IL); excluding the 34 dialogues flagged for event denial (7 of which involved classified believers): $N$ = 1{,}035 (523 / 257 / 255). DD = Debunking Dialogue; ID = Irrelevant Dialogue; IL = Information List. $|\Delta b|$ = absolute difference between the two displayed coefficients (0--100 scale).\par}
\medskip

\clearpage
\begin{scriptsize}
\setlength{\tabcolsep}{3.5pt}
\begin{longtable}{llrccccc}
\toprule
\textbf{Outcome} & \textbf{Contrast} & \textbf{Unadj.\ diff.} & \textbf{Identified set} & \textbf{Outer 95\% CI} & \textbf{Retention DD / ctl} & \textbf{Trim (arm)} & \textbf{Sig.} \\
\midrule
\endfirsthead
\toprule
\textbf{Outcome} & \textbf{Contrast} & \textbf{Unadj.\ diff.} & \textbf{Identified set} & \textbf{Outer 95\% CI} & \textbf{Retention DD / ctl} & \textbf{Trim (arm)} & \textbf{Sig.} \\
\midrule
\endhead
\bottomrule
\endlastfoot
Own belief & vs ID & $-8.14$ & $[-8.32,\ -7.72]$ & $[-11.71,\ -4.17]$ & 97.1\% / 96.5\% & 0.7\% (DD) & yes \\
Own belief & vs IL & $-6.33$ & $[-7.74,\ -5.77]$ & $[-11.17,\ -2.38]$ & 97.1\% / 99.6\% & 2.5\% (IL) & yes \\
Cover-up & vs ID & $-7.44$ & $[-7.72,\ -7.13]$ & $[-11.90,\ -2.94]$ & 97.1\% / 96.5\% & 0.7\% (DD) & yes \\
Cover-up & vs IL & $-6.19$ & $[-7.60,\ -5.20]$ & $[-11.66,\ -1.18]$ & 97.1\% / 99.6\% & 2.5\% (IL) & yes \\
Hidden factors & vs ID & $-7.41$ & $[-7.65,\ -7.06]$ & $[-11.62,\ -3.06]$ & 97.1\% / 96.5\% & 0.7\% (DD) & yes \\
Hidden factors & vs IL & $-7.72$ & $[-9.20,\ -6.92]$ & $[-13.23,\ -2.93]$ & 97.1\% / 99.6\% & 2.5\% (IL) & yes \\
Trust official expl. & vs ID & $5.19$ & $[4.86,\ 5.45]$ & $[0.57,\ 9.74]$ & 97.1\% / 96.5\% & 0.7\% (DD) & yes \\
Trust official expl. & vs IL & $3.19$ & $[2.18,\ 4.56]$ & $[-2.15,\ 8.89]$ & 97.1\% / 99.6\% & 2.5\% (IL) & no \\
\end{longtable}
\end{scriptsize}
\vspace{-6pt}
\smcap{Table SM11.2.}{Lee bounds on the Experiment~2 immediate effects when the 34 event-denial dialogues are excluded (compare Table SM9.1).}
\vspace{-10pt}
\noindent{\scriptsize\emph{Note.} Same estimator and conventions as Table SM9.1, computed on the believer sample that excludes the 7 believers whose dialogues were flagged for event denial ($N = 1{,}035$; 523 DD / 257 ID / 255 IL). DD = Debunking Dialogue; ID = Irrelevant Dialogue; IL = Information List.\par}
\medskip

Excluding the 34 flagged dialogues changes no coefficient by more than 0.12 points on the 0--100
scale: the seven contrasts that are significant in the main text remain significant at
$p < .001$, and trust in the official explanation versus the Information List remains
non-significant ($p = .17$ vs.\ $p = .16$). The Lee bounds are likewise unchanged in every
respect; the identified sets and outer intervals of all seven significant contrasts exclude
zero.

\clearpage
\section{SM12. Experiment~2 follow-up: cohort, covariates, persistence, and mediation}
\label{sec:sm12}

The follow-up analyses use the intention-to-treat (ITT) follow-up cohort, as in the main text:
every believer in the main-experiment sample who returned to the follow-up survey,
$N = 872$ (454 Debunking Dialogue / 211 Information List / 207 Irrelevant Dialogue). Table
SM12.1 also reports the per-protocol follow-up cohort, which excludes the six returning believers
whose dialogues denied the event (SM11; $N = 866$; 448 / 211 / 207), for
comparison; across the two cohorts no estimate differs by more than 0.2 points on the 0--100
scale and no persistence slope by more than $.01$.

\begin{scriptsize}
\setlength{\tabcolsep}{3pt}
\begin{longtable}{>{\raggedright\arraybackslash}p{1.55in}>{\raggedright\arraybackslash}p{1.50in}>{\raggedright\arraybackslash}p{1.50in}}
\toprule
\textbf{Quantity} & \textbf{ITT (872; main text)} & \textbf{Per-protocol (866)} \\
\midrule
\endfirsthead
\toprule
\textbf{Quantity} & \textbf{ITT (872; main text)} & \textbf{Per-protocol (866)} \\
\midrule
\endhead
\bottomrule
\endlastfoot
Cohort ($N$; DD / IL / ID) & 872; 454 / 211 / 207 & 866; 448 / 211 / 207 \\
\midrule
LDS shooting beliefs $\sim$ DD (ANCOVA) &  &  \\
\quad prereg.\ covariate (own-belief rating) & $-0.46$ $[-3.30,\ 2.37]$, $p=.748$, $d=-.022$ & $-0.27$ $[-3.12,\ 2.57]$, $p=.851$, $d=-.013$ \\
\quad Kirk composite covariate & $-0.07$ $[-2.68,\ 2.55]$, $p=.961$, $d=-.003$ & $0.07$ $[-2.55,\ 2.69]$, $p=.959$, $d=.003$ \\
Generic conspiracy beliefs $\sim$ DD (ANCOVA) &  &  \\
\quad prereg.\ covariate (own-belief rating) & $-3.33$ $[-6.57,\ -0.09]$, $p=.044$, $d=-.135$ & $-3.21$ $[-6.47,\ 0.04]$, $p=.053$, $d=-.131$ \\
\quad Kirk composite covariate & $-2.92$ $[-5.88,\ 0.04]$, $p=.053$, $d=-.119$ & $-2.89$ $[-5.86,\ 0.09]$, $p=.057$, $d=-.117$ \\
Persistence slope $b_1$, LDS shooting beliefs &  &  \\
\quad Kirk composite as $t_0$, $t_1$ & $.117$ $[.02,\ .22]$, $p=.019$ & $.117$ $[.02,\ .22]$, $p=.019$ \\
\quad\quad worst-case return imputation & $.113$ $[.02,\ .20]$, $p=.015$ & $.114$ $[.02,\ .21]$, $p=.014$ \\
\quad own-belief rating as $t_0$, $t_1$ & $.073$ $[-.000,\ .146]$, $p=.050$ & $.075$ $[.002,\ .148]$, $p=.043$ \\
Persistence slope $b_1$, generic conspiracy beliefs &  &  \\
\quad Kirk composite as $t_0$, $t_1$ & $.194$ $[.08,\ .30]$, $p<.001$ & $.195$ $[.09,\ .31]$, $p<.001$ \\
\quad\quad worst-case return imputation & $.179$ $[.08,\ .28]$, $p<.001$ & $.182$ $[.08,\ .28]$, $p<.001$ \\
\quad own-belief rating as $t_0$, $t_1$ & $.172$ $[.090,\ .255]$, $p<.001$ & $.173$ $[.091,\ .256]$, $p<.001$ \\
\end{longtable}
\end{scriptsize}
\vspace{-6pt}
\smcap{Table SM12.1.}{The four preregistered Experiment~2 follow-up tests in the intention-to-treat follow-up cohort (as reported in the main text) and the per-protocol cohort, with both covariate versions of the assignment-based tests, both measures of immediate change for the persistence slopes, and the worst-case return imputation.}
\vspace{-10pt}
\noindent{\scriptsize\emph{Note.} DD = Debunking Dialogue; ID = Irrelevant Dialogue; IL = Information List. ``ITT'' = all believers of the intention-to-treat main-experiment sample (SM11) who returned to the follow-up survey; the follow-up preregistration specifies no exclusions, and this is the cohort reported in the main text. ``Per-protocol'' = returnees from the per-protocol main-experiment sample (i.e., excluding the six returnees whose dialogues were flagged for event denial). ANCOVA rows: follow-up outcome regressed on a DD dummy (Information List and Irrelevant Dialogue pooled) plus either the preregistered covariate (the pre-treatment rating of the participant's own conspiracy belief) or the pre-treatment three-item Kirk composite, entered as a linear covariate without interaction; HC2 SEs; $d$ = $b$ / SD of the outcome. Persistence rows: Lin et al.\ estimator $(t_2 - t_0) \sim b_1 (t_1 - t_0) + t_0$, with $t_2$ the follow-up outcome and $t_0$, $t_1$ either the pre- and post-treatment Kirk composites (as reported in the main text) or the pre- and post-treatment ratings of the participant's own conspiracy belief; ``worst-case return imputation'' sets $t_2 = t_0$ for every believer who did not return (complete decay of belief change) and refits the composite model on all believers with $t_0$ and $t_1$ observed.\par}
\medskip

The follow-up preregistration (AsPredicted \#251092) names ``pre-treatment ratings of
respondents' own conspiracy views'' as the covariate for the assignment-based tests, and the
main text reports that model. Table SM12.1 also reports each test with the pre-treatment
three-item Kirk composite as the covariate (the composite is the baseline measure that enters
the persistence and mediation models); each covariate is entered linearly, without interaction
terms. In the ITT cohort, LDS shooting beliefs give $b = -0.46$ (95\% CI
$[-3.30, 2.37]$, $p = .748$) with the preregistered covariate and $b = -0.07$ ($[-2.68, 2.55]$,
$p = .961$, $d = -.003$) with the composite; generic conspiracy beliefs give $b = -3.33$
($[-6.57, -0.09]$, $p = .044$) and $b = -2.92$ ($[-5.88, 0.04]$, $p = .053$, $d = -.119$),
respectively. The minimum detectable effect on generic conspiracy beliefs at
80\% power, computed from the standard error of the assignment coefficient in the ITT cohort as
$(1.96 + 0.84) \times SE$, is 4.6 points with the preregistered covariate ($SE = 1.65$ points;
$d = 0.19$; power of roughly 33\% for a true effect of 2.5 points and 44\% for 3 points) and
4.2 points with the composite covariate ($SE = 1.51$ points; $d = 0.17$; power roughly 38\% and
51\%). The computation is reproduced by the deposited R pipeline.

Following Lin et al., the persistence analysis regresses follow-up change on immediate change
with baseline controlled: $(t_2 - t_0) = b_1 (t_1 - t_0) + b_2\, t_0 + \varepsilon$, where $t_0$
and $t_1$ are the pre- and post-treatment measures of belief in the Kirk conspiracy and $t_2$ is
the follow-up outcome. Random assignment does not enter this model, so $b_1$ is described as the
persistence of immediate belief change (identified under the assumption that, given $t_0$, no
unmeasured factor confounds post-treatment and follow-up belief), not as a treatment effect.

In the ITT cohort, with the three-item Kirk composite as $t_0$ and $t_1$, $b_1 = .117$ (95\% CI
$[.02, .22]$, $p = .019$) for LDS shooting beliefs and $.194$ ($[.08, .30]$, $p < .001$) for
generic conspiracy beliefs ($n = 851$ returnees with all three measures). With the pre- and
post-treatment ratings of the participant's own conspiracy belief as $t_0$ and $t_1$ instead
(the measure named as the covariate in the preregistration), the slopes are $.073$
($[-.000, .146]$, $p = .050$) and $.172$ ($[.090, .255]$, $p < .001$), respectively (Table SM12.1).

Table SM12.2 fits the persistence model separately within the Debunking Dialogue condition and
within the pooled control conditions and tests whether the slope differs by condition. In the
ITT cohort, the carry-forward rate for generic conspiracy beliefs is nearly identical across
conditions ($b_1 = .193$ in the Debunking Dialogue condition, $p = .003$; $.197$ in the control
conditions, $p = .153$; slope $\times$ condition interaction $b = -.056$, $p = .71$). For LDS shooting beliefs the
control-condition slope is numerically larger ($b_1 = .104$ in the Debunking Dialogue condition,
$p = .062$; $.262$ in the control conditions, $p = .037$; slope $\times$ condition interaction
$b = -.199$, $p = .14$), although the control-condition estimate is imprecise (95\% CI
$[.02, .51]$) and the interaction is not significant.

\begin{scriptsize}
\setlength{\tabcolsep}{3.5pt}
\begin{longtable}{llcccc}
\toprule
\textbf{Cohort} & \textbf{Follow-up outcome} & \textbf{Pooled $b_1$} & \textbf{Within DD $b_1$} & \textbf{Within controls $b_1$} & \textbf{$b_1 \times$ DD [95\% CI]} \\
\midrule
\endfirsthead
\toprule
\textbf{Cohort} & \textbf{Follow-up outcome} & \textbf{Pooled $b_1$} & \textbf{Within DD $b_1$} & \textbf{Within controls $b_1$} & \textbf{$b_1 \times$ DD [95\% CI]} \\
\midrule
\endhead
\bottomrule
\endlastfoot
ITT (872; main text) & Generic beliefs & $.194$ ($p<.001$) & $.193$ ($p=.003$) & $.197$ ($p=.153$) & $-.056$ $[-.35,\ .24]$ ($p=.711$) \\
 & LDS shooting beliefs & $.117$ ($p=.019$) & $.104$ ($p=.062$) & $.262$ ($p=.037$) & $-.199$ $[-.46,\ .06]$ ($p=.138$) \\
\midrule
Per-protocol (866) & Generic beliefs & $.195$ ($p<.001$) & $.194$ ($p=.002$) & $.197$ ($p=.153$) & $-.052$ $[-.35,\ .24]$ ($p=.730$) \\
 & LDS shooting beliefs & $.117$ ($p=.019$) & $.106$ ($p=.058$) & $.262$ ($p=.037$) & $-.198$ $[-.46,\ .07]$ ($p=.141$) \\
\end{longtable}
\end{scriptsize}
\vspace{-6pt}
\smcap{Table SM12.2.}{Persistence slopes estimated on all returnees and separately within the Debunking Dialogue and control conditions, with the slope $\times$ condition interaction (intention-to-treat cohort, as reported in the main text; per-protocol cohort for comparison).}
\vspace{-10pt}
\noindent{\scriptsize\emph{Note.} Lin et al.\ persistence model $(t_2 - t_0) \sim b_1 (t_1 - t_0) + t_0$ fit on all returnees (pooled), separately within the Debunking Dialogue (DD) condition and within the pooled control conditions, and with a DD dummy and its interaction with $(t_1 - t_0)$ added (last column). Conventional OLS SEs, matching the published estimator. Returnees with $t_0$, $t_1$, and $t_2$ observed: ITT (872; main text): $n = 851$ pooled (442 DD, 409 controls); Per-protocol (866): $n = 845$ pooled (436 DD, 409 controls).\par}
\medskip

Selective return of the participants whose beliefs changed most would bias $b_1$ upward. The
imputation that addresses this concern from the conservative side assumes complete decay of
belief change for every non-returner: we set $t_2 = t_0$ (a follow-up change of zero) for each
believer with observed $t_0$ and $t_1$ who did not return (or who returned without a value on
that follow-up outcome), and refit the model on all such believers ($N = 1{,}017$ in the ITT
sample, of whom 851 are returnees with all three measures). Under this assumption every
non-returner contributes evidence \emph{against} persistence, however much their belief changed
immediately. The slopes are essentially unchanged: $b_1 = .113$ (95\% CI $[.02, .20]$,
$p = .015$) for LDS shooting beliefs and $.179$ ($[.08, .28]$, $p < .001$) for generic
conspiracy beliefs.

Persistence was present in both arms, but far more change was induced in the treatment arm:
among ITT returnees the post-treatment composite fell by $7.55$ points on average in the
Debunking Dialogue condition and by $1.31$ points in the control conditions. Table SM12.3
therefore reports an exploratory (not preregistered) mediation model that combines the two
preregistered analyses. With $t_1$ the post-treatment Kirk composite and $t_0$ the pre-treatment
composite, the $a$-path model is $t_1 \sim \text{DD} + t_0$, the $b$-path model is
$y \sim \text{DD} + t_1 + t_0$ (whose DD coefficient is the direct effect), and the total-effect
model is $y \sim \text{DD} + t_0$; the indirect effect is $a \times b$, with a percentile
bootstrap 95\% CI (5,000 resamples). In the ITT cohort ($n = 851$ returnees with $t_0$, $t_1$,
and the follow-up outcome observed), assignment to the Debunking Dialogue lowered the
post-treatment composite by $6.36$ points ($[-8.10, -4.62]$), and the post-treatment composite
predicted both follow-up outcomes given assignment and baseline ($b = .182$, $p = .005$, for
generic conspiracy beliefs; $b = .123$, $p = .038$, for LDS shooting beliefs). The indirect
effects are $-1.16$ points (95\% bootstrap CI $[-2.11, -0.33]$) on generic conspiracy beliefs
and $-0.78$ points ($[-1.54, -0.07]$) on LDS shooting beliefs; the direct effects are $-1.33$
($p = .39$) and $+0.61$ ($p = .66$), and the total effects $-2.49$ ($p = .10$) and $-0.17$
($p = .90$), estimated on the 851 rows with an observed $t_1$ and therefore slightly different
from the 872-row ANCOVAs of Table SM12.1. The per-protocol cohort gives indirect effects of
$-1.18$ ($[-2.06, -0.34]$) and $-0.79$ ($[-1.51, -0.04]$). The indirect effect is the immediate
treatment effect scaled by the persistence slope: the $a$-path is experimental, but the $b$-path
is not, so the indirect effect is identified under sequential ignorability---no unmeasured
confounding of the relation between post-treatment belief and follow-up belief, given assignment
and $t_0$---which is the same assumption as the persistence analysis, and it shares its
observational character.

\clearpage
\begin{scriptsize}
\setlength{\tabcolsep}{2.5pt}
\begin{longtable}{>{\raggedright\arraybackslash}p{0.70in}>{\raggedright\arraybackslash}p{0.62in}r>{\centering\arraybackslash}p{0.78in}>{\centering\arraybackslash}p{0.95in}>{\centering\arraybackslash}p{0.78in}>{\centering\arraybackslash}p{0.90in}>{\centering\arraybackslash}p{0.90in}}
\toprule
\textbf{Cohort} & \textbf{Follow-up outcome} & $\boldsymbol{n}$ & \textbf{$a$: DD $\to$ post-treatment composite [95\% CI]} & \textbf{$b$: post-treatment composite $\to$ follow-up [95\% CI]} & \textbf{Indirect $a \times b$ [bootstrap 95\% CI]} & \textbf{Direct $c'$ [95\% CI]} & \textbf{Total $c$ [95\% CI]} \\
\midrule
\endfirsthead
\toprule
\textbf{Cohort} & \textbf{Follow-up outcome} & $\boldsymbol{n}$ & \textbf{$a$: DD $\to$ post-treatment composite [95\% CI]} & \textbf{$b$: post-treatment composite $\to$ follow-up [95\% CI]} & \textbf{Indirect $a \times b$ [bootstrap 95\% CI]} & \textbf{Direct $c'$ [95\% CI]} & \textbf{Total $c$ [95\% CI]} \\
\midrule
\endhead
\bottomrule
\endlastfoot
ITT (872; main text) & Generic conspiracy beliefs & 851 & $-6.36$ $[-8.10,\ -4.62]$ & $.182$ $[.05,\ .31]$, $p=.005$ & $-1.16$ $[-2.11,\ -0.33]$ & $-1.33$ $[-4.37,\ 1.71]$, $p=.390$ & $-2.49$ $[-5.47,\ 0.49]$, $p=.101$ \\
 & LDS shooting beliefs & 851 & $-6.36$ $[-8.10,\ -4.62]$ & $.123$ $[.01,\ .24]$, $p=.038$ & $-0.78$ $[-1.54,\ -0.07]$ & $0.61$ $[-2.10,\ 3.31]$, $p=.660$ & $-0.17$ $[-2.83,\ 2.49]$, $p=.898$ \\
\midrule
Per-protocol (866) & Generic conspiracy beliefs & 845 & $-6.38$ $[-8.14,\ -4.63]$ & $.184$ $[.06,\ .31]$, $p=.005$ & $-1.18$ $[-2.06,\ -0.34]$ & $-1.28$ $[-4.33,\ 1.77]$, $p=.409$ & $-2.46$ $[-5.45,\ 0.53]$, $p=.107$ \\
 & LDS shooting beliefs & 845 & $-6.38$ $[-8.14,\ -4.63]$ & $.124$ $[.01,\ .24]$, $p=.036$ & $-0.79$ $[-1.51,\ -0.04]$ & $0.75$ $[-1.96,\ 3.47]$, $p=.585$ & $-0.04$ $[-2.71,\ 2.63]$, $p=.979$ \\
\end{longtable}
\end{scriptsize}
\vspace{-6pt}
\smcap{Table SM12.3.}{Indirect effect of assignment to the Debunking Dialogue on each follow-up outcome through immediate post-treatment belief in the Kirk conspiracy, with the direct and total effects.}
\vspace{-10pt}
\noindent{\scriptsize\emph{Note.} DD = Debunking Dialogue (versus the pooled Information List and Irrelevant Dialogue conditions). Mediator = the post-treatment three-item Kirk composite ($t_1$); all models control for the pre-treatment composite ($t_0$). $a$-path: $t_1 \sim$ DD $+ t_0$; $b$-path and direct effect: follow-up outcome $\sim$ DD $+ t_1 + t_0$; total effect: follow-up outcome $\sim$ DD $+ t_0$. Path coefficients, direct and total effects with HC2 SEs; indirect effect $= a \times b$ with a percentile bootstrap 95\% CI (5,000 resamples, seed 20260901); bootstrap CIs for the direct and total effects are nearly identical to the analytic ones and are stored in \texttt{analysis/r2\_mediation.json}. Rows use the returnees with $t_0$, $t_1$, and the follow-up outcome observed, so the total effects are estimated on 851 (ITT) or 845 (per-protocol) rows and differ slightly from the full-cohort ANCOVAs of Table SM12.1. Mean immediate change in the composite, ITT cohort: $-7.55$ points (DD) versus $-1.31$ points (controls). The indirect effect is identified under the same assumption as the persistence slope (no unmeasured confounding of the $t_1 \to$ follow-up relation given $t_0$ and assignment) and is observational in character; the proportion mediated is not reported because the total effects are near zero.\par}
\medskip

\end{document}